\documentclass[aip,jcp,nobibnotes,twocolumn,floatfix,reprint]{revtex4-1}%
\usepackage{amsmath}
\usepackage{graphicx}
\usepackage{amsfonts}
\usepackage{amssymb}
\usepackage{color}
\usepackage{comment}%
\providecommand{\U}[1]{\protect\rule{.1in}{.1in}}
\begin{document}
\title{Analysis of quantum mechanics with real-valued Schr\"{o}dinger equation,
single-event quantum-path dynamics, Mauprtuis path in parameter space, and
branching paths beyond semiclassics}
\author{Kazuo Takatsuka}
\thanks{kaztak@fukui.kyoto-u.ac.jp}
\affiliation{Fukui Institute for Fundamental Chemistry, Kyoto University, 606-8103 Kyoto, Japan.}

\begin{abstract}
We analyze the Schr\"{o}dinger dynamics and the Schr\"{o}dinger function (or
the so-called wavefunction), from the viewpoint of conflict and compatibility
between a distribution function and dynamical paths. The following four
analyses will be made in due order. (1) The Schr\"{o}dinger equation is
reconstructed from scratch in the real field only, without referring to
Newtonian mechanics nor optics. Only the very simple conditions such as the
space-time translational symmetry and the conservation of flux and energy are
imposed on the factorization of the density distribution in configuration
space. The real-valued Schr\"{o}dinger function attained is naturally
interpreted as a coherent distribution function of a two-dimensional vector.
On returning to the original Schr\"{o}dinger equation, it is readily
understood why and how the imaginary number should arise in the very
fundamental equation, and we directly confirm that the Schr\"{o}dinger
equation is physically consistent. (2) The Langevin dynamics gives a path of a
Brownian particle in contrast to the Gaussian distribution function subject to
the diffusion equation. Likewise in quantum dynamics, we formulate a
single-event path dynamics, contrasting with the Schr\"{o}dinger distribution
function. The path thus attained is referred to as one-world path (or quantum stochastic path), which
represents, for instance, a path of a singly launched electron in the
double-slit experiment that leaves a spot at the measurement board, while many
of accumulated spots give rise to the fringe pattern. To formulate the one-world path, we start from
the Feynman-Kac formula to draw a relation between a stochastic dynamics and
the parabolic differential equations, to one of which\ the Schr\"{o}dinger
equation is transformed. We thus extract the basic stochastic dynamics for
individual single-event (one-world) dynamics. (3) To highlight the roles of
the flux and energy conservation as the central pillars for the
Schr\"{o}dinger dynamics, we next study a path dynamics in an infinite
dimensional parameter space. We find that the quantum Maupertuis-Hamilton
principle reveals manifolds of symplectic structure in the parameter space on
which the energy and flux to be conserved. The classical-trajectory-like paths
are driven by the coupled ordinary differential equations in a parameter
space, which is actually equivalent to propagating the Schr\"{o}dinger
function in space-time. (4) A detailed physical analysis over the space-time
propagation of the Schr\"{o}dinger function is made from the viewpoint of the
ultimate roles of classical paths. It is well known that the standard
semiclassical theory can be very well represented in terms of classical
trajectories. Yet, we show that such trajectory components are demanded to
branch into many coherent pieces beyond the semiclassical regime and dissolve
into the deep dynamics of genuine full quantum dynamics. We track how the
inherent quantum nature like the Huygens-principle-like properties is built.

\end{abstract}
\date{\today}
\maketitle

%\tableofcontents

\color{blue}\color{black}

\section{Introduction\label{sec:Introduction}}

Towards a deeper understanding about quantum
mechanics,\cite{dirac1981principles, schwinger2001quantum, Bohm-text, Messiah,
Schiff, Feynman-Hibbs,ruetsche2011interpreting,omnes1992consistent} we discuss
the physical meaning of the Schr\"{o}dinger dynamics and the Schr\"{o}dinger
function (we avoid the use of the common word \textquotedblleft
wavefunction\textquotedblright\ in this particular article), in the context of
relationship between the dynamics of a space-time distribution function and
deterministic paths.

Despite the great success of quantum mechanics, many interpretations have been
proposed without converging to a consensus about the genuine nature of quantum
mechanics.\cite{freire2022oxford} In particular, the understanding of the physical meaning of
the Schr\"{o}dinger equation is at the center of controversy. Well-known among
others are the Copenhagen interpretation, Einstein-Bohr arguments, de
Broglie--Bohm idea that the Schr\"{o}dinger function serves as a pilot wave to
guide particle paths, many-world interpretation, the various philosophical
interpretations and epistemology about the quantum world such as the problems and issues of the Schr\"{o}dinger cat, quantum
nonlocality, hidden variables, Bell's inequality, and so on. Besides, some
of are now in the stage of experimental
examinations.\cite{auletta2001foundations,home2007einstein,whitaker2012new,freire2022oxford,schirber2022nobel}
In addition, so many mathematical theories about the algebraic structure of
operators and the Hilbert space theories have been
published.\cite{von2018mathematical,dirac1981principles,schwinger2001quantum,heinosaari2011mathematical}
Also, the so-called measurement
theory\cite{schlosshauer2004decoherence,wilde2013quantum,jacobs2014quantum,schlosshauer2019quantum,schlosshauer2022decoherence}
and quantum information
theory\cite{jaeger2009entanglement,schumacher2010quantum} are among the most
exciting subjects in current quantum physics.

Yet, the present article does not intend to make a comprehensive review of the
theories and ideas so far proposed, and these topics are not even a main issue
of the present article. Rather, the
present work is composed of new outcomes from basic, general, and sometimes
practical questions about the intrinsic nature of quantum dynamics raised by a
scientist who has long been involved in application studies of the
Schr\"{o}dinger dynamics. In place of the philosophical arguments like
ontology versus epistemology about \textquotedblleft quantum
substances\textquotedblright, every claims in this article is backed with
explicit mathematical expressions. Those questions include:

i) What is the physical foundation of the Schr\"{o}dinger equation to begin
with. How is it derived? Does it really need the help of classical
mechanics?\cite{yourgrau2012variational,auletta2001foundations,freire2022oxford}
Why is this fundamental equation complex-valued\ by the presence of
$i=\sqrt{-1}$? What is the physical meaning of the Schr\"{o}dinger spatially
broadening function after all?

ii) It is well known in the double slit experiment\cite{SCDiffractionl} that
the so-called interference pattern is to be formed after many particles like
electrons launched one by one. However, there is no way to predict which one
of slits each particle passes through, let alone to determine a path for each
particle to run on. Does a quantum mechanical equation of motion exist for
such a single-event (one-world) path? If any, what is the relation to the
Schr\"{o}dinger equation?

iii) Theoretical foundation of classical mechanics begins with the Maupertuis
variational principle\cite{yourgrau2012variational} and is completed by the
symplectic structure.\cite{Arnold} Have the Schr\"{o}dinger distribution
function lost such a beautiful structure by abandoning the dynamical path formalism?

iv) Semiclassical mechanics is indeed a powerful approximation to quantum
mechanics.\cite{U-Miller1,U-Miller2,berry1972semiclassical,Smilansky,ChildBook91,Lasser,de2016principles,Brack}
It is designed to take as much account of quantum effect as possible along the
classical trajectories (\textquotedblleft quantum flesh on classical
bones\textquotedblright\ due to Berry and Mount\cite{berry1972semiclassical}).
Then, what makes the ultimate difference between quantum and classical
mechanics? What is necessary for us to do to go beyond semiclassics and
converge to quantum mechanics? How and when can the quantum wave-like nature
as in the Huygens principle emerge?

Through all the above questions, we see the mathematical conflict or
incompatibility lying between the Schr\"{o}dinger function as a spatially
broad solution of partial differential equation and path objects (or
trajectories) to represent ray solutions of well-posed ordinary differential
equations. As for the path representation in quantum mechanics, Feynman's path
integral is best known,\cite{Feynman,Feynman-Hibbs,Schulman,kleinert2006path}
which is designed to represent the kernel (the Green function of the
Schr\"{o}dinger operator) as a democratic sum of infinite number of possible
paths. However it does not care about the dynamics of each path, and the
problem about missing of the integral measure is not yet fully
resolved.\cite{davison1954feynmann,Klauder2} In the studies of heavy particle
dynamics like chemical reactions, semiclassical mechanics is frequently
adopted with use of classical
trajectories.\cite{U-Miller1,U-Miller2,berry1972semiclassical,Smilansky,ChildBook91,Lasser,de2016principles,Brack}
The quantum trajectory method based on the Bohm representation\cite{Bohm-text}
of the Schr\"{o}dinger equation are now adopted as a practical tool by Wyatt
and his
colleague.\cite{lopreore1999quantum,wyatt1999quantum,Wyatttext,sanz2013trajectory}
Thus the path concept is seen to be essentially important not only for
intuitive understanding of the essential nature of quantum mechanics but in
possible practical applications. 

The goal of this article is to analyze the Schr\"{o}dinger equation and
Schr\"{o}dinger function from the view point of space-time distribution and
path dynamics. Here in this article, we study the Schr\"{o}dinger equation and
its function in the following four stages.

1) (in Sec. \ref{sec:RealValue}) Derivation (reconstruction) of the
Schr\"{o}dinger equation from scratch in the real-valued space without
referring to classical mechanics nor optics: Since quantum mechanics is in the
deeper level than classical mechanics, the former should exist without help of the latter such as the Hamilton-Jacobi
equation. Through the present construction, we firmly identify the physical
meaning of the Schr\"{o}dinger function as a quantum distribution function.
The validity and theoretical limitations of the Schr\"{o}dinger dynamics is
also mentioned to.

2) (in Sec. \ref{sec:Stochastic}) The dynamics of a stochastic path (referred
to as one-world path) to figure out a single event quantum dynamics, in
contrast to the coherent superposition of \textquotedblleft
many-world\ events\textquotedblright\ by a Schr\"{o}dinger function: We start
a single event dynamics from the Feyman-Kac formula. As in the relationship
between a Brownian path subjected to the Langevin equation and the Gaussian
distribution function to the diffusion equation, we discuss the one-world path
versus the Schr\"{o}dinger distribution function.

3) (in Sec. \ref{sec:Variational}) Paths on symplectic manifolds to propagate
the Schr\"{o}dinger function: It is well known that classical mechanics
emerges from quantum mechanics in the small Planck constant limit. This
implies that there can remain a mathematical structure common both to quantum
and classical mechanics in their background. We show that just as in classical
mechanics, the quantum Maupertuis-Hamilton variational principle\cite{Arnold}
appears to drive the paths on a manifold of symplectic structure in a
parameter space and thereby time-propagates the Schr\"{o}dinger functions.

4) (in Sec. \ref{sec:ADFpath}) Path dynamics in the realm beyond semiclassical
mechanics to see the ultimate difference between quantum and classical
mechanics: In the real time propagation of the Schr\"{o}dinger function in
configuration space, we scrutinize the ultimate roles of classical paths. If
the trajectories survive after all up to the final stage of quantum theory, the interference
pattern\ in the double-slit experiment will fail to appear. We show that as
the quantum nature deepens, the \textquotedblleft deterministic
paths\textquotedblright\ are replaced by a sum of countless branching coherent
non-classical paths. We thus see the classical paths be ultimately dissolved
into the deep sea of the Schr\"{o}dinger functions and thereby eventually
terminate the series of their roles in quantum dynamics.

Of the four pillars mentioned above, the first and the second ones appear new
in this article, whereas the third and fourth are ones recaptured under a
consistent view from the author's previously developed theories.

\section{Schr\"{o}dinger equation without the imaginary number: Minimal
quantum mechanics from scratch\label{sec:RealValue}}

\subsection{Introductory remarks}

It is widely
documented\cite{yourgrau2012variational,auletta2001foundations,home2007einstein,whitaker2012new}
that Schr\"{o}dinger assumed his function $\psi(q,t)$ in the form $W=K\log
\psi(q,t)$, and substituted it into the Hamilton-Jacobi equation and then
applied a variational method to arrive at the stationary-state Schr\"{o}dinger
equation $\hat{H}\tilde{\psi}(q)=E\tilde{\psi}(q)$. Later he noticed the
correspondence $\hat{p}\leftrightarrow\hbar/i\nabla$ and $\hat{H}%
\leftrightarrow i\hbar\partial/\partial t$, thereby \textquotedblleft
creating\textquotedblright\ the time-dependent Schr\"{o}dinger equation. The
constant $K$ is afterward identified to be the Planck constant $\hbar$ by
comparison with the experimental discoveries like the Compton
effect\cite{Messiah}. The similarity of the Schr\"{o}dinger equation with the
geometrical optics has been also widely pointed
out.\cite{simeonov2024derivation} The \textquotedblleft
derivation\textquotedblright\ led Schr\"{o}dinger himself to an interpretation
that $\psi(q,t)$ should represent a physical wave (packet). However, this view
was quickly refuted by Lorentz and later by Schr\"{o}dinger
himself.\cite{freire2022oxford} The notion of the physical wave is not
consistent with the treatment of electrons as point masses in the
Schr\"{o}dinger equation to begin with.

Instead of tracking the historical progress of quantum mechanics,
\cite{auletta2001foundations,home2007einstein,whitaker2012new,freire2022oxford}
we here only attempt to derive the real-valued Schr\"{o}dinger equation in a
simple manner without any reference to classical mechanics nor optics. In
doing so, we impose only a small number of simple and general conditions with
respect to the space-time translational symmetry and a conservation law on a
real-valued function, which is actually a two component vector function. Then
it is equivalently transformed to the canonical Schr\"{o}dinger equation with
the imaginary unit $i$ in it, confirming that the Schr\"{o}dinger equation is
indeed free of theoretical inconsistency. We discuss the physical meaning of
the Schr\"{o}dinger function on the basis of the present derivation.

\subsection{Real-valued Schr\"{o}dinger equation from scratch}

\subsubsection{From phase space to configuration space\label{Sec:ConfigSpace}}

We start with $\rho(q,t)$, a density function of particles in configuration
space $q$ in a quantum level. First, we revisit the physical interpretation of
its classical counterpart. The classical Liouville phase-space distribution
function $\Gamma(q,p,t)$ can be represented in classical trajectories (ray
solutions), each never mutually crossing in $(q,p,t)$-space. It represents an
ensemble of independent dynamics of different initial conditions. It is not
necessarily a probability distribution function, unless a probabilistic
framework is additionally introduced. As a simple example of classical
dynamics, we consider an ensemble of free particles on a zero (or constant)
potential function $V(q)$. The Liouville distribution function is assumed to
be
\begin{equation}
\Gamma(q,p,0)=1.0\text{ in }q\in\left[  q_{1}(t),q_{2}(t)\right]  \text{ and
}p\in\left[  p_{0}-\varepsilon,p_{0}+\varepsilon\right]  \text{,}
\label{Constant}%
\end{equation}
with $\varepsilon\rightarrow0$. Remember that this distribution function
represents an ensemble of mutually non-interacting independent particle
dynamics in a free space, all with the same momentum $p_{0}$.

The $q$-$p$ uncertainty relation inherent to quantum mechanics \cite{dirac1981principles, schwinger2001quantum, Bohm-text, Messiah,
Schiff, Feynman-Hibbs,ruetsche2011interpreting,omnes1992consistent} (and we will discuss this aspect later in Eq. (\ref{nop}) and the relevant part below it) basically
requires the classical phase-space view of particle dynamics to be abandoned.
\begin{comment}
We also discuss in Section \ref{Sec. Wiener} that the stochasticity of the
quantum paths leads to a divergence in the classical definition of velocity,
that is, $\lim_{\Delta q\rightarrow0,\Delta t\rightarrow0}\Delta q/\Delta t$
does not necessarily exist and $q(t)$ is not differentiable almost everywhere.
\end{comment}
We are hence forced to give up the phase-space description, (we are not
talking about the Wigner phase-space
representation\cite{wigner1932quantum,zachos2005quantum} in this stage).
However, the configuration space distribution function $\rho(q,t)$ reduced by%

\begin{equation}
\rho(q,t)=%
%TCIMACRO{\dint }%
%BeginExpansion
{\displaystyle\int}
%EndExpansion
dp\Gamma(q,p,t) \label{rho}%
\end{equation}
remains as a meaningful quantity as an observable. It is natural to suppose
that $\rho(q,t)$ induces an incompressible flow in $q$-space due to the
particle number conservation, and $\partial\rho(q,t)/\partial t$ gives rise to
an equation of continuity, which is one of the fundamental requirements that
$\rho(q,t)$ should fulfill.

Incidentally, the Wigner phase-space distribution function $W(q,p,t)$%
,\cite{wigner1932quantum} which is a quantity nonlinearly transformed from the
Schr\"{o}dinger function, also gives%
\begin{equation}
\rho(q,t)=%
%TCIMACRO{\dint }%
%BeginExpansion
{\displaystyle\int}
%EndExpansion
dpW(q,p,t) \label{Wigq}%
\end{equation}
and%
\begin{equation}
\tilde{\rho}(p,t)=%
%TCIMACRO{\dint }%
%BeginExpansion
{\displaystyle\int}
%EndExpansion
dqW(q,p,t). \label{Wigp}%
\end{equation}
Equation (\ref{Wigq}) suggests that $\rho(q,t)$ be regarded as a quantity to
remain meaningful both in classical and quantum mechanics.

\subsubsection{The Schr\"{o}dinger vectors
\label{subsec:factoraliztion}}

Suppose the presence of a distribution $\rho(q,t)$ at hand. We then factorize
it in what we refer to as the Schr\"{o}dinger vectors as%

\begin{align}
\rho(q,t)  &  =\left(
\begin{array}
[c]{cc}%
\phi_{r}(q,t) & \phi_{c}(q,t)
\end{array}
\right)  \left(
\begin{array}
[c]{c}%
\phi_{r}(q,t)\\
\phi_{c}(q,t)
\end{array}
\right) \nonumber\\
&  =\bar{\psi}(q,t)^{T}\bar{\psi}(q,t) \label{fact}%
\end{align}
with%
\begin{equation}
\bar{\psi}(q,t)\equiv\left(
\begin{array}
[c]{c}%
\phi_{r}(q,t)\\
\phi_{c}(q,t)
\end{array}
\right)  . \label{fact2}%
\end{equation}
Since $\rho(q,t)$ is positive semidefinite everywhere, we may choose both
$\phi_{r}(q,t)$ and $\phi_{c}(q,t)$ to be real-valued functions and they can
be negative. Incidentally, the configuration space distribution $\rho(q,t)dq$
in a small volume element is also factored to%
\begin{equation}
\rho(q,t)dq=\left[  \bar{\psi}(q,t)^{T}dq^{1/2}\right]  \left[  \bar{\psi
}(q,t)dq^{1/2}\right]
\end{equation}
with the half density $dq^{1/2}$, the implication of which will be discussed
in the context of semiclassical mechanics later in Sec. \ref{sec:Rescaling}.
It would be appropriate to refer to $\bar{\psi}(q,t)$ as a
distribution amplitude function. Thus, after we have lost the classical momentum
coordinates $p$ as the independent degrees of freedom, the corresponding
dimensionality is retrieved in the vector space as in Eq. (\ref{fact2}).

Incidentally, the dimensionality of the factorizing vectors $\bar{\psi}(q,t)$
is not mathematically limited to two. However, it will actually turn out later
that the very basic Schr\"{o}dinger dynamics emerges from this frame.

\subsubsection{A simple example}

Back to the trivial example of Eq. (\ref{Constant}) we may consider an almost
\textquotedblleft pure distribution\textquotedblright\
\begin{equation}
\rho(q,t)=1.0\text{ in }q\in\left[  q_{1}(t),q_{2}(t)\right]  , \label{exrho}%
\end{equation}
after the $p$\ integration in Eq. (\ref{rho}). The intrinsic information about
the particle velocity is therefore formally lost. It is obvious that there can
exist many sets of factorization for any given $\rho(q,t)$. Here in this
example, it is natural to chose $\bar{\psi}(q,t)$ as
\begin{equation}
\phi_{r}(q,t)=\cos(\frac{p_{0}}{K}q+\chi)\text{ \ \ and \ }\phi_{c}%
(q,t)=\sin(\frac{p_{0}}{K}q+\chi), \label{pln}%
\end{equation}
in $q\in\left[  q_{1}(t),q_{2}(t)\right]  $ and zero otherwise, since
$\phi_{r}^{2}(q,t)+$ $\phi_{c}^{2}(q,t)=\rho(q,t)=1.0$. $K$ is a constant
having the physical dimension of action, and $\chi$ is an arbitrary constant.
$K$ will be determined later.

\subsubsection{Basic conditions to establish the dynamics of $\bar{\psi}%
(q,t)$}

To capture the dynamics of $\bar{\psi}(q,t)$ in the real-value field, we
impose the following basic and universal constraints on $\rho(q,t)$ of Eq.
(\ref{fact}).

1) The uniformity of the Euclidean space with respect to translational invariance.

2) Time translational and reversal symmetry and the energy conservation. The
Lorentz invariance is not imposed.

3) Conservation of the number of particles. The creation-annihilation process
in vacuum is out of the scope.

\subsection{Translational symmetry of free configuration space}

We first consider a primitive translational symmetry of $\bar{\psi}(q,t)$ in
the free space by shifting the coordinate $q\rightarrow q+\Delta q$%
,$\,\ $which results in%
\begin{equation}
\bar{\psi}(q,t)\rightarrow\tilde{\psi}(q+\Delta q,t).
\end{equation}
It is possible for $\tilde{\psi}(q+\Delta q,t)$ to take a vector rotation in
addition to the space shifting in such a way that%

\begin{align}
\tilde{\psi}(q+\Delta q,t)  &  =\exp\left(  \Delta q\text{\textbf{A}}%
\vec{\nabla}_{q}\right)  \bar{\psi}(q,t)\nonumber\\
&  =\left(  \text{\textbf{I}}+\Delta q\text{\textbf{A}}\vec{\nabla}%
_{q}\right)  \left(
\begin{array}
[c]{c}%
\phi_{r}(q,t)\\
\phi_{c}(q,t)
\end{array}
\right)  +\left(  \text{higher}\right)  ,
\end{align}
where \textbf{A} is a $2\times2$ real matrix, and%
\begin{equation}
\tilde{\psi}(q+\Delta q,t)^{T}=\left(
\begin{array}
[c]{cc}%
\phi_{r}(q,t) & \phi_{c}(q,t)
\end{array}
\right)  \exp\left(  \Delta q\text{\textbf{A}}^{T}\vec{\nabla}_{q}\right)  .
\end{equation}
The vector rotation matrix \textbf{A }is included, since the vector has a
freedom of rotation even in the simple spatial translation. Then the identity%

\begin{equation}
\tilde{\psi}(q+\Delta q,t)^{T}\tilde{\psi}(q+\Delta q,t)=\bar{\psi}%
(q,t)^{T}\bar{\psi}(q,t)
\end{equation}
demands in the first order of $\Delta q$ that%

\begin{align}
&  \tilde{\psi}(q+\Delta q,t)^{T}\tilde{\psi}(q+\Delta q,t)-\bar{\psi
}(q,t)^{T}\bar{\psi}(q,t)\nonumber\\
&  =\left(
\begin{array}
[c]{cc}%
\phi_{r}(q,t) & \phi_{c}(q,t)
\end{array}
\right)  \left[  \text{\textbf{A}}^{T}\vec{\nabla}_{q}+\text{\textbf{A}}%
\vec{\nabla}_{q}\right]  \left(
\begin{array}
[c]{c}%
\phi_{r}(q,t)\\
\phi_{c}(q,t)
\end{array}
\right)  \Delta q. \label{q-trans}%
\end{align}
The invariance requires%

\begin{equation}
\text{\textbf{A}}^{T}=-\text{\textbf{A,}}%
\end{equation}
and therefore we may set%
\begin{equation}
\text{\textbf{A}}=c_{p}\text{\textbf{J,}}%
\end{equation}
where \textbf{J} is a 2$\times$2 unit symplectic matrix (or called the
standard symplectic matrix)\cite{Arnold,de2006symplectic} defined as%
\begin{equation}
\text{\textbf{J}}\mathbf{=}\left(
\begin{array}
[c]{cc}%
0 & -1\\
1 & 0
\end{array}
\right)  .
\end{equation}
The basic properties are%

\begin{equation}
\text{\textbf{J}}^{2}=-\text{\textbf{I,}} \label{JJ}%
\end{equation}
and
\begin{equation}
\text{\textbf{J}}^{-1}=-\text{\textbf{J}}=\text{\textbf{J}}^{T}.
\end{equation}
The constant $c_{p}$ is to be determined later.

According to the spirit of Noether theorem\cite{Arnold} and more precisely to
the discussion about the displacement operator and linear momentum by
Dirac,\cite{dirac1981principles} we may assign the momentum operator as%

\begin{equation}
\hat{p}=c_{p}\text{\textbf{J}}\vec{\nabla}_{q}. \label{p-op}%
\end{equation}
Equation (\ref{q-trans}) suggests that $\hat{p}$ should be operated in such a
manner that%
\begin{equation}
\left(
\begin{array}
[c]{cc}%
\phi_{r}(q,t) & \phi_{c}(q,t)
\end{array}
\right)  \hat{p}\left(
\begin{array}
[c]{c}%
\phi_{r}(q,t)\\
\phi_{c}(q,t)
\end{array}
\right)  .
\end{equation}
The constant $c_{p}$ is to be determined using the example presented back in
Eq. (\ref{pln}). The sine-cosine factorization along with the definition Eq.
(\ref{p-op}) gives%

\begin{align}
&  \left(
\begin{array}
[c]{cc}%
\phi_{r}(q,t) & \phi_{c}(q,t)
\end{array}
\right)  \hat{p}\left(
\begin{array}
[c]{c}%
\phi_{r}(q,t)\\
\phi_{c}(q,t)
\end{array}
\right) \nonumber\\
&  =-\frac{p_{0}}{K}c_{p}. \label{plane}%
\end{align}
It is natural that this value is made equivalent to $p_{0}$, that is,
\begin{equation}
-\frac{p_{0}}{K}c_{p}=p_{0}%
\end{equation}
and therefore
\begin{equation}
c_{p}=-K.
\end{equation}
Further, the quantum experiments like the Compton effect\cite{Messiah} demands
that
\begin{equation}
K=\hbar
\end{equation}
and we have%
\begin{equation}
\hat{p}=-\hbar\text{\textbf{J}}\nabla. \label{momentum}%
\end{equation}

\subsection{Translational symmetry in time\label{subsec:variation}}

As for the translational symmetry in time%

\begin{equation}
\tilde{\psi}(q,t+\Delta t)^{T}\tilde{\psi}(q,t+\Delta t)=\bar{\psi}%
(q,t)^{T}\bar{\psi}(q,t),
\end{equation}
the similar procedure in Eq. (\ref{q-trans}) to Eq. (\ref{p-op}) is applied
and after all gives a quantum Hamiltonian $\hat{H}$ in the form%
\begin{equation}
\hat{H}=c_{t}\text{\textbf{J}}\frac{\partial}{\partial t}\text{\textbf{,}}
\label{ct}%
\end{equation}
where $\hat{H}$ is to be operated as in%
\begin{equation}
\left(
\begin{array}
[c]{cc}%
\phi_{r} & \phi_{c}%
\end{array}
\right)  \left(  c_{t}\text{\textbf{J}}\frac{\partial}{\partial t}\right)
\left(
\begin{array}
[c]{c}%
\phi_{r}\\
\phi_{c}%
\end{array}
\right)  =\left(
\begin{array}
[c]{cc}%
\phi_{r} & \phi_{c}%
\end{array}
\right)  \hat{H}\left(
\begin{array}
[c]{c}%
\phi_{r}\\
\phi_{c}%
\end{array}
\right)  . \label{ct2}%
\end{equation}
The explicit form of $\hat{H}$ will be determined later along with the
constant value of $c_{t}$.

\subsection{The equation of continuity for $\rho(q,t)$}

We next scrutinize the direct consequence of incompressibility of $\rho(q,t)$
in terms of the equation of continuity. Consider an analogy to classical
incompressible flow%
\begin{equation}
\frac{\partial}{\partial t}\rho(q,t)=-\nabla\cdot\vec{j}(q,t)
\label{Econtinuity}%
\end{equation}
with the flux being defined%
\begin{equation}
\vec{j}=\vec{v}\rho,
\end{equation}
with the local velocity $\vec{v}$. Our quantum flux starts based on the
factorization of $\rho$ should be defined as%
\begin{align}
\vec{j}  &  =\left(
\begin{array}
[c]{cc}%
\phi_{r}(q,t) & \phi_{c}(q,t)
\end{array}
\right)  \frac{\hat{p}}{m}\left(
\begin{array}
[c]{c}%
\phi_{r}(q,t)\\
\phi_{c}(q,t)
\end{array}
\right) \nonumber\\
&  =-\frac{\hbar}{m}\left(
\begin{array}
[c]{cc}%
\phi_{r}(q,t) & \phi_{c}(q,t)
\end{array}
\right)  \text{\textbf{J}}\left(
\begin{array}
[c]{c}%
\vec{\nabla}\phi_{r}(q,t)\\
\vec{\nabla}\phi_{c}(q,t)
\end{array}
\right) \nonumber\\
&  =\frac{\hbar}{m}\left(  \phi_{r}\vec{\nabla}\phi_{c}-\phi_{c}\vec{\nabla
}\phi_{r}\right)  \label{primeflux}%
\end{align}
with the help of Eq. (\ref{momentum}). The continuity equation in this context
is%
\begin{align}
\frac{\partial}{\partial t}\rho &  =2\left(
\begin{array}
[c]{cc}%
\phi_{r} & \phi_{c}%
\end{array}
\right)  \frac{\partial}{\partial t}\left(
\begin{array}
[c]{c}%
\phi_{r}\\
\phi_{c}%
\end{array}
\right) \nonumber\\
&  =-\vec{\nabla}\cdot\vec{j},
\end{align}
which is also referred to as the law of flux conservation. Combining all these
with $\vec{j}$ of Eq. (\ref{primeflux}), we have%
\begin{align}
&  \left(
\begin{array}
[c]{cc}%
\phi_{r} & \phi_{c}%
\end{array}
\right)  \frac{\partial}{\partial t}\left(
\begin{array}
[c]{c}%
\phi_{r}\\
\phi_{c}%
\end{array}
\right) \nonumber\\
&  =-\frac{\hbar}{2m}\vec{\nabla}\cdot\left(  \phi_{r}\vec{\nabla}\phi
_{c}-\phi_{c}\vec{\nabla}\phi_{r}\right) \nonumber\\
&  =-\frac{1}{\hbar}\left(
\begin{array}
[c]{cc}%
\phi_{r} & \phi_{c}%
\end{array}
\right)  \text{\textbf{J}}\left(  -\frac{\hbar^{2}}{2m}\nabla^{2}\right)
\left(
\begin{array}
[c]{c}%
\phi_{r}\\
\phi_{c}%
\end{array}
\right) \nonumber\\
&  =\left(
\begin{array}
[c]{cc}%
\phi_{r} & \phi_{c}%
\end{array}
\right)  \left(  -\text{\textbf{J}}\frac{\hat{p}^{2}}{2m\hbar}\right)  \left(
\begin{array}
[c]{c}%
\phi_{r}\\
\phi_{c}%
\end{array}
\right)  , \label{flux0}%
\end{align}
where we have used%
\begin{equation}
\hat{p}^{2}=-\hbar^{2}\text{\textbf{J}}^{2}\nabla^{2}=\hbar^{2}\nabla
^{2}\text{\textbf{I}}.
\end{equation}
\newline

\subsection{Variational principle leading to the real-valued Schr\"{o}dinger
equation}

Suppose we have a scalar function $f(q)$. Back in Eq. (\ref{flux0}), we notice
that the presence of $f(q)$\textbf{J} does not affect it%
\begin{align}
&  \left(
\begin{array}
[c]{cc}%
\phi_{r} & \phi_{c}%
\end{array}
\right)  \left(  \hbar\frac{\partial}{\partial t}\right)  \left(
\begin{array}
[c]{c}%
\phi_{r}\\
\phi_{c}%
\end{array}
\right) \nonumber\\
&  =\left(
\begin{array}
[c]{cc}%
\phi_{r} & \phi_{c}%
\end{array}
\right)  \left(  -\text{\textbf{J}}\frac{\hat{p}^{2}}{2m}+\text{\textbf{J}%
}f(q)\right)  \left(
\begin{array}
[c]{c}%
\phi_{r}\\
\phi_{c}%
\end{array}
\right)  , \label{Sch1}%
\end{align}
due to the skew orthogonality%
\begin{align}
&  \left(
\begin{array}
[c]{cc}%
\phi_{r} & \phi_{c}%
\end{array}
\right)  \left(  \text{\textbf{J}}f(q)\right)  \left(
\begin{array}
[c]{c}%
\phi_{r}\\
\phi_{c}%
\end{array}
\right) \nonumber\\
&  =f(q)\left(
\begin{array}
[c]{cc}%
\phi_{r} & \phi_{c}%
\end{array}
\right)  \text{\textbf{J}}\left(
\begin{array}
[c]{c}%
\phi_{r}\\
\phi_{c}%
\end{array}
\right)  =0
\end{align}
for any scalar function $f(q)$. It is obvious by construction that $f(q)$ can
be time dependent, namely $f(q,t)$. Therefore the variational condition is
extended such that%
\begin{align}
&  \left(
\begin{array}
[c]{cc}%
\delta\phi_{r} & \delta\phi_{c}%
\end{array}
\right)  \left(  \hbar\frac{\partial}{\partial t}\right)  \left(
\begin{array}
[c]{c}%
\phi_{r}\\
\phi_{c}%
\end{array}
\right) \nonumber\\
&  =\left(
\begin{array}
[c]{cc}%
\delta\phi_{r} & \delta\phi_{c}%
\end{array}
\right)  \left(  -\text{\textbf{J}}\frac{\hat{p}^{2}}{2m}+\text{\textbf{J}%
}f(q,t)\right)  \left(
\begin{array}
[c]{c}%
\phi_{r}\\
\phi_{c}%
\end{array}
\right)  , \label{var}%
\end{align}
which imposes an external condition depending on the choice of $f(q,t)$. A
natural choice is
\begin{equation}
f(q,t)=-V(q,t) \label{fchoice}%
\end{equation}
with $V(q,t)$ being a relevant system potential function. Thus the variational
principle demands the following expression
\begin{equation}
\hbar\text{\textbf{J}}\frac{\partial}{\partial t}\left(
\begin{array}
[c]{c}%
\phi_{r}\\
\phi_{c}%
\end{array}
\right)  =\left(  \frac{\hat{p}^{2}}{2m}+V(q,t)\right)  \left(
\begin{array}
[c]{c}%
\phi_{r}\\
\phi_{c}%
\end{array}
\right)  , \label{Sch2}%
\end{equation}
and the quantum mechanical Hamiltonian is defined as%
\begin{equation}
\hat{H}=\frac{\hat{p}^{2}}{2m}+V(q,t),
\end{equation}
and thereby Eq. (\ref{Sch2}) can be rewritten as%
\begin{equation}
\hbar\text{\textbf{J}}\frac{\partial}{\partial t}\left(
\begin{array}
[c]{c}%
\phi_{r}\\
\phi_{c}%
\end{array}
\right)  =\hat{H}\left(
\begin{array}
[c]{c}%
\phi_{r}\\
\phi_{c}%
\end{array}
\right)  , \label{Sch3}%
\end{equation}
or
\begin{equation}
\hbar\text{\textbf{J}}\frac{d}{dt}\left(
\begin{array}
[c]{c}%
\phi_{r}\\
\phi_{c}%
\end{array}
\right)  =\left(  -\frac{\hbar^{2}}{2m}\nabla^{2}+V\right)  \left(
\begin{array}
[c]{c}%
\phi_{r}\\
\phi_{c}%
\end{array}
\right)  . \label{Sch30}%
\end{equation}
By comparing this expression with Eq. (\ref{ct}), we find%
\begin{equation}
c_{t}=1.
\end{equation}

The Schr\"{o}dinger equation in the present representation is summarized as
follows.%
\begin{equation}
\hat{H}=\text{\textbf{J}}\hbar\frac{\partial}{\partial t} \label{H}%
\end{equation}
and
\begin{equation}
\hat{p}=-\text{\textbf{J}}\hbar\vec{\nabla} \label{p}%
\end{equation}
for the real-valued vector%
\begin{equation}
\bar{\psi}(q,t)=\left(
\begin{array}
[c]{c}%
\phi_{r}(q,t)\\
\phi_{c}(q,t)
\end{array}
\right)  ,
\end{equation}
which may be referred to as the Schr\"{o}dinger vector, and should be
distinguished from the Dirac state vector $\left\vert \Psi\right\rangle
$.\cite{dirac1981principles}

\bigskip Incidentally, we confirm that
\begin{equation}
\left(
\begin{array}
[c]{cc}%
\delta\phi_{r} & \delta\phi_{c}%
\end{array}
\right)  \left[  \left(  \hbar\frac{\partial}{\partial t}\right)  -\left(
\text{\textbf{J}}\frac{\hat{p}^{2}}{2m}+\text{\textbf{J}}V\right)  \right]
\left(
\begin{array}
[c]{c}%
\phi_{r}\\
\phi_{c}%
\end{array}
\right)  =0 \label{v1}%
\end{equation}
is never equivalent to
\begin{equation}
\left(
\begin{array}
[c]{cc}%
\phi_{r} & \phi_{c}%
\end{array}
\right)  \left[  \left(  \hbar\frac{\partial}{\partial t}\right)  -\left(
\text{\textbf{J}}\frac{\hat{p}^{2}}{2m}+\text{\textbf{J}}V\right)  \right]
\left(
\begin{array}
[c]{c}%
\phi_{r}\\
\phi_{c}%
\end{array}
\right)  =0, \label{v2}%
\end{equation}
since the latter holds irrespective of $V(q)$ as long as the flux conservation
along with the translational symmetries in space-time for $\left(
\begin{array}
[c]{cc}%
\phi_{r} & \phi_{c}%
\end{array}
\right)  ^{T}$ are satisfied, while Eq. (\ref{v1}) demands that $\left(
\begin{array}
[c]{cc}%
\phi_{r} & \phi_{c}%
\end{array}
\right)  ^{T}$ should follow a dynamics on the specified $V(q,t)$.

\subsection{The Heisenberg equation}

From Eq. (\ref{Sch3}) it follows that for an arbitrary time-independent
operator $\hat{A}$%
\begin{align}
&  \frac{\partial}{\partial t}\left(
\begin{array}
[c]{cc}%
\phi_{r}(q,t) & \phi_{c}(q,t)
\end{array}
\right)  \hat{A}\left(
\begin{array}
[c]{c}%
\phi_{r}(q,t)\\
\phi_{c}(q,t)
\end{array}
\right) \nonumber\\
&  =\frac{1}{\hbar}\left(
\begin{array}
[c]{cc}%
\phi_{r} & \phi_{c}%
\end{array}
\right)  \left(  \text{\textbf{J}}^{-1T}\hat{H}\hat{A}+\hat{A}\hat
{H}\text{\textbf{J}}^{-1}\right)  \left(
\begin{array}
[c]{c}%
\phi_{r}\\
\phi_{c}%
\end{array}
\right) \nonumber\\
&  \equiv\frac{1}{\hbar}\left(
\begin{array}
[c]{cc}%
\phi_{r} & \phi_{c}%
\end{array}
\right)  \left[  \hat{H},\hat{A}\right]  \text{\textbf{J}}\left(
\begin{array}
[c]{c}%
\phi_{r}\\
\phi_{c}%
\end{array}
\right)  \label{Heisenberg}%
\end{align}
with $\left[  \hat{H},\hat{A}\right]  =\hat{H}\hat{A}-\hat{A}\hat{H}$, where
the Hamiltonian $\hat{H}=$ $\frac{\hat{p}^{2}}{2m}+V(q)$ is assumed to be
Hermitian for square integrable (L$^{2}$) functions $\left(
\begin{array}
[c]{cc}%
\phi_{r} & \phi_{c}%
\end{array}
\right)  ^{T}.$ Equation (\ref{Heisenberg}) gives just the Heisenberg equation
of motion in the real-valued space. Putting $\hat{A}=$ $\hat{H}\,$, the
conservation of energy is seen.

\subsection{The canonical
Schr\"{o}dinger equation in the complex~number field}

The symplectic form of two component vectors in the real number field is
essentially equivalent to the complex number scalar field (see Eq.
(\ref{JJ})), under setting%

\begin{equation}
\text{\textbf{J}}\rightarrow i.
\end{equation}
Then the Hamiltonian, Eq. (\ref{H}), and the momentum operator, Eq. (\ref{p}),
respectively read%

\begin{equation}
\hat{H}=i\hbar\frac{\partial}{\partial t} \label{HH}%
\end{equation}
and
\begin{equation}
\hat{p}=-i\hbar\vec{\nabla}, \label{pp}%
\end{equation}
which are to be operated on a scalar function%
\begin{equation}
\psi(q,t)=\phi_{r}(q,t)+i\phi_{c}(q,t), \label{Sf}%
\end{equation}
with $\bar{\psi}(q,t)\leftrightarrow\psi(q,t)$. The real-valued vector
Schr\"{o}dinger equation (\ref{Sch3}) is transformed to the canonical
Schr\"{o}dinger equation%
\begin{align}
i\hbar\frac{\partial}{\partial t}\psi(q,t)  &  =\hat{H}\psi(q,t)\nonumber\\
&  =\left(  -\frac{\hbar^{2}}{2m}\nabla^{2}+V\right)  \psi(q,t). \label{Sch4}%
\end{align}
Conversely, the separation of Eq. (\ref{Sch4}) to the real and imaginary parts
gives Eq. (\ref{Sch30}).

\begin{comment}
Also from the expression of Eq. (\ref{t-reverse}), we simply have%

\begin{equation}
i\hbar\frac{\partial}{\partial t}\psi^{\ast}(q,-t)=\hat{H}\psi^{\ast}(q,-t),
\label{Sch5}%
\end{equation}
or simply%
\begin{equation}
-i\hbar\frac{\partial}{\partial s}\psi^{\ast}(q,s)=\hat{H}\psi^{\ast}(q,s).
\end{equation}
Equation (\ref{t-reverse}) is of course the time-reversal counterpart of Eq.
(\ref{Sch2}).
\end{comment}

It is obvious that the complex-valued equation of motion, Eq. (\ref{Sch4}) is
far easier to handle both mathematically and numerically. In particular, the
electron spin and the extension to the Dirac equation is virtually impossible
without the complex algebra with the higher order factorization than the inner
product of two dimensional vectors. Nevertheless, the way leading to the
formulation of the real-valued Schr\"{o}dinger equation, Eq. (\ref{Sch2}),
with use of neither classical nor optics is somewhat instructive for the
simpler understanding what the Schr\"{o}dinger functions is.

\subsection{Normalization}

\begin{comment}
The normalization of the Schr\"{o}dinger vector $\bar{\psi}(q,t)$ is rather
straightforward. It satisfies%
\begin{align}
\left(  \text{\textbf{J}}\bar{\psi}(q,t)\right)  ^{T}\bar{\psi}(q,t)  &
=\left(
\begin{array}
[c]{cc}%
-\phi_{c}(q,t) & \phi_{r}(q,t)
\end{array}
\right)  \left(
\begin{array}
[c]{c}%
\phi_{r}(q,t)\\
\phi_{c}(q,t)
\end{array}
\right) \nonumber\\
&  =0 \label{ort}%
\end{align}
and%
\begin{align}
-\left(  \text{\textbf{J}}\bar{\psi}(q,t)\right)  ^{T}\wedge\bar{\psi}(q,t)
&  =\left\vert
\begin{array}
[c]{cc}%
\phi_{c}(q,t) & -\phi_{r}(q,t)\\
\phi_{r}(q,t) & \phi_{c}(q,t)
\end{array}
\right\vert \nonumber\\
&  =\phi_{r}(q,t)^{2}+\phi_{c}(q,t)^{2}.
\end{align}
Hence, it is normalizable as%
\begin{equation}
\int dq\bar{\psi}(q,t)^{T}\wedge\left(  \text{\textbf{J}}\bar{\psi
}(q,t)\right)  =1, \label{psi-wedge}%
\end{equation}
the functional form of Eq. (\ref{psi-wedge}) reminds of the
Poincar\'{e}-Cartan theorem of integral invariance in classical
mechanics.\cite{Arnold}
\end{comment}

There are basically two ways of normalization. One is to resort to the usual
average (expectation value), for instance%
\begin{equation}
E^{\text{av}}=\left\langle \hat{H}\right\rangle =\frac{\int dq\left(
\begin{array}
[c]{cc}%
\phi_{r}(q,t) & \phi_{c}(q,t)
\end{array}
\right)  \hat{H}\left(
\begin{array}
[c]{c}%
\phi_{r}(q,t)\\
\phi_{c}(q,t)
\end{array}
\right)  }{\int\rho(q,t)dq}.
\end{equation}
The other one is the space-time distribution%
\begin{align}
E^{\text{local}}(q,t)  &  =\frac{1}{\rho}\left(
\begin{array}
[c]{cc}%
\phi_{r}(q,t) & \phi_{c}(q,t)
\end{array}
\right)  \hat{H}\left(
\begin{array}
[c]{c}%
\phi_{r}(q,t)\\
\phi_{c}(q,t)
\end{array}
\right) \nonumber\\
&  =\operatorname{Re}\frac{\hat{H}\psi}{\psi}, \label{e-density}%
\end{align}
from which the energy eigenvalue problem is already seen. Likewise the
space-time distribution of the velocity is%

\begin{align}
\vec{v}^{\text{local}}(q,t)  &  =\frac{1}{\rho}\left(
\begin{array}
[c]{cc}%
\phi_{r}(q,t) & \phi_{c}(q,t)
\end{array}
\right)  \frac{\hat{p}}{m}\left(
\begin{array}
[c]{c}%
\phi_{r}(q,t)\\
\phi_{c}(q,t)
\end{array}
\right) \nonumber\\
&  =\frac{\hbar}{m\rho}\left(  \phi_{r}\left(  \vec{\nabla}\phi_{c}\right)
-\phi_{c}\left(  \vec{\nabla}\phi_{r}\right)  \right) \nonumber\\
&  =\frac{\hbar}{m}\operatorname{Im}\frac{\vec{\nabla}\psi}{\psi}.
\label{v-density}%
\end{align}
This class of normalization arises because the physical quantities like
$E^{\text{av}}$ and $\vec{v}^{\text{local}}(q,t)$ should be homogeneous of
degree zero in $\psi$. The notion of quantum velocity is hidden behind
$\bar{\psi}$ in this manner and is retrieved as above. This form of $\vec
{v}^{\text{local}}(q,t)$ will be used in the next section (Sec.
\ref{sec:Stochastic}) for the discussion of the one-world Schr\"{o}dinger
dynamics. The comparison of Eq. (\ref{e-density}) and Eq. (\ref{v-density})
under setting $\phi_{c}(q,t)\equiv0$ highlights why the real-valued
Schr\"{o}dinger function should be a two component vector and equivalently why
the original Schr\"{o}dinger function has to be complex-valued.

\subsection{Summary of this section: Meaning of the Schr\"{o}dinger functions
and vectors}

We here summarize the route to the real-valued Schr\"{o}dinger equation and
what lie behind the procedures.\ \ 

\subsubsection{The process to the real-valued Schr\"{o}dinger equation
itemized}

1) Consider a function $\rho(q,t)$ in the Euclidean space-time, which is
positive semidefinite everywhere. ($q$ specifies a multidimensional
configuration space.)

\bigskip

2) Factorize $\rho(q,t)$ as

$\rho(q,t)=\bar{\psi}(q,t)^{T}\bar{\psi}(q,t)$,

with the Schr\"{o}dinger vector

$\bar{\psi}(q,t)=\left(
\begin{array}
[c]{cc}%
\phi_{r}(q,t) & \phi_{c}(q,t)
\end{array}
\right)  ^{T}$,

where both $\phi_{r}(q,t)$ and $\phi_{c}(q,t)$ are real-valued functions.

\bigskip

3) Translational symmetry of $q$-space gives a generator and associated
momentum operator as in

$\bar{\psi}(q,t)^{T}(-\hbar$\textbf{J}$\vec{\nabla})\bar{\psi}(q,t)$. The
constant $\hbar$ emerges only from comparison with the experimental facts.

\bigskip

4) Translational symmetry of $t$-space gives a generator

$\bar{\psi}(q,t)^{T}(\hbar$\textbf{J}$\partial_{t})\bar{\psi}(q,t).$

\bigskip

5) $\rho(q,t)$ is assumed to induce an incompressible flow satisfying the
equation of continuity

$\partial_{t}\rho(q,t)=-\vec{\nabla}\cdot\vec{j}(q,t)$ because $\rho(q,t)$ is
assumed to conserve the content, which is actually the number of particles involved.

\bigskip

6) The flux $\vec{j}(q,t)$ is found to be

$\vec{j}(q,t)=-\left(  \hbar/m\right)  \bar{\psi}(q,t)^{T}$\textbf{J}%
$\vec{\nabla}\bar{\psi}(q,t)$,

and the equation of continuity reads

$\bar{\psi}(q,t)^{T}\left(  \hbar\partial_{t}+\text{\textbf{J}}\hat{p}%
^{2}/2m\right)  \bar{\psi}(q,t)=0.$

\bigskip

7) The continuity equation is automatically extended to that under a potential
function $V(q,t)$ as

$\bar{\psi}(q,t)^{T}\left(  \hbar\partial_{t}+\text{\textbf{J}}\left(  \hat
{p}^{2}/2m+V\right)  \right)  \bar{\psi}(q,t)=0$, and thereby the quantum
Hamiltonian is defined as

$\hat{H}=\hat{p}^{2}/2m+V$ .

\bigskip

8) Among many possible $\rho(q,t)$ that satisfy the equation of continuity, we
choose the generic (most probable in this context) one so that $\bar{\psi
}(q,t)$ behind $\rho(q,t)$ satisfies a variational condition

$\left(  \delta\bar{\psi}(q,t)^{T}\right)  \left(  \hbar\partial
_{t}+\text{\textbf{J}}\left(  \hat{p}^{2}/2m+V\right)  \right)  \bar{\psi
}(q,t)=0$

or simply

$\left(  \text{\textbf{J}}\hbar\partial_{t}-\left(  \hat{p}^{2}/2m+V\right)
\right)  \bar{\psi}(q,t)=0$,

which is equivalent to the original Schr\"{o}dinger equation that is
decomposed to the real and imaginary parts.

\section{Stochastic path dynamics to represent a single quantum incident and
the Schr\"{o}dinger distribution function\label{sec:Stochastic}}

In order to capture the intrinsic physical nature of the Schr\"{o}dinger
dynamics, we below continue a series of detailed analyses from three different
perspectives of \textquotedblleft continuous and smooth distribution
function\textquotedblright\ versus \textquotedblleft path (lay)
dynamics.\textquotedblright\ In doing do, we need to distinguish the
path-representation of the Schr\"{o}dinger function and a physical quantum
path, a continuous line along which a point mass may track. This section
discusses the notion of possible physical path, and the two subsequent
sections, Sec. \ref{sec:Variational} and Sec. \ref{sec:ADFpath}, treat the
path representations of the Schr\"{o}dinger function.

\subsection{Introductory remarks}

\subsubsection{Distribution and paths in quantum mechanics}

\begin{comment}
As an illustrative guide in considering a conflict between a continuos
coherent distribution and path dynamics, we suppose a thought double-slit
experiment in what follows. A packet \textquotedblleft mimicking a particle
dynamics\textquotedblright\ is initially launched to a direction normal to the
slit panel. The packet is supposed to bifurcate into three pieces at the slit
panel after the time propagation in a free space; the one entering into one of
the slits, another into the other slit, and the rest reflecting back from the
panel. Those packets entering into the slits should feel suddenly a narrow
potential, the range of which is supposed to be shorter than the wavepacket
width. The packet has to deform to fit in. At each exit of the slits, a
significant Huygens-like diffraction should take place to resume free
propagation towards a wider range of the measurement panel. A
\textquotedblleft particle\textquotedblright\ hitting the panel (measurement
apparatus) is supposed to leave a spot at each event.\cite{Tonomura1989} We
here do not attempt to determine which one of the slits a \textquotedblleft
particle\textquotedblright\ passes through.
\end{comment}

The Schr\"{o}dinger equation and its function do not represent a single
incident (occurrence of an event) but a distribution of coherent superposition
over an ensemble of the possible incidents. Therefore, the Schr\"{o}dinger
equation is unlikely to be capable of giving direct (unsuperposed) information
for individual one-world experiment. Take an example again from a one-particle
launching in the double slit experiment. It is often misunderstood that a
particle may be able to simultaneously pass through the two slits, in view
that a Schr\"{o}dinger function $\psi(q,t)$ mathematically does. An essential
question about quantum mechanics is therefore whether one can single out a
dynamical path corresponding to an individual event. If yes, it must be
consistent with the prediction by the Schr\"{o}dinger function. That is, we
need to answer a question on how such an ensemble of independent single event
paths, if any, can create the interference pattern. A single-event path, which
we refer to as one-world path in what follows, serves as a path of a single
dice \textquotedblleft God\textquotedblright\ may cast.\cite{freire2022oxford}
Here in this section we clarify the difference and relationship between the
one-world path and the coherent sum of \textquotedblleft
many-worlds\textquotedblright\ reflected in the Schr\"{o}dinger function.

\subsubsection{Paths and flow lines}

One of the most beautiful and significant path-representations of the
Schr\"{o}dinger function is due to Bohm with the so-called Bohmian
trajectories.\cite{Bohm-text,Wyatttext,sanz2013trajectory} The Bohmian
trajectories are actually the smooth and continuous flow lines of the
Schr\"{o}dinger function.\cite{sanz2013trajectory} Due to the presence of the
Laplacian in the Schr\"{o}dinger equation, the Schr\"{o}dinger function itself
and the associated trajectories are required to be smooth (differentiable
everywhere) in general. This is the intrinsic property of distribution
functions. However, the smoothness is not necessarily applied to the possible
quantum paths. Notably, the Laplacian is translated to the generator of the
Wiener process in the Ito
formula,\cite{van1976stochastic,risken1996fokker,Gardiner,oksendal2013stochastic}
thereby giving birth to indifferentiable paths (an example as such is seen in
Eq. (\ref{ItoA})).

\subsubsection{Feynman path integrals}

In discussing the roles of paths in quantum mechanics, it is natural to start
from the Feynman kernel\cite{Feynman,Feynman-Hibbs,Schulman,kleinert2006path}%

\begin{align}
&  K(q,t)=\left\langle q\left\vert \exp\left(  \frac{1}{i\hbar}\hat
{H}t\right)  \right\vert 0\right\rangle \nonumber\\
&  =\lim_{N\rightarrow\infty}\int d^{3}q_{1}\cdots\int d^{3}q_{N}\left(
\frac{m}{2\pi i\hbar\Delta t}\right)  ^{3(N+1)/2}\nonumber\\
&  \times\exp\left[  \frac{i}{\hbar}\sum_{k=0}^{N}\left(  \frac{m}{2}%
\frac{\left(  q_{k+1}-q_{k}\right)  ^{2}}{\Delta t}-V(q_{k})\Delta t\right)
\right]  \label{Fkernel}%
\end{align}
with%
\begin{equation}
\Delta t=\frac{t}{N+1}.
\end{equation}
As seen in Eq. (\ref{Fkernel}), it is expressed in term of the democratic
summation of continuous polylines (broken lines), each connecting two
neighboring positions $q_{k}$ and $q_{k+1}$ of an infinitessimal distance.
This kernel is somewhat similar to the Wiener path-integrals for the Brownian
motion (shown later in order in Eq. (\ref{WienerG})). However, the integral
measure is not well defined in Eq. (\ref{Fkernel}). As seen explicitly, no
dynamics is imposed on each line or \textquotedblleft path\textquotedblright.
Yet, the stationary-phase paths are found to satisfy the Lagrange equation of
motion, thus giving rise to classical trajectories. Otherwise, no specific
dynamics is assumed.

\subsubsection{Nagasawa theory}

Nagasawa\cite{nagasawa1989transformations,Nagasawa2}
established his extensive theory of quantum stochastic path dynamic based on the Kolmogorov and Ito theories, which is
mathematically rigorous: A Markov process having the transition probability
density consisting of the Green function for the following parabolic partial
differential equation (or a backward Fokker-Planck equation)%

\begin{equation}
\frac{\partial u}{\partial t}+\frac{1}{2}\sigma^{2}\nabla^{2}u+b(q,t)\cdot
\nabla u=0 \label{Kol}%
\end{equation}
is associated with the stochastic process satisfying%
\begin{equation}
X_{t}=X_{0}+\sigma B_{t}+\int_{0}^{t}b(X_{s},t)ds, \label{Kol2}%
\end{equation}
with $B_{t}$ being the Brownian process. Nagasawa transforms Eq. (\ref{Kol})
to a set of
\begin{equation}
\frac{\partial\psi_{N}}{\partial t}+\frac{1}{2}\sigma^{2}\nabla^{2}\psi
_{N}+V(q,t)\psi_{N}=0
\end{equation}
and
\begin{equation}
-\frac{\partial\tilde{\psi}_{N}}{\partial t}+\frac{1}{2}\sigma^{2}\nabla
^{2}\tilde{\psi}_{N}+V(q,t)\tilde{\psi}_{N}=0
\end{equation}
each having a pair of real-valued solutions
\begin{equation}
\psi_{N}(q,t)=\exp(R(q,t)+S(q,t))\text{ \ and \ }\tilde{\psi}_{N}=\exp(R-S),
\label{evolutionF}%
\end{equation}
where the real-valued functions $R(q,t)$ and $S(q,t)$ are equivalent to those
in the complex valued Schr\"{o}dinger function
\begin{equation}
\psi=\exp\left(  R+iS\right)  , \label{Naga}%
\end{equation}
(note the position of $R$ in Eqs. (\ref{evolutionF}) and (\ref{Naga})).
Nagasawa calls $\psi_{N}(x,t)$ and $\tilde{\psi}_{N}(x,t)$, respectively,
Evolution function and Backward evolution function, neither of which is the
direct solution of the Schr\"{o}dinger equation. Then, after Eq. (\ref{Kol2}),
he gives the following stochastic process
\begin{equation}
dX_{t}=\frac{\hbar}{m}\nabla(R+S)dt+\sqrt{\frac{\hbar}{m}}dW. \label{Nagasawa}%
\end{equation}
Note again that this process is to represent his $\psi_{N}=\exp(R+S)$ of Eq.
(\ref{evolutionF}) but not the Schr\"{o}dinger function itself. Nagasawa
claims that his theory thus formulated has exceeded the Schr\"{o}dinger
framework in many aspects, including the uncertainty
principle.\cite{nagasawa1989transformations,Nagasawa2} He has thus established
the relationship between stochastic theory and quantum dynamics.

\subsection{Stochastic paths and the Schr\"{o}dinger equation}

We below consider a possible one-world Schr\"{o}dinger dynamics in our own
way, which is based on the relation between the Feynman-Kac formula for
statistical theory and a stochastic differential
equation.\cite{kac1949distributions,del2004feynman} We first outline the
Feynman-Kac formula and the relevant Green function.

\subsubsection{Feynman-Kac formula and stochastic differential equation}

To single out a dynamical path from quantum dynamics, we revisit the standard
relation between the stochastic process and the associated forward diffusion
equation under a potential function $V(q,t)$\cite{kac1949distributions,Ezawa}%
\begin{equation}
\frac{\partial}{\partial t}u(q,t)=D\frac{\partial^{2}}{\partial q^{2}%
}u(q,t)-\lambda V(q,t)u(q,t). \label{Diff00}%
\end{equation}
The Feynman-Kac formula for the Green function of this diffusion equation is%
\begin{align}
&  G(q,t)\nonumber\\
&  =\int_{\Omega\left(  q,t:0.0)\right)  }\exp\left[  -\lambda\int_{0}%
^{t}V(s,X(s,\omega))ds\right]  dP_{W\left(  q,t:0.0)\right)  }(\omega),
\label{FC}%
\end{align}
where $dP_{W\left(  q,t:0.0)\right)  }(\omega)$ is the Wiener measure of the
Brownian motion, and $G(q,t)$ can be written more explicitly as%

\begin{align}
G(q,t)  &  =\lim_{N\rightarrow\infty}\frac{1}{\left(  4\pi D\Delta t\right)
^{N/2}}\prod_{k=1}^{\infty}\int_{-\infty}^{\infty}dq_{k}\nonumber\\
&  \times\exp\left[  -\sum_{k=0}^{N-1}\Delta t\left(  \frac{1}{4D}%
\frac{\left(  \Delta q_{k}\right)  ^{2}}{\Delta t}+\lambda V(q_{k}%
,t_{k})\right)  \right]  \label{WienerG}%
\end{align}
with $t_{k}=k\Delta t,$ $q_{k}=q(t_{k})$, $q_{0}=0$, \ $q_{n}=q,$ $\Delta
q_{k}=q_{k+1}-q_{k}$, and $D$ being the diffusion constant. $G(q,t)$ in Eq.
(\ref{WienerG}) looks similar to the Feynman kernel of Eq. (\ref{Fkernel}),
but no imaginary number is included here. Behind the distribution function
$u(q,t)$ and the Green function lies a dynamical equation in the form of the
Ito stochastic differential
equation\cite{oksendal2013stochastic,van1976stochastic}%

\begin{equation}
dX_{t}=\alpha(X_{t},t)dt+dW\left(  t,\omega\right)  \label{St1}%
\end{equation}
with
\[
X_{t}=X\left(  t,\omega\right)
\]
and $W\left(  t,\omega\right)  $ being the Wiener process.

Consider a transition probability for
\begin{equation}
\left(  t,X_{t}\right)  \rightarrow\left(  t+\Delta t,X_{t+\Delta t}\right)
\text{ with a small }\Delta t.
\end{equation}
From Eq. (\ref{St1}) the Wiener process is retrieved out as%
\begin{equation}
W\left(  t+\Delta t\right)  -W\left(  t\right)  =X\left(  t+\Delta t\right)
-X\left(  t\right)  -\alpha(X_{t},t)\Delta t,
\end{equation}
and since $W\left(  t+\Delta t\right)  -W\left(  t\right)  $ is subjected to
the transition probability of the Wiener process, we define a probability for
a path starting from $\left(  X\left(  t\right)  ,t\right)  $ to arrive at a
place nearby $X\left(  t+\Delta t\right)  $ at time $t+\Delta t$ as
\begin{align}
P_{X}(\Delta t)  &  =\frac{\Delta x}{\left(  4\pi D\Delta t\right)  ^{1/2}%
}\exp\left[  -\frac{\left(  W\left(  t+\Delta t\right)  -W\left(  t\right)
\right)  ^{2}}{4D\Delta t}\right] \nonumber\\
&  =\frac{\Delta x}{\left(  4\pi D\Delta t\right)  ^{1/2}}\exp\left[
-\frac{\left(  X\left(  t+\Delta t\right)  -X\left(  t\right)  \right)  ^{2}%
}{4D\Delta t}\right] \nonumber\\
&  \times\exp[\frac{1}{2D}\alpha(X_{t},t)\left(  X\left(  t+\Delta t\right)
-X\left(  t\right)  \right)  ^{2}\nonumber\\
&  -\frac{1}{4D}\alpha(X_{t},t)^{2}\Delta t].
\end{align}
Using the transition probability for the pure Wiener process (with $V=0$)%
\begin{equation}
P_{W}(\Delta t)=\frac{\Delta x}{\left(  4\pi D\Delta t\right)  ^{1/2}}%
\exp\left[  -\frac{\left(  W\left(  t+\Delta t\right)  -W\left(  t\right)
\right)  ^{2}}{4D\Delta t}\right]  ,
\end{equation}
$P_{X}(\Delta t)$ is further rewritten in a compact form as%
\begin{align}
&  P_{X}(\Delta t)=P_{W}(\Delta t)\nonumber\\
&  \times\exp\left[  \frac{\alpha(X_{t},t)}{2D}\left(  X\left(  t+\Delta
t\right)  -X\left(  t\right)  \right)  ^{2}-\frac{\alpha(X_{t},t)^{2}}%
{4D}\Delta t\right]  .
\end{align}

Here, as usual, we introduce a function $A(q,t)$ that satisfies%
\begin{equation}
\frac{\partial A(q,t)}{\partial q}=\alpha(q,t). \label{alpha}%
\end{equation}
Then the Ito formula for $A$ reads
\begin{align}
dA(X,t)  &  =\frac{\partial A}{\partial t}dt+\frac{\partial A}{\partial
X}dX+\frac{\partial^{2}A}{\partial X^{2}}Ddt\nonumber\\
&  =\left(  \frac{\partial A}{\partial t}+D\frac{\partial^{2}A}{\partial
X^{2}}\right)  dt+\alpha(X,t)dX, \label{ItoA}%
\end{align}
in which
\begin{equation}
dXdX=2Ddt \label{Ito2}%
\end{equation}
has been used. Hence%
\begin{align}
\int_{0}^{T}\alpha(x,t)dX  &  =A(T)-A(0)\nonumber\\
&  -\int_{0}^{T}\left(  \frac{\partial A}{\partial t}+D\frac{\partial^{2}%
A}{\partial x^{2}}\right)  dt
\end{align}
and the probability for a stochastic path to pass through the possible range
$\Phi$ is
\begin{align}
&  P_{X}(\Phi)=\int_{\Phi}\exp\left[  \frac{1}{2D}\left(  A(T)-A(0)\right)
\right] \nonumber\\
&  \times\exp\left[  -\int_{0}^{T}\frac{1}{4D}\left(  2\frac{\partial
A}{\partial t}+2D\frac{\partial^{2}A}{\partial X^{2}}+\left(  \frac{\partial
A}{\partial X}\right)  ^{2}\right)  dt\right] \nonumber\\
&  \times dP_{W}(\omega).
\end{align}
Now by setting%
\begin{equation}
2\frac{\partial A}{\partial t}+2D\frac{\partial^{2}A}{\partial q^{2}}+\left(
\frac{\partial A}{\partial q}\right)  ^{2}=4\lambda V(q,t). \label{Aeq}%
\end{equation}
we reach the Feynman-Kac formula of Eq. (\ref{FC}).

Further, defining a real-valued function $\psi(q,t)$ as%
\begin{equation}
\psi_{f}(q,t)=\exp\left(  \frac{1}{2D}A(q,t)\right)  , \label{Apsi}%
\end{equation}
Eq. (\ref{Aeq}) is transformed to\ a backward diffusion equation%
\begin{equation}
\frac{\partial\psi_{f}(q,t)}{\partial t}=\left(  -D\frac{\partial^{2}%
}{\partial q^{2}}+\lambda V(q,t)\right)  \psi_{f}(q,t).
\end{equation}
With further changes of variables,\cite{Ezawa} we after all reach the standard
Feynman-Kac formula, Eq. (\ref{FC}), which is followed by the Green function
of a forward diffusion equation%
\begin{equation}
\frac{\partial G(q,t)}{\partial t}=\left(  D\frac{\partial^{2}}{\partial
q^{2}}-\lambda V(q,t)\right)  G(q,t) \label{Bdiff}%
\end{equation}
with
\begin{equation}
G(q,0)=\delta(q),
\end{equation}
which is to be used for propagation of a forward diffusion function $\psi
_{f}(q,t)$ from time zero to the future ($t>0$) such that%
\begin{equation}
\psi_{f}(q,t)=\int_{0}^{t}dsG(q,s)\psi_{f}(q,s).
\end{equation}
The Markov process of Eq. (\ref{St1}) is therefore followed by a stochastic
dynamics%
\begin{align}
dX_{t}  &  =\alpha(X_{t},t)dt+dW\left(  t,\omega\right) \nonumber\\
&  =\frac{\partial A(X_{t},t)}{\partial x}dt+dW\left(  t,\omega\right)
\nonumber\\
&  =2D\frac{\nabla\psi_{f}}{\psi_{f}}dt+dW\left(  t,\omega\right)  ,
\label{FCdX}%
\end{align}
which sets a foundation on which to consider the stochastic paths in harmony
with the Schr\"{o}dinger dynamics. We thus have seen a relation between a
stochastic process and the parabolic partial differential equation. We next
consider the similar treatment for the Schr\"{o}dinger dynamics.

It is interesting to compare the velocity drift term of Eq. (\ref{FCdX})
\begin{equation}
\alpha(X_{t},t)=2D\frac{\nabla\psi_{f}}{\psi_{f}} \label{velv1}%
\end{equation}
and the quantum local velocity of Eq. (\ref{v-density})
\begin{equation}
\frac{1}{\rho}v(q,t)=\frac{\hbar}{m}\operatorname{Im}\frac{\nabla\psi}{\psi},
\label{velv2}%
\end{equation}
where $\psi$ in the latter equation is the Schr\"{o}dinger function. They
already show an intrinsic similarity between stochastic and quantum dynamics
and $D\leftrightarrow\hbar/2m$.

\subsubsection{Quantum stochastic paths driven by the Schr\"{o}dinger velocity field}

We intend to implant the information of the Schr\"{o}dinger equation into the
drift term of the stochastic process, $\alpha(X_{t},t)$ of Eq. (\ref{FCdX}).
The real-valued Schr\"{o}dinger equation Eq. (\ref{Sch30}) in the form%

\begin{equation}
\frac{d}{dt}\left(
\begin{array}
[c]{c}%
\phi_{r}(q,t)\\
\phi_{c}(q,t)
\end{array}
\right)  =\left(  \frac{\hbar}{2m}\nabla^{2}-\frac{V}{\hbar}\right)
\text{\textbf{J}}\left(
\begin{array}
[c]{c}%
\phi_{r}(q,t)\\
\phi_{c}(q,t)
\end{array}
\right)  . \label{Sch45}%
\end{equation}
seems to be fine to apply Eq. (\ref{Bdiff}). However, Eq. (\ref{Sch45}) is
actually composed of a pair of coupled equations, and therefore we need to
detour via the complex-valued Schr\"{o}dinger equations to formally uncoupled
ones as
\begin{equation}
i\frac{\partial}{\partial t}(\phi_{r}+i\phi_{c})=\left(  \frac{\hbar}%
{2m}\nabla^{2}-\frac{V}{\hbar}\right)  (\phi_{r}+i\phi_{c}) \label{Sch50}%
\end{equation}
and its complex conjugate%
\begin{equation}
-i\frac{\partial}{\partial t}(\phi_{r}-i\phi_{c})=\left(  \frac{\hbar}%
{2m}\nabla^{2}-\frac{V}{\hbar}\right)  (\phi_{r}-i\phi_{c}). \label{Sch51}%
\end{equation}
These are further transformed so to mimic the form of Eq. (\ref{Bdiff}) by the
rotation of time coordinate to%

\begin{equation}
\frac{\partial}{\partial s^{+}}\psi^{+}(q,s^{+})=\left(  \frac{\hbar}%
{2m}\nabla^{2}-\frac{V}{\hbar}\right)  \psi^{+}(q,s^{+}) \label{Sch40}%
\end{equation}
with%

\begin{equation}
s^{+}=-it
\end{equation}
and $\psi^{+}=\phi_{r}+i\phi_{c}$. Likewise we have another one from Eq.
(\ref{Sch51})%

\begin{equation}
\frac{\partial}{\partial s^{-}}\psi^{-}(q,s^{-})=\left(  \frac{\hbar}%
{2m}\nabla^{2}-\frac{V}{\hbar}\right)  \psi^{-}(q,s^{-}) \label{Sch41}%
\end{equation}
by another rotation of time to the direction opposite to $s^{+}$%

\begin{equation}
s^{-}=it
\end{equation}
and $\psi^{-}=\phi_{r}-i\phi_{c}$. Although both Eqs. (\ref{Sch40}) and
(\ref{Sch41}) are similar to Eq. (\ref{Bdiff}) with $D=\hbar/2m$, $\psi
^{+}(q,s^{+})$ and $\psi^{-}(q,s^{-})$ are complex functions. Hence, the
direct application of \ Eq. (\ref{alpha})\ or Eq. (\ref{FCdX}) end up with%
\begin{equation}
\alpha^{+}(X_{t},s^{+})ds^{+}=2D\frac{\nabla\psi^{+}}{\psi^{+}}ds^{+}
\label{alpplus}%
\end{equation}
and
\begin{equation}
\alpha^{-}(X_{t},s^{-})ds^{-}=2D\frac{\nabla\psi^{-}}{\psi^{-}}ds^{-},
\label{alpminus}%
\end{equation}
which are complex valued, too, and necessarily make $X_{t}$ in Eq. (\ref{St1}) complex-valued.

To proceed further, we make an analytic continuation for $A,\alpha,$ and
$\psi$ from Eq. (\ref{Aeq}) to Eq. (\ref{Bdiff}). It is obvious that the
quantities in Eqs. (\ref{alpplus}) and (\ref{alpminus}) are mutually complex
conjugate, and we may define%
\begin{equation}
\alpha^{+}(X_{t},s^{+})ds^{+}=\left[  \alpha^{\text{Real}}(X_{t}%
,t)+i\alpha^{\text{Imag}}(X_{t},t)\right]  dt
\end{equation}
and%
\begin{equation}
\alpha^{-}(X_{t},s^{-})ds^{-}=\left[  \alpha^{\text{Real}}(X_{t}%
,t)-i\alpha^{\text{Imag}}(X_{t},t)\right]  dt.
\end{equation}
$\alpha^{\text{Real}}(X_{t},t)$ and $\alpha^{\text{Imag}}(X_{t},t)$ are
readily obtained such that%

\begin{align}
\alpha^{\text{Real}}(X_{t},t)dt  &  =\frac{1}{2}2D\left(  \frac{\nabla\psi
^{+}}{\psi^{+}}ds^{+}+\frac{\nabla\psi^{-}}{\psi^{-}}ds^{-}\right) \nonumber\\
&  =D\left(  \frac{\nabla\psi^{+}}{\psi^{+}}(-idt)+\frac{\nabla\psi^{-}}%
{\psi^{-}}(idt)\right)
\end{align}
which is followed by a simple manipulation
\begin{align}
&  \alpha^{\text{Real}}(X_{t},t)dt=-idtD\left(  \frac{\nabla\psi^{+}}{\psi
^{+}}-\frac{\nabla\psi^{-}}{\psi^{-}}\right) \nonumber\\
&  =\frac{2D}{\rho}dt\left(  \phi_{r}\left(  \nabla\phi_{c}\right)  -\phi
_{c}\left(  \nabla\phi_{r}\right)  \right) \nonumber\\
&  =-\frac{2D}{\rho}\left(
\begin{array}
[c]{cc}%
\phi_{r} & \phi_{c}%
\end{array}
\right)  \text{\textbf{J}}\nabla\left(
\begin{array}
[c]{c}%
\phi_{r}\\
\phi_{c}%
\end{array}
\right)  dt, \label{Realalpha}%
\end{align}
with $\rho=\phi_{r}^{2}+\phi_{c}^{2}$. Noting the expression $\hat{p}%
=-$\textbf{J}$\hbar\vec{\nabla}$ as found in Eq. (\ref{primeflux}) and Eq.
(\ref{momentum}), we see that the physical meaning of $\alpha^{\text{Real}}$
turns out to be the locally normalized velocity at $X_{t}$ and $t$ (see Eq.
(\ref{v-density})). This velocity is well-defined in terms of the
Schr\"{o}dinger distribution function. We also note that $\alpha^{\text{Real}%
}$ thus attained is invariant with respect to any rotation of the vector
$\left(
\begin{array}
[c]{cc}%
\phi_{r}(q,t) & \phi_{c}(q,t)
\end{array}
\right)  ^{T}$. Likewise we have%

\begin{align}
&  \alpha^{\text{Imag}}(X_{t},t)dt=\frac{1}{2i}2D\left(  \frac{\nabla\psi^{+}%
}{\psi^{+}}ds^{+}-\frac{\nabla\psi^{-}}{\psi^{-}}ds^{-}\right) \nonumber\\
&  =-\frac{2D}{\rho}dt(\phi_{r}\left(  \nabla\phi_{r}\right)  +\phi_{c}\left(
\nabla\phi_{c}\right)  )\nonumber\\
&  =-\frac{D}{\rho}\left(  \nabla\rho\right)  dt.
\end{align}

Back to Eqs. (\ref{alpplus}) and (\ref{alpminus}), we may formally regard
$\alpha^{+}(X_{t},s^{+})ds^{+}$ as the drift velocity term, which is driven by
the dynamics for $\psi^{+}=\phi_{r}+i\phi_{c}$, and $\alpha^{-}(X_{t}%
,s^{-})ds^{-}$ is another drift term obtained through $\psi^{-}=\phi_{r}%
-i\phi_{c}$. On the other hand, the real-valued Schr\"{o}dinger equation
(\ref{Sch45}) does not have distinction between Eqs. (\ref{Sch40}) and
(\ref{Sch41}), and therefore we need to take account of these two channels in
an even manner by taking an average of the two drift terms%
\begin{equation}
\frac{1}{2}\left[  \alpha^{+}(X_{t},s^{+})ds^{+}+\alpha^{-}(X_{t},s^{-}%
)ds^{-}\right]  ,
\end{equation}
which is none other than $\alpha^{\text{Real}}(X_{t},t)dt$. Thus we have
effectively returned to the real-valued drift term, because the imaginary
parts are cancelled after all.

Besides, defining the normalized Wiener process%

\begin{equation}
dW_{0}\left(  t,\omega\right)  dW_{0}\left(  t,\omega\right)  =2Ddt=\frac
{\hbar}{m}dt \label{Dyndiff}%
\end{equation}
with
\begin{equation}
D=\frac{\hbar}{2m} \label{Dconst}%
\end{equation}
we have%

\begin{align}
dX_{t}  &  =\alpha^{\text{Real}}(X_{t},t)dt+\sqrt{\frac{\hbar}{m}}%
dW_{0}\left(  t,\omega\right) \nonumber\\
&  =-\frac{\hbar}{\rho m}\left(
\begin{array}
[c]{cc}%
\phi_{r} & \phi_{c}%
\end{array}
\right)  \text{\textbf{J}}\nabla\left(
\begin{array}
[c]{c}%
\phi_{r}\\
\phi_{c}%
\end{array}
\right)  dt+\sqrt{\frac{\hbar}{m}}dW_{0}\left(  t,\omega\right) \nonumber\\
&  =\frac{1}{\rho}\left(
\begin{array}
[c]{cc}%
\phi_{r} & \phi_{c}%
\end{array}
\right)  \frac{\hat{p}}{m}\left(
\begin{array}
[c]{c}%
\phi_{r}\\
\phi_{c}%
\end{array}
\right)  dt+\sqrt{\frac{\hbar}{m}}dW_{0}\left(  t,\omega\right)  .
\label{Stchastic1}%
\end{align}
The stochastic path $X_{t}$ thus remains to run in the real-valued space.
However, notice that the drift term thus found is nonlocal in that no direct
interaction drives the momentum but only the Schr\"{o}dinger function is
involved in.

A precise study shows that the drift term of Eq. (\ref{Stchastic1}) and
Nagasawa's one in Eq. (\ref{Nagasawa}) are essentially the same, since both
represent the quantum mechanical velocity locally normalized at $(q,t)$. We
note however that the two expressions have been derived through the different
pathways. Particularly Nagasawa's real-valued evolution function defined in
Eq. (\ref{evolutionF}) are not the Schr\"{o}dinger functions in themselves.
Equation (\ref{Stchastic1}) seems rather compact and more intuitively appealing.

The imaginary part of the drift term disappears from the practice of the
calculation of the Schr\"{o}dinger paths. However, we have no idea at present
whether it plays no role at all. For instance, The imaginary part of $X_{t}$,
say $X_{t}^{\text{Imag}}$ to be determined by%
\begin{equation}
dX_{t}^{\text{Imag}}=i\alpha^{\text{imag}}(X_{t},t)dt
\end{equation}
might serve as a part of a dynamics that is not imagined thus far.

\subsubsection{Hamilton canonical equations in quantum mechanics}

The dynamics of the momentum $\left(
\begin{array}
[c]{cc}%
\phi_{r} & \phi_{c}%
\end{array}
\right)  \hat{p}\left(
\begin{array}
[c]{cc}%
\phi_{r} & \phi_{c}%
\end{array}
\right)  ^{T}$ in Eq. (\ref{Stchastic1}) is tracked with the Heisenberg
equation of motion of Eq. (\ref{Heisenberg}) such that%

\begin{align}
&  \frac{d}{dt}\left(
\begin{array}
[c]{cc}%
\phi_{r} & \phi_{c}%
\end{array}
\right)  \hat{p}\left(
\begin{array}
[c]{c}%
\phi_{r}\\
\phi_{c}%
\end{array}
\right)  =\frac{1}{\hbar}\left(
\begin{array}
[c]{cc}%
\phi_{r} & \phi_{c}%
\end{array}
\right)  \left[  \hat{H},\hat{p}\right]  \text{\textbf{J}}\left(
\begin{array}
[c]{c}%
\phi_{r}\\
\phi_{c}%
\end{array}
\right) \nonumber\\
&  =-\left(
\begin{array}
[c]{cc}%
\phi_{r} & \phi_{c}%
\end{array}
\right)  \left[  V,\text{\textbf{J}}\vec{\nabla}\right]  \text{\textbf{J}%
}\left(
\begin{array}
[c]{c}%
\phi_{r}\\
\phi_{c}%
\end{array}
\right) \nonumber\\
&  =\left(
\begin{array}
[c]{cc}%
\phi_{r} & \phi_{c}%
\end{array}
\right)  \left(  -\vec{\nabla}V\right)  \left(
\begin{array}
[c]{c}%
\phi_{r}\\
\phi_{c}%
\end{array}
\right)  , \label{Ehrenfestp}%
\end{align}
where the momentum operator $\hat{p}$ is taken from Eq. (\ref{pp}). Equation
(\ref{Ehrenfestp}) is just an alternative expression of the Ehrenfest
theorem\cite{Messiah} (notice however that the integration over the
$q-$coordinates is not taken). Thus the drift velocity implicitly reflects the
dynamics driven by the Schr\"{o}dinger equation. Therefore, we may formally
combine Eqs. (\ref{Stchastic1}) and (\ref{Ehrenfestp}) into a set of the
quantum canonical equations of motion%
\begin{equation}
\left\{
\begin{array}
[c]{c}%
dX_{t}=\frac{1}{m\rho}P_{X_{t}}dt+\sqrt{\frac{\hbar}{m}}dW_{0}\left(
t,\omega\right) \\
\\
dP_{X_{t}}=-\left[  \vec{\nabla}V\right]  _{X_{t}}dt\text{
\ \ \ \ \ \ \ \ \ \ \ \ \ \ \ }%
\end{array}
\right.  \label{HM12}%
\end{equation}
with%
\begin{equation}
P_{X_{t}}=\left.  \left(
\begin{array}
[c]{cc}%
\phi_{r} & \phi_{c}%
\end{array}
\right)  \hat{p}\left(
\begin{array}
[c]{c}%
\phi_{r}\\
\phi_{c}%
\end{array}
\right)  \right\vert _{X_{t}}%
\end{equation}
and%
\begin{equation}
\text{\ }\left[  \vec{\nabla}V\right]  _{X_{t}}=\left.  \left(
\begin{array}
[c]{cc}%
\phi_{r} & \phi_{c}%
\end{array}
\right)  \left(  \vec{\nabla}V\right)  \left(
\begin{array}
[c]{c}%
\phi_{r}\\
\phi_{c}%
\end{array}
\right)  \right\vert _{X_{t}}.
\end{equation}
(Notice that $\left[  A\right]  _{X_{t}}$ is not the standard quantum
expectation value of $A,$ since the integration over $q-$coordinates is not
performed.) Having formulated as above, we should note that the second
equation of Eq. (\ref{HM12}) (for $dP_{X_{t}}$) is actually not necessary to
integrate the first one (for $dX_{t}$), as long as $\bar{\psi}(q,t)=\left(
\begin{array}
[c]{cc}%
\phi_{r}(q,t) & \phi_{c}(q,t)
\end{array}
\right)  $ is given.

\subsection{No direct interaction but indirect correlation among the
one-world paths \label{Sec:correlation}}

We stress again that the one-world paths are supposed to run on different
occasions, because each of them is launched independently. In this
regard, we recall that each path involved in the Brownian process represents
an individual sample of a single Brownian motion in real space. An infinite
number of possible Brownian paths are theoretically supposed to run on
different occasions in a manner one after another one. (We are not discussing
the collision among the Brownian particles.) Therefore, no mutual direct
interaction among the paths is expected.

On the other hand, the Feynman path integral is subtle in that each path in it
is not subject to any dynamics and is associated with the imaginary
\textquotedblleft weighting factor\textquotedblright. Since they are
democratically superposed in the kernel, each of the paths may be regarded as
a \textquotedblleft basis function\textquotedblright\ to expand the
kernel.\cite{davison1954feynmann} The simultaneous interference among the
paths through the coherent summation is the very core of the theory.\bigskip

\color{black}Nevertheless, a one-world path can \textquotedblleft
indirectly\textquotedblright\ correlate with other one-world paths through the
velocity drift term, Eq. (\ref{Realalpha}), which is a function of the
Schr\"{o}dinger function. We mean by indirect correlation that (i) we have no
machinery through which two or more one-world paths, launched on different
occasions, mechanically interact each other, but (ii) as long as a same
Schr\"{o}dinger function lies behind the drift velocity of Eq.
(\ref{Stchastic1}), not only they conform to the same global trend but also
each is mathematically responsible to form the $\alpha(X_{t},t)$ through the
Schr\"{o}dinger function. Therefore, even a slight change of the surroundings
like an experimental setting in the level of the one-world path can affect the
Schr\"{o}dinger function, and thereby it can modulate the way of correlation
among the paths. Therefore $\alpha(X_{t},t)$ may (yet doesn't have to) be
regarded as a nonlocal hidden variable, and it varies in general depending on
experimental contexts such as the initial conditions, boundary conditions,
switching the measurement channels, and so on. The one-world paths are thus
supposed to run under the mutual indirect correlation, which is nonlinear in
the sense of self-referring. We will be back to this aspect later with a
precise discussion.\color{black}

\begin{comment}
\color{red}=================
On the permutation symmetry%
\begin{align}
&  \psi(q_{1},\omega_{1},q_{2},\omega_{2})\nonumber\\
&  =\left(  \phi_{r}(q_{1},q_{2})+i\phi_{c}(q_{1},q_{2})\right)  +\left(
\phi_{r}(q_{2},q_{1})+i\phi_{c}(q_{2},q_{1})\right) \nonumber\\
&  \times\frac{1}{\sqrt{2}}\left(  \alpha(\omega_{1})\beta(\omega_{2}%
)-\beta(\omega_{1})\alpha(\omega_{2})\right)
\end{align}
%
\begin{align}
&  \alpha(X_{t},t)\nonumber\\
&  =-\frac{\hbar}{m\rho}[\left(  \phi_{r}(q_{1},q_{2})+\phi_{r}(q_{2}%
,q_{1}\right)  )\nonumber\\
&  \times\left(  \nabla_{1}+\nabla_{2}\right)  \left(  \phi_{c}(q_{1}%
,q_{2})+\phi_{c}(q_{2},q_{1})\right) \nonumber\\
&  -\left(  \phi_{c}(q_{1},q_{2})+\phi_{c}(q_{2},q_{1})\right) \nonumber\\
&  \times\left(  \nabla_{1}+\nabla_{2}\right)  \left(  \phi_{r}(q_{1}%
,q_{2})+\phi_{r}(q_{2},q_{1})\right)  ]
\end{align}
==================== \color{black}
\end{comment}

\subsubsection{Interference fringe intensity pattern in the double slit experiment}

Each path can pass through only one of the two slits in the double-slit
experiment. On the other hand, a single Schr\"{o}dinger function can bifurcate
and pass through the two slits simultaneously as a coherent distribution
function. This is well represented by the Feynman path integrals, since the
packet consists of many independent geometrical paths in the sense of Eq.
(\ref{Fkernel}). Note that it is not $\left\vert \psi(q,t)\right\vert ^{2}$
that physically makes a set of spots on the measurement board, but each
one-world path one by one with an amplitude $\left(
\begin{array}
[c]{cc}%
\phi_{r}(X_{0},0) & \phi_{c}(X_{0},0)
\end{array}
\right)  ^{T}$. Nevertheless, each one-world
path is \textquotedblleft driven and guided\textquotedblright\ by the drift
velocity term, which is composed of the relevant Schr\"{o}dinger function
$\psi(q,t)$. It is therefore not very mysterious that the interference pattern
is shaped after many frequencies of launching of single particles. It is a
great mystery how nature manages to materialize the nonlinear relation between
the parts (one-world paths) and the whole (the Schr\"{o}dinger function).\ 

\color{black}

\subsubsection{A tie to and difference from the Bohm trajectory}

It is interesting in the present context to recall the idea of de Broglie and
Bohm that the Schr\"{o}dinger function should serve as a pilot wave that
guides the particle motion.\cite{jaeger2009entanglement,schumacher2010quantum}
As a matter of fact, the Schr\"{o}dinger function serves in the drift term as
in Eq. (\ref{Stchastic1}), but not as a direct \textquotedblleft
wave\textquotedblright\ to guide. We need\ to time-propagate the
Schr\"{o}dinger function simultaneously or before integration of the
stochastic equation. Then we have quantum Hamilton equations of motion, Eq.
(\ref{HM12}), in which the velocity drift term is installed. On the other
hand, the Bohm representation of the Schr\"{o}dinger
function\cite{de2016principles} reads in the form%
\begin{equation}
\psi\left(  q,t\right)  =R(q,t)\exp\left(  \frac{i}{\hbar}S_{B}\left(
q,t\right)  \right)  , \label{Bohm0}%
\end{equation}
and the quantum Hamilton-Jacobi (HJ) equation%
\begin{equation}
\frac{\partial S_{B}}{\partial t}+\frac{1}{2m}\left(  \nabla S_{B}\right)
^{2}+V-\frac{\hbar^{2}}{2m}\frac{\nabla^{2}R}{R}=0, \label{Bohm2}%
\end{equation}
is derived along with the equation of continuity for $R(q,t)^{2}$. If
$\psi\left(  q,t\right)  $ is given beforehand, the quantum velocity is given
by%
\begin{equation}
v_{B}=\frac{1}{m}\nabla S_{B}=\frac{\hbar}{m}\operatorname{Im}\frac{\nabla
\psi}{\psi}. \label{Bohm3}%
\end{equation}
Therefore $v_{B}$ is exactly the same as the drift velocity of Eq.
(\ref{Stchastic1}), as long as a common Schr\"{o}dinger function is resorted
to (see also Eq. (\ref{v-density})). Hence, the mathematical difference
between the Bohmian trajectory and the one-world path is the presence of the
Wiener process or not. Meanwhile, it is known already that the interference
(fringe intensity) pattern in the double-slit experiment is numerically
realized by the set of the Bohmian trajectories,\cite{philippidis1979quantum}
and therefore a sufficient large set of the one-world paths should reproduce
the similar interference pattern. Physically, however, the Bohmian
trajectories do not represent a one-world physical path but only flow-lines of
the $S_{B}$-plane.\cite{sanz2013trajectory} Mathematically they are a set of
integral curves of Eq. (\ref{Bohm2}). Since the Bohmian representations, Eq.
(\ref{Bohm0}), Eq. (\ref{Bohm2}), and the equation of continuity for
$R(q,t)^{2}$, are all equivalent to the Schr\"{o}dinger equation itself, it
must be logically hard for the Bohm theory to make an essentially novel
interpretation beyond the limit of the Schr\"{o}dinger equation. Yet, it sheds
a new light on the hidden properties of the Schr\"{o}dinger dynamics as in Eq.
(\ref{Bohm3}).

\color{black}

Recalling how the real-valued Schr\"{o}dinger equation has been made up in
Sec. \ref{sec:RealValue}, we notice that there is no theoretical mechanism
that can bring about stochasticity into the Schr\"{o}dinger dynamics. To
single out a one-world path, therefore, we made use of the Feynman-Kac formula
as a starting point, and the stochastic nature of a one-world path has been
identified in such a process. Therefore, the Bohmian trajectories have no way
to take account of the Wiener process. $S_{B}$-plane is similar to the
classical action plane $S_{cl}$, and indeed in the limit of $\hbar
\rightarrow0$, $S_{B}\rightarrow S_{cl}$. \color{black}Nonetheless, this
convergence does not warrant that the Bohmian paths at $\hbar\neq0$ represent
the one-world quantum dynamics, because of the lack of stochasticity. Also,
Eq. (\ref{Bohm2}) claims that the continuous set of Bohmian paths run
simultaneously, not one after another. This is therefore not realistic as a
physical path but each is regarded as a probe of the
flow-lines.\cite{sanz2013trajectory}

\begin{comment}
\subsubsection{Time-irreversibility}

The Schr\"{o}dinger equation is time reversal. However, each one-world path
does not have time-reversal property due to the presence of the Wiener process
in it. Nonetheless, an average over the Wiener process leads to%

\begin{equation}
\left\langle \frac{dW_{0}}{dt}\right\rangle =0,
\end{equation}
resulting in%
\begin{equation}
\left\langle \frac{dX_{t}}{dt}\right\rangle =\left\langle \frac{1}{\rho
}\left(
\begin{array}
[c]{cc}%
\phi_{r} & \phi_{c}%
\end{array}
\right)  \frac{\hat{p}}{m}\left(
\begin{array}
[c]{c}%
\phi_{r}\\
\phi_{c}%
\end{array}
\right)  \right\rangle , \label{avrX}%
\end{equation}
and of course we have
\begin{equation}
\left\langle \frac{dP_{X_{t}}}{dt}\right\rangle =-\left[  \vec{\nabla
}V\right]  _{X_{t}}. \label{avrP}%
\end{equation}
from Eq. (\ref{HM12}). The Schr\"{o}dinger dynamics as an average over the
accumulated one-world paths is therefore time-reversal. This in turn suggests
that one cannot pick (separate to single out) a physical path from the
Schr\"{o}dinger equation directly.
\end{comment}
\color{black}

\subsection{Classical limit of the quantum Hamilton canonical equations}

Let us consider the classical limit ($\hbar\rightarrow0$) in the quantum
canonical equations of motion in Eq. (\ref{HM12}). First, since the Wiener
process is simply linear in $\hbar$, it can be simply reduced to zero as
$\hbar\rightarrow0$. We then rewrite Eq. (\ref{HM12}) in the polar coordinate.
It simply follows that%

\begin{equation}
P_{X_{t}}=\hbar\rho\nabla\theta(X_{t},t) \label{PX}%
\end{equation}
and%

\begin{align}
\left[  \vec{\nabla}V\right]  _{X_{t}}  &  =\left.  \left(
\begin{array}
[c]{cc}%
\phi_{r} & \phi_{c}%
\end{array}
\right)  \left(  \vec{\nabla}V\right)  \left(
\begin{array}
[c]{c}%
\phi_{r}\\
\phi_{c}%
\end{array}
\right)  \right\vert _{X_{t}}\nonumber\\
&  =\rho(X_{t},t)\vec{\nabla}V(X_{t},t).
\end{align}
Since the Wiener process is nullified, a one-world path is now a smooth path
that does not have a chance of branching in its direction. 
This implies that $X_{t}$ is determined uniquely as soon
as $X_{0}$ is prepared under an initial velocity field. Since there is no path
branching, we may expect that magnitude of $\rho(X_{0},0)$ is to be maintained
constant throughout the propagation as
\begin{equation}
\rho(X_{t},t)=\rho(X_{0},0)\equiv\rho_{0}. \label{ClassicalC}%
\end{equation}
Then we have
\begin{equation}
P_{X_{t}}=\hbar\rho_{0}\nabla\theta(X_{t},t).
\end{equation}
in Eq. (\ref{PX}). (Note that $\hbar\rightarrow0$ should not be taken in this
stage, because $\nabla\theta(X_{t},t)$ gives a term proportional to
$\hbar^{-1}$.) Further, we define%

\begin{equation}
\hbar\nabla\theta(X_{t},t)=p(X_{t},t) \label{thetap}%
\end{equation}
and
\begin{equation}
P_{X_{t}}=\hbar\rho_{0}\nabla\theta(X_{t},t)=\rho_{0}p(X_{t},t). \label{Pp}%
\end{equation}
Insertion of Eq. (\ref{Pp}) into the first equation of Eq. (\ref{HM12}) gives
\begin{equation}
dX_{t}=\frac{1}{m}p(X_{t},t)dt. \label{CLH1}%
\end{equation}
On the other hand, the second equation in Eq. (\ref{HM12}) is rewritten as%

\begin{align}
dP_{X_{t}}  &  =-\left[  \vec{\nabla}V\right]  _{X_{t}}dt=-\rho_{0}\vec
{\nabla}V\left(  X_{t}\right)  dt\nonumber\\
&  =\rho_{0}dp(X_{t},t) \label{Vclassical}%
\end{align}
and therefore%
\begin{equation}
dp(X_{t},t)=-\vec{\nabla}V\left(  X_{t}\right)  dt. \label{CLH2}%
\end{equation}
It is obvious to see the coupled equations of (\ref{CLH1}) and (\ref{CLH2}) be
just the classical Hamilton canonical equations by identifying $p(X_{t},t)$ of
Eq. (\ref{thetap}) as the classical momentum. Because there is no stochastic
term in these expressions, we can take the simple limit $dX_{t}\rightarrow0$
and $dp(X_{t},t)\rightarrow0$ as $dt\rightarrow0$ and thereby have the
Hamilton canonical equations of motion%

\begin{equation}
\left\{
\begin{array}
[c]{c}%
\frac{dX_{t}}{dt}=\frac{1}{m}p(X_{t},t)\text{ \ \ }\\
\\
\frac{dp(X_{t},t)}{dt}=-\vec{\nabla}V\left(  X_{t}\right)  .
\end{array}
\right.  \label{Canonical}%
\end{equation}

It is well known that the Schr\"{o}dinger equation converges to (more
precisely, correspond to) the Hamilton-Jacobi equation through WKB
theory\cite{Schiff, Messiah} (see also Sec. \ref{sec:ADFpath}) and also
through the Bohm representation as in Eq. (\ref{Bohm2}). The equation of
motion for the Wigner phase-space distribution function is reduced to the
classical Liouville
equation.\cite{wigner1932quantum,zachos2005quantum,polkovnikov2010phase} These
classical limits are seen in the distribution functions. On the other hand,
since we are treating the quantum path dynamics, its classical limit should be
the Hamilton canonical equations of motion. This indicates that the classical
trajectories are a classical approximation to the one-world paths.

Incidentally, the canonical equations of Eq. (\ref{Canonical}) are readily
reduced to the Newtonian equation%
\begin{equation}
m\frac{d^{2}X_{t}}{dt^{2}}=-\vec{\nabla}V\left(  X_{t}\right)  .
\end{equation}
On the other hand, the quantum canonical equations, Eq. (\ref{HM12}), are
reduced to%
\begin{align}
d^{2}X_{t}  &  =\frac{1}{m\rho}dP_{X_{t}}dt+\frac{1}{m\rho}P_{X_{t}}%
d^{2}t+\sqrt{\frac{\hbar}{m}}d^{2}W_{0}\left(  t,\omega\right) \nonumber\\
&  =-\frac{1}{m\rho}\left[  \vec{\nabla}V\right]  _{X_{t}}\left(  dt\right)
^{2}+\frac{1}{m\rho}P_{X_{t}}d^{2}t+\sqrt{\frac{\hbar}{m}}d^{2}W_{0}\left(
t,\omega\right)
\end{align}
under an assumption that $\rho$ is nearly constant at $(X_{t},t)$. 

The study of the classical limit of the one-world path dynamics highlights the
intrinsic quantum effects.

1. The Wiener process is found to be indeed critical to quantum mechanics. Equation
(\ref{ClassicalC}) and those following it have shown that the one-world path
is immediately fallen down to a classical path, indicating that
\textquotedblleft if a one-world path could be tracked deterministically due
to the absence of stochasticity, it should be subject to the law of classical
mechanics.\textquotedblright\ The contraposition of this statement is that
\textquotedblleft if a one-world path does not satisfy classical mechanics, it
cannot be tracked in a deterministic manner due to the presence of
stochasticity.\textquotedblright\ This last statement is consistent with why
Nelson began his theory with the stochastic equations.\cite{Nelson,nelson2012review}

2. The quantum effects are expected to appear more significantly when the
condition of Eq. (\ref{ClassicalC}) is violated and $\rho(X_{t},t)$ has a
broader spatial distribution around $X_{t}$, which indicates that the
stochastic path can wander about the area covered by $\psi(q,t)$. This aspect
is critical in semiclassical mechanics and will be discussed in Sec.
\ref{sec:ADFpath}.

3. By the stochastic dynamics is activated the main quantum effect on the
one-world paths, which arises from the fact that the drift velocity is a
function of $\psi(q,t)$. Besides, the time derivative of the velocity term
itself depends on $\rho(X_{t},t)$ in the term of $\left[  \vec{\nabla
}V\right]  _{X_{t}}$ as in Eq. (\ref{HM12}) and (\ref{Vclassical}). Meanwhile,
the one-world paths in turn contribute to the formation of $\rho(X_{t},t)$. Therefore, it should be stressed again that there is
a self-referring nonlinear relation between the parts (the one-world path) and
whole ($\rho(X_{t},t)$ and $\psi(q,t)$). By quantum nonlocality we mean the
present nonlinear relation and the resultant correlation among one-world paths
on separated occasions, which was discussed in Sec. \ref{Sec:correlation}.
Hence, the presupposition for the Bell inequality that has been figured out to
test the existence of the nonlocal hidden variables behind quantum dynamics is
unlikely to be applied to the present theoretical structure. Bell
assumed\cite{bell1964einstein} that a correlation between two observed
quantities $a$ and $b$ can be represented in the following simple expression
\begin{equation}
C(a,b)=\int d\lambda A\left(  a,\lambda\right)  \rho_{B}(\lambda)B\left(
b,\lambda\right)  , \label{Bell}%
\end{equation}
where $\rho_{B}(\lambda)$ is a probability distribution function for a
possible hidden parameter $\lambda$, satisfying $\int\rho_{B}(\lambda
)d\lambda=1$. He proved that
\begin{equation}
S=C(a,b)-C(a,b^{\prime})+C(a^{\prime},b)+C(a^{\prime},b^{\prime})
\end{equation}
should satisfy%
\begin{equation}
\left\vert S\right\vert \leq2.
\end{equation}
It is quite unlikely that the simple structure of Eq. (\ref{Bell}) can well
take account of the self-referring nonlinearity between the parts and the
whole, besides the effect of switching of the experimental circuit to
distinguish, for example, $(a,b)$ and $(a,b^{\prime})$ (the context of the
measurement). However, this aspect is beyond the scope of this article and
will be discussed elsewhere.

\subsection{Summary of this section: Single-event quantum path}

The Schr\"{o}dinger function $\psi(q,t)$ is supposed to represent a
\textit{coherent} sum of numerous physical events, each of which cannot be
singled out from it in general. The fringe (interference) pattern in the
double-slit experiment made up with an ensemble of the singly launched
electrons can be reproduced (by destruction of the quantum phase) in terms of
$\left\vert \psi(q,t)\right\vert ^{2}=\rho(q,t)$ on the measurement board.
Meanwhile, we have formulated in this section the quantum stochastic path
(one-world path) for each of electron, which connects $\rho(X_{0},0)$ to
$\rho(X_{t},t)$. It seems that many of confusion in the interpretation of the
Schr\"{o}dinger dynamics emerges from the lack of the discussion of the
quantum stochastic path (one-world path) and instead resorting to $\psi(q,t)$
alone to explain everything.

The dynamics of the one-world paths is given as%
\begin{equation}
dX_{t}=-\frac{\hbar}{m\rho}\left(
\begin{array}
[c]{cc}%
\phi_{r} & \phi_{c}%
\end{array}
\right)  \mathbf{J}\nabla\left(
\begin{array}
[c]{c}%
\phi_{r}\\
\phi_{c}%
\end{array}
\right)  dt+\sqrt{\frac{\hbar}{m}}dW_{0}\left(  t,\omega\right)  ,
\label{Stoch}%
\end{equation}
which includes the two conflicting factors: The drift term Eq.
(\ref{Realalpha})%
\begin{equation}
\bar{v}(q,t)\equiv-\frac{\hbar}{m\rho}\left(
\begin{array}
[c]{cc}%
\phi_{r} & \phi_{c}%
\end{array}
\right)  \text{\textbf{J}}\nabla\left(
\begin{array}
[c]{c}%
\phi_{r}\\
\phi_{c}%
\end{array}
\right)  \label{vel}%
\end{equation}
offers a particle velocity taken out of the smooth Schr\"{o}dinger
distribution, and the Wiener process resulting in
\begin{equation}
\left\langle \left(  \Delta q\right)  ^{2}\right\rangle =\frac{\hbar}{m}\Delta
t, \label{QB}%
\end{equation}
refuses the classical definition of the velocity
\begin{equation}
v=\lim_{\Delta t\rightarrow0}\frac{\Delta q}{\Delta t}. \label{nop}
\end{equation}
Thus the paths are indifferentiable almost everywhere. The presence of
$\psi(q,t)$ in $\bar{v}(x,t)$ and the Wiener process are both the origins of
the quantum nature in the one-word path dynamics: If the velocity drift term
$\bar{v}(q,t)$ is a local quantity, Eq. (\ref{Stoch}) is reduced merely to a
classical stochastic motion like the Brownian motion. If the Wiener process is
removed from Eq. (\ref{Stoch}), the path equation is eventually reduced to the
classical Hamilton canonical equations of motion due to the property of Eq.
(\ref{ClassicalC}). Thus each one-world path is regarded as a
\textquotedblleft physical path\textquotedblright, and the assumption of
sudden collapse of the Schr\"{o}dinger function (wave packet) has no place to
play a role.

A serious question remains on how the individual one-world paths can capture
the information of the Schr\"{o}dinger vector $\bar{\psi}(q,t)$ involved in
the velocity drift term of Eq. (\ref{Stoch}), in spite of the fact that the
total $\rho(q,t)$ is supposed to be a sum of all the information from those
individual stochastic paths and $\bar{\psi}(q,t)$ is a factorization of
$\rho(q,t)$. How does nature resolve such a nonlinear relationship between the
parts and the whole? This is identified to be the deepest mystery in quantum
mechanics. However, if we
deem that nature creates $\bar{v}(q,t)$ of Eq. (\ref{vel}) as a (classical)
field depending on the physical context, the question will be somewhat
relaxed. Besides, it would be exciting if the quantum Hamilton equations of
motion Eq. (\ref{HM12}) inspires a new development of semiclassical ideas.
This aspect will be discussed elsewhere.

\color{black}Before closing this section, we would like to touch upon the
characteristics of the dices God may play with. As for our ordinary dices, one
can roll a large set of dices at a time, or single dice many times. The law of
large numbers shows that the probability for a given role to appear should
converge to 1/6 in both ways. In quantum dynamics we can imagine an ensemble
of experimental events, but we cannot practically perform those experiments at
a time. (A double slit experiments in which to launch many electrons at a time
is different from the experiment to launch electrons one by one.)
Nevertheless, the Schr\"{o}dinger equation can mathematically realize the
situation as a coherent superposition of the events. Meanwhile, we can run a
one-world path one after another separately. However, in contrast to the
ordinary dices, the one-world paths indirectly correlate with one another
through the velocity field made up with the Schr\"{o}dinger function. Besides,
the density (or the density matrix) carried by all those one-world paths is expected to converge to the density
(density matrix) predicted by the relevant Schr\"{o}dinger function. This is
the rule of the quantum dice play.\ \color{black}

\section{Path dynamics on the manifolds of symplectic structure in parameter
spaces to time-evolve the Schr\"{o}dinger functions \label{sec:Variational}}

Here in this section and Sec. \ref{sec:ADFpath}, we study the space-time
propagation of the Schr\"{o}dinger function from the view point of path
dynamics. The physical meaning of the paths to be considered below is totally
different from that of the one-world stochastic paths studied above.

\subsection{Introductory remarks: The Schr\"{o}dinger partial differential
equation to coupled ordinary differential equations}

As we saw in Sec. \ref{sec:RealValue}, the real-valued Schr\"{o}dinger
dynamics stands only on space-time translational symmetry, and the equation of
continuity (also referred to as flux conservation), along with a variational
principle of Dirac-Frenkel type.\cite{broeckhove1988equivalence} No classical
mechanics was referred to in the derivation. Nevertheless we see a formal
proximity between the real-valued vector Schr\"{o}dinger equation (\ref{Sch3})
and the Hamilton canonical equations of motion such that%
\begin{equation}
\frac{\partial}{\partial t}\left(
\begin{array}
[c]{c}%
\phi_{r}\\
\phi_{c}%
\end{array}
\right)  =-\text{\textbf{J}}\frac{\hat{H}}{\hbar}\left(
\begin{array}
[c]{c}%
\phi_{r}\\
\phi_{c}%
\end{array}
\right)  \label{QE}%
\end{equation}
and%
\begin{equation}
\frac{d}{dt}\left(
\begin{array}
[c]{c}%
q\\
p
\end{array}
\right)  =-\text{\textbf{J}}\left(
\begin{array}
[c]{c}%
\partial/\partial q\\
\partial/\partial p
\end{array}
\right)  H_{cl}(q,p), \label{CE}%
\end{equation}
where $\hat{H}$ and $H_{cl}(q,p)$ are the quantum and classical Hamiltonian.
We also recall the quantum Hamilton equations of motion, Eqs. (\ref{HM12}).
The prominent difference between Eq. (\ref{QE}) and Eq. (\ref{CE}) is that the
former is partial differential equation with its solutions extending in
space-time, while the latter is composed of coupled ordinary differential
equations (ODE) having ray solutions (paths, trajectories) that are supposed
to run on the Lagrangian manifolds of the symplectic property in phase-space.
\textbf{J} in Eq. (\ref{CE}) highlights such a symplectic property. The
symplectic property is closely relevant to the conservation of total energy as
readily seen in Eq. (\ref{CE}) and moreover to the conservation of phase-space
volume elements as highlighted by the Poincar\'{e}-Cartan
theorem.\cite{Arnold} The Schr\"{o}dinger equation of Eq. (\ref{QE}) is also
characterized by \textbf{J}, which suggests that symplectic property should be
relevant to quantum dynamics, too. In fact, de
Gosson\cite{de2006symplectic,de2016principles} has extensively exhibited the
symplectic structure particularly of phase-space quantum mechanics. Such
phase-space quantum
mechanics,\cite{carruthers1983quantum,takatsuka1988phase,takatsuka1989phase,lee1995theory,zachos2005quantum,polkovnikov2010phase}
the dynamics of the Wigner phase-space distribution
function\cite{wigner1932quantum} being the first and most well-known example,
has a beautiful reduction to semiclassical solutions, which are represented in
terms of classical trajectories.\ However, the exact equation of motion for
the Wigner phase-space functions turns out to be a high order partial
differential equation, which is ironically far more complicated than the
Schr\"{o}dinger equation itself is.\cite{wigner1932quantum,zachos2005quantum}

In order to study another aspect of the symplectic property of the
Schr\"{o}dinger dynamics, we track the space-time deformation of the
Schr\"{o}dinger function $\psi(q,t)$ by mapping it onto an (infinitely) many
dimensional parameter space in such a way that $\psi(q,t)\rightarrow$
$\psi(q,t:\mathbf{u},\mathbf{v})$ or $\psi(\mathbf{u}(t),\mathbf{v}(t))$. The
dynamics of those paths $(\mathbf{u}(t),\mathbf{v}(t))$ is found to emerge
from the quantum mechanical Maupertuis-Hamilton variational
principle\cite{takatsuka2020maupertuis} with the functionals $\left\langle
\psi(\mathbf{u}(t),\mathbf{v}(t))\left\vert \hat{H}\right\vert \psi
(\mathbf{u}(t),\mathbf{v}(t))\right\rangle $ and $\left\langle \psi
(\mathbf{u}(t),\mathbf{v}(t))\left\vert i\hbar\frac{\partial}{\partial
t}\right\vert \psi(\mathbf{u}(t),\mathbf{v}(t))\right\rangle $. The
trajectories $(\mathbf{u}(t),\mathbf{v}(t))$ are actually driven by coupled
ordinary differential equations for $(\mathbf{u}(t),\mathbf{v}(t))$ and are
supposed to run on the manifolds of symplectic property. Thus the present path
dynamics reveals another mathematical and physical structure of quantum
Schr\"{o}dinger dynamics.

\subsection{Variational principles in quantum mechanics}

Before presenting the quantum mechanical Maupertuis-Hamilton variational
principle, we briefly mention to other variational methods developed for
quantum mechanics. The variational principles are inevitable for practical
studies in quantum mechanics: The Rayleigh--Ritz principle dominates the
eigenvalue problems for the stationary-sate Schr\"{o}dinger equation,
particularly in molecular electronic structure
theory.\cite{helgaker,pyykko2012introduction,dunning2023nature} In the
stationary-state quantum scattering problems, the so-called Kohn-type
variational principles based on the Schr\"{o}dinger
equation\cite{kohn1948variational,kato1950variational} and Schwinger principle
based on the Lippmann-Schwinger
equation\cite{lippmann1950variational,Takatsuka1984,nesbet2013variational,McKoy}
have been actively studied.

As a variational principle devised for the time-dependent Schr\"{o}dinger
dynamics, we briefly outline the theory of Kan.\cite{Kan} It starts from%
\begin{equation}
\delta\int\left\langle \psi(\mathbf{\alpha})\left\vert \left(  i\hbar
\frac{\partial}{\partial t}-\hat{H}\right)  \right\vert \psi(\mathbf{\alpha
})\right\rangle dt=0, \label{QEL}%
\end{equation}
where $\mathbf{\alpha}$ is a vector of \textit{real-valued} parameters. The
Euler-Lagrange principle immediately gives%

\begin{equation}
\frac{\partial}{\partial\mathbf{\alpha}}\mathcal{F}-\frac{d}{dt}\frac
{\partial}{\partial\dot{\alpha}}\mathcal{F}=0,
\end{equation}
with%
\begin{equation}
\mathcal{F}=\left\langle \psi(\mathbf{\alpha})\left\vert \left(  i\hbar
\frac{\partial}{\partial t}-\hat{H}\right)  \right\vert \psi(\mathbf{\alpha
})\right\rangle . \label{KF}%
\end{equation}
After some manipulation, coupled ODEs for $\mathbf{\alpha}(t)$ are given as
\begin{equation}
\frac{d\alpha_{k}}{dt}=\sum_{j}B_{kj}^{-1}\frac{\partial}{\partial\alpha_{j}%
}\left\langle \psi(\mathbf{\alpha})\left\vert \hat{H}\right\vert
\psi(\mathbf{\alpha})\right\rangle , \label{Kandyn}%
\end{equation}
with
\begin{equation}
\mathbf{B}_{jk}=i\hbar\left(  \left\langle \left.  \frac{\partial
\psi(\mathbf{\alpha})}{\partial\alpha_{j}}\right\vert \frac{\partial
\psi(\mathbf{\alpha})}{\partial\alpha_{k}}\right\rangle -\left\langle \left.
\frac{\partial\psi(\mathbf{\alpha})}{\partial\alpha_{k}}\right\vert
\frac{\partial\psi(\mathbf{\alpha})}{\partial\alpha_{j}}\right\rangle \right)
. \label{B}%
\end{equation}
In this study Kan concludes\cite{Kan} that there is no Legendre transformation
that allows $d\alpha_{k}/dt$ to be treated as independent variables as momenta
in classical mechanics, and thereby the theory is regarded to be closed. This
is because the right hand side of Eq. (\ref{Kandyn}) does not contain
momentum-like variables like $d\alpha_{k}/dt$ \cite{Kan}. Matrix $\mathbf{B}$
in Eq. (\ref{B}) can be singular in practice. The Kan variational principle
emerges from a family of the similar ideas.
\cite{mclachlan1964variational,kramer1981geometry,broeckhove1988equivalence,lavenda2003classical}%

Being beautiful, the variational principles of the above kinds do not seem to
reveal the symplectic structure behind the Schr\"{o}dinger dynamics. We
therefore consider below another class of the variational theory.

\subsection{Maupertuis-Hamilton principle in quantum parameter space}

In what follows we consider trial Schr\"{o}dinger functions in a form of
$\psi\left(  \mathbf{u,v}\right)  $, in which a pair of real-valued vectors
$\mathbf{u}$ and $\mathbf{v}$ having pairwise components $\left(  u_{i}%
,v_{i}\right)  $ are supposed to serve as parameters such that%

\begin{equation}
\left(  \mathbf{u},\mathbf{v}\right)  =\left(
\begin{array}
[c]{cccc}%
u_{1} & \cdots & u_{k} & \cdots\\
v_{1} & \cdots & v_{k} & \cdots
\end{array}
\right)  . \label{pairs}%
\end{equation}
We here consider only time-independent Hamiltonian, yet it is straightforward
to extend to time-dependent problems.

As in the classical Maupertuis least action principle\cite{Arnold}%
\begin{equation}
\delta\int pdq=\delta\int p\dot{q}dt=0,
\end{equation}
we consider the similar variational principle in the parameter space%
\begin{equation}
\delta\int\mathbf{v}\cdot d\mathbf{u}=\delta\int\mathbf{v}\cdot\mathbf{\dot
{u}}dt\mathbf{=}0
\end{equation}
under the following constraints.

\subsubsection{Variational principle under energy conservation}

A variational functional that conserves the energy can be readily made up in
analogy to the classical least action principle by simply writing
as\cite{takatsuka2020maupertuis}%
\begin{equation}
S_{H}=\int_{\mathit{C}}\left(  \mathbf{v\cdot\dot{u}}-\left\langle
\psi\left\vert \hat{H}\right\vert \psi\right\rangle \right)  dt, \label{varH}%
\end{equation}
where $\mathit{C}$ indicates an arbitrary time interval under study, and
$\int_{\mathit{C}}\mathbf{v\cdot\dot{u}}dt=$ $\int_{\mathit{C}}\mathbf{v\cdot
}d\mathbf{\dot{u}}$ is the reduced action. The variation of $S_{H}$ gives
%\onecolumngrid%
\begin{align}
&  \delta S_{H}=\left[  \mathbf{v\cdot}\delta\mathbf{u}\right]  _{end1}%
^{end2}\nonumber\\
&  +\int_{\mathit{C}}\left(  -\frac{d\mathbf{v}}{dt}-\frac{\partial
\left\langle \psi\left\vert \hat{H}\right\vert \psi\right\rangle }%
{\partial\mathbf{u}}\right)  \mathbf{\cdot}\delta\mathbf{u}dt\nonumber\\
&  \mathbf{+}\int_{\mathit{C}}\left(  \frac{d\mathbf{u}}{dt}-\frac
{\partial\left\langle \psi\left\vert \hat{H}\right\vert \psi\right\rangle
}{\partial\mathbf{v}}\right)  \mathbf{\cdot}\delta\mathbf{v}dt=0 \label{delSH}%
\end{align}
under the fixed boundary condition
\begin{equation}
\left[  \mathbf{v\cdot}\delta\mathbf{u}\right]  _{end1}^{end2}=0.
\label{boundary}%
\end{equation}
It immediately follows that%
\begin{equation}
\frac{d\mathbf{u}}{dt}=\frac{\partial\left\langle \psi\left\vert \hat
{H}\right\vert \psi\right\rangle }{\partial\mathbf{v}} \label{LA21}%
\end{equation}
and%
\begin{equation}
\frac{d\mathbf{v}}{dt}=-\frac{\partial\left\langle \psi\left\vert \hat
{H}\right\vert \psi\right\rangle }{\partial\mathbf{u}}. \label{LA22}%
\end{equation}
Since both $\delta\mathbf{u}$ and $\delta\mathbf{v}$ in Eq. (\ref{delSH}) are
individually arbitrary, $\delta S_{H}=0$ requires Eqs. (\ref{LA21}) and
(\ref{LA22}) to identically hold. Thus we have%
\begin{equation}
\frac{d}{dt}\left(
\begin{array}
[c]{c}%
\mathbf{v}\\
\mathbf{u}%
\end{array}
\right)  =\text{\textbf{J}}\left(
\begin{array}
[c]{c}%
\partial/\partial\mathbf{v}\\
\partial/\partial\mathbf{u}%
\end{array}
\right)  \left\langle \psi\left\vert \hat{H}\right\vert \psi\right\rangle
\label{pHamil}%
\end{equation}
which is mathematically of the same form as Eq. (\ref{CE}).

Along the flow lines $\left(  \mathbf{u}(t)\mathbf{,v}(t)\right)  $ thus
determined, the energy conservation is readily ensured by Eq. (\ref{pHamil}).
An advantage of this simple method is that quantum effects such as the
zero-point energy along the path are naturally taken into account, and yet
their resultant paths can be similar to those of the classical counterpart.
Yet, it is obvious that the energy conservation alone is not enough to
determine the correct Schr\"{o}dinger dynamics. In other words, $\delta
S_{H}=0$ is not sufficient. \ \ 

\subsubsection{Variational principle under flux conservation}

As in the energy conservation, we need to consider the flux conservation (the
equation of continuity) in terms of $\left\langle \psi(\mathbf{u}%
(t),\mathbf{v}(t))\left\vert \frac{\partial}{\partial t}\right\vert
\psi(\mathbf{u}(t),\mathbf{v}(t))\right\rangle $ as an independent condition.
The energy conservation and flux conservation are mutually independent as we
saw in Eqs. (\ref{var}) and (\ref{Sch2}) and the discussion above them. We
therefore consider the least action principle under the condition of flux
conservation with a variational functional%
\begin{align}
S_{W}(\mathbf{u,v})  &  =\int_{\mathit{C}}\left(  \mathbf{v\cdot\dot{u}%
}-\left\langle \psi\left\vert i\hbar\frac{\partial}{\partial t}\right\vert
\psi\right\rangle \right)  dt\nonumber\\
&  =\int_{\mathit{C}}\left(  \mathbf{v\cdot\dot{u}}-i\hbar\left\langle \left.
\psi\right\vert \dot{\phi}\right\rangle \right)  dt \label{varW}%
\end{align}
demanding%
\begin{equation}
\delta S_{W}(\mathbf{u,v})=0, \label{delSW}%
\end{equation}
which gives rise to another coupled ODE, that is
\begin{equation}
\frac{d\mathbf{u}}{dt}=i\hbar\frac{\partial\left\langle \left.  \psi
\right\vert \dot{\psi}\right\rangle }{\partial\mathbf{v}} \label{Wu}%
\end{equation}
and
\begin{equation}
\frac{d\mathbf{v}}{dt}=-i\hbar\frac{\partial\left\langle \left.
\psi\right\vert \dot{\psi}\right\rangle }{\partial\mathbf{u}}, \label{Wv}%
\end{equation}
or
\begin{equation}
\frac{d}{dt}\left(
\begin{array}
[c]{c}%
\mathbf{v}\\
\mathbf{u}%
\end{array}
\right)  =\text{\textbf{J}}\left(
\begin{array}
[c]{c}%
\partial/\partial\mathbf{v}\\
\partial/\partial\mathbf{u}%
\end{array}
\right)  \left\langle \psi\left\vert i\hbar\frac{\partial}{\partial
t}\right\vert \psi\right\rangle .
\end{equation}
A flux conservation law follows directly as%
\begin{equation}
\frac{d}{dt}\left\langle \left.  \psi\right\vert \dot{\psi}\right\rangle
=\frac{d\mathbf{v}}{dt}\cdot\frac{\partial\left\langle \left.  \psi\right\vert
\dot{\psi}\right\rangle }{\partial\mathbf{v}}+\frac{d\mathbf{u}}{dt}\cdot
\frac{\partial\left\langle \left.  \psi\right\vert \dot{\psi}\right\rangle
}{\partial\mathbf{u}}=0. \label{fcon}%
\end{equation}
In Sec. \ref{sec:RealValue} we have stressed the critical importance of the
flux conservation arising from the incompressible flow that $\left\vert
\psi(q,t)\right\vert ^{2}$ induces in the $(q,t)$ space. Likewise, we need to
impose the flux conservation in the $\left(  \mathbf{u},\mathbf{v}%
,\mathbf{\,}t\right)  $ space in addition to the energy conservation of Eq.
(\ref{pHamil}). It is not clear if the variational functional of Eq.
(\ref{QEL}) conserves the norm in any $\mathbf{\alpha}$-parameter space unconditionally.

We analyze $S_{W}$ in a little more precisely, resuming from%
\begin{equation}
i\hbar\left\langle \left.  \psi\right\vert \dot{\psi}\right\rangle
=\frac{i\hbar}{2}\left(  \left\langle \left.  \psi\right\vert \dot{\psi
}\right\rangle -\left\langle \left.  \dot{\psi}\right\vert \psi\right\rangle
\right)  , \label{jjj}%
\end{equation}
which is real-valued. Rewrite the right hand side as%

\begin{align}
&  \left\langle \left.  \psi\right\vert \dot{\psi}\right\rangle -\left\langle
\left.  \dot{\psi}\right\vert \psi\right\rangle \nonumber\\
&  =\frac{d\mathbf{u}}{dt}\cdot\int\left(  \psi^{\ast}\frac{\partial\psi
}{\partial\mathbf{u}}-\psi\frac{\partial\psi^{\ast}}{\partial\mathbf{u}%
}\right)  d\mathbf{q}\\
&  +\frac{d\mathbf{v}}{dt}\cdot\int\left(  \psi^{\ast}\frac{\partial\psi
}{\partial\mathbf{v}}-\phi\frac{\partial\psi^{\ast}}{\partial\mathbf{v}%
}\right)  d\mathbf{q,}%
\end{align}
in which the fluxes in the parameter space have manifested themselves. For
instance,
\begin{equation}
\int\left(  \psi^{\ast}\frac{\partial\psi}{\partial\mathbf{u}}-\psi
\frac{\partial\psi^{\ast}}{\partial\mathbf{u}}\right)  d\mathbf{q}%
\end{equation}
corresponds to the gradient of the field of fluid in $\mathbf{u}$-direction.
Let us recall the definition of quantum mechanical flux (current of
probability density) \cite{Schiff} which is defined as
\begin{equation}
\mathbf{j}\left(  \mathbf{q}\right)  \mathbf{=}\frac{i\hbar}{2m}\left(
\psi^{\ast}\nabla\psi-\psi\nabla\psi^{\ast}\right)  ,
\end{equation}
with scaling the mass to $m=1$. By analogy, we may define fluxes (probability
current) in the parameter space in the $\mathbf{u}$ and $\mathbf{v}$
directions, respectively, such that%
\begin{equation}
\mathbf{j}_{\mathbf{u}}\equiv\frac{i\hbar}{2}\int\left(  \phi^{\ast}%
\frac{\partial\psi}{\partial\mathbf{u}}-\phi\frac{\partial\psi^{\ast}%
}{\partial\mathbf{u}}\right)  d\mathbf{q}=-\hbar\operatorname{Im}\left\langle
\left.  \psi\right\vert \frac{\partial\psi}{\partial\mathbf{u}}\right\rangle ,
\label{Ju}%
\end{equation}
and%
\begin{equation}
\mathbf{j}_{\mathbf{v}}\equiv\frac{i\hbar}{2}\int\left(  \psi^{\ast}%
\frac{\partial\psi}{\partial\mathbf{v}}-\psi\frac{\partial\psi^{\ast}%
}{\partial\mathbf{v}}\right)  d\mathbf{q}=-\hbar\operatorname{Im}\left\langle
\left.  \psi\right\vert \frac{\partial\psi}{\partial\mathbf{v}}\right\rangle .
\label{Jv}%
\end{equation}
After all we have
\begin{equation}
i\hbar\left\langle \left.  \psi\right\vert \dot{\psi}\right\rangle
=\frac{d\mathbf{u}}{dt}\cdot\mathbf{j}_{\mathbf{u}}+\frac{d\mathbf{v}}%
{dt}\cdot\mathbf{j}_{\mathbf{v}}, \label{Juv}%
\end{equation}
which gives an alternative interpretation of $\left\langle \left.
\psi\right\vert \dot{\psi}\right\rangle $ in terms of the flow in the
parameter space. Flux of such mathematical form exists in classical mechanics, and therefore Eq.
(\ref{Juv}) represents a genuine quantum nature.

Back to Eqs. (\ref{Wu})-(\ref{Wv}), they are rewritten as
\begin{equation}
\frac{d\mathbf{u}}{dt}=\frac{\partial}{\partial\mathbf{v}}\left(
\frac{d\mathbf{u}}{dt}\cdot\mathbf{j}_{\mathbf{u}}+\frac{d\mathbf{v}}{dt}%
\cdot\mathbf{j}_{\mathbf{v}}\right)  \label{WWu}%
\end{equation}
and
\begin{equation}
\frac{d\mathbf{v}}{dt}=-\frac{\partial}{\partial\mathbf{u}}\left(
\frac{d\mathbf{u}}{dt}\cdot\mathbf{j}_{\mathbf{u}}+\frac{d\mathbf{v}}{dt}%
\cdot\mathbf{j}_{\mathbf{v}}\right)  . \label{WWv}%
\end{equation}
Since $d\mathbf{u/}dt$ and $d\mathbf{v/}dt$ appear in both Eqs. (\ref{WWu})
and (\ref{WWv}), they are essentially nonlinear and hard to integrate in these
forms. We return to this aspect later.

\subsubsection{A simple example bringing back to classical
dynamics\label{Sec:coherentstate}}

For a later purpose, we here show a simple example from the so-called coherent
state wavepacket
\begin{equation}
\psi\left(  q,t\right)  =N\exp\left(  -\alpha\left(  q-q_{0}(t)\right)
^{2}+\frac{i}{\hbar}p_{0}(t)\left(  q-q_{0}(t)\right)  \right)  . \label{coh}%
\end{equation}
with $\alpha>0$ being deliberately fixed. This is frequently used in practice
as an expansion basis (see for instance ref. \cite{shalashilin2008gaussian}).
Set $q_{0}(t)\rightarrow\mathbf{u}(t)$ and $p_{0}(t)\rightarrow\mathbf{v}$. Noting%

\begin{equation}
\left\langle \psi\left(  q,t\right)  \left\vert \frac{d^{2}}{dq^{2}%
}\right\vert \psi\left(  q,t\right)  \right\rangle =-\alpha-\left(
\frac{p_{0}(t)}{\hbar}\right)  ^{2} \label{KIN}%
\end{equation}
we see%
\begin{equation}
\frac{dq_{0}}{dt}=\frac{\partial}{\partial p_{0}}\left\langle \psi\left(
q,t\right)  \left\vert \hat{H}\right\vert \psi\left(  q,t\right)
\right\rangle =\frac{p_{0}}{m}. \label{HLq}%
\end{equation}
On the other hand, since $N^{2}\exp\left(  -2\alpha\left(  q-q_{0}(t)\right)
^{2}\right)  \rightarrow\delta(q-q_{0}(t))$ as $\alpha\rightarrow\infty$
\begin{align}
\left\langle \psi\left(  q,t\right)  \left\vert V\right\vert \psi\left(
q,t\right)  \right\rangle  &  =N^{2}\int dqV(q)\exp\left(  -2\alpha\left(
q-q_{0}(t)\right)  ^{2}\right) \nonumber\\
&  \rightarrow V(q_{0}(t))\text{ \ \ as \ }\alpha\rightarrow\infty
\end{align}
and
\begin{equation}
\frac{\partial}{\partial q_{0}}\left\langle \psi\left(  q,t\right)  \left\vert
\hat{H}\right\vert \psi\left(  q,t\right)  \right\rangle \rightarrow
\frac{\partial}{\partial q_{0}}V(q_{0}(t))\text{ \ \ as \ }\alpha
\rightarrow\infty. \label{HLp}%
\end{equation}
Thus the combination of Eqs. (\ref{HLq}) and (\ref{HLp}) brings us back to the
canonical equations Eq. (\ref{CE}) in this limit. The above feature is well
known in the studies of practical application of the coherent
states.\cite{frozenG,shalashilin2008gaussian}

Hence, it turns out that this simple example bring us back to the classical
Maupertuis-Hamilton principle
\begin{equation}
\delta\int\left(  p_{0}d\dot{q}-H_{cl}(q_{0},p_{0})\right)  dt=0,
\end{equation}
which is the simplest example of $\delta S_{H}=0$ only. In more general cases,
we need to consider $\delta S_{W}=0$.

Next we see the flux in the parameter space according to Eqs. (\ref{Ju}) and
(\ref{Jv}). The results are simple%

\begin{equation}
\mathbf{j}_{q_{0}}=-\hbar\operatorname{Im}\left\langle \left.  \psi\right\vert
\frac{\partial\psi}{\partial q_{0}}\right\rangle =p_{0}, \label{jq}%
\end{equation}
and%
\begin{equation}
\mathbf{j}_{p_{0}}\equiv-\hbar\operatorname{Im}\left\langle \left.
\psi\right\vert \frac{\partial\psi}{\partial p_{0}}\right\rangle =0.
\label{jp}%
\end{equation}
There is no flux in the $p_{0}$ space. On the other hand, $\mathbf{j}_{q_{0}}$
in Eq. (\ref{jq}) seems natural and is consistent with the quantum momentum average%

\begin{equation}
\left\langle p\right\rangle =\frac{\hbar}{i}\left\langle \left.
\psi\right\vert \frac{\partial\psi}{\partial q}\right\rangle , \label{dq}%
\end{equation}
This trivial-looking example will show later why the flux-conservation
condition $\delta S_{W}(\mathbf{u,v})=0$ is not necessary in this particular case.

\bigskip

\subsection{Intersection of two independent manifolds of symplectic structure
\label{sec:practice}}

We now have two variational conditions $\delta S_{H}\left(  \mathbf{u}\left(
t\right)  \mathbf{,v}\left(  t\right)  \right)  =0$ and $\delta S_{W}\left(
\mathbf{u}\left(  t\right)  \mathbf{,v}\left(  t\right)  \right)  =0$, which
are mathematically independent from each other. For them to physically
represent the Schr\"{o}dinger dynamics, we need to combine them.

\subsubsection{Dual least action principle with the symplectic structure in
the parameter space \label{sec:real}}

It is obvious that two conditions both should be fulfilled simultaneously such
that
\begin{equation}
\delta S_{W}(\mathbf{u,v})=\delta S_{W}(\mathbf{u,v})=0, \label{dual2}%
\end{equation}
which naturally brings us back
\begin{equation}
\delta\int_{\mathit{C}}\left\langle \psi(\mathbf{u,v})\left\vert \left(
i\hbar\frac{\partial}{\partial t}-\hat{H}\right)  \right\vert \psi
(\mathbf{u,v})\right\rangle dt=0. \label{totalvar}%
\end{equation}
This one is obviously weaker a condition than that of Eq. (\ref{dual2}). To
summarize, we need to study the simultaneous variational principles
\begin{equation}
\left\{
\begin{array}
[c]{c}%
\delta S_{H}\left(  \mathbf{u}\left(  t\right)  \mathbf{,v}\left(  t\right)
\right)  -\delta S_{W}\left(  \mathbf{u}\left(  t\right)  \mathbf{,v}\left(
t\right)  \right)  =0\\
\delta S_{H}\left(  \mathbf{u}\left(  t\right)  \mathbf{,v}\left(  t\right)
\right)  =0\\
\delta S_{W}\left(  \mathbf{u}\left(  t\right)  \mathbf{,v}\left(  t\right)
\right)  =0,
\end{array}
\right.  \text{ \ \ \ } \label{dual}%
\end{equation}
which is a dual least action principle.

An obvious reason why the functional $S_{W}$ appears to be necessary in
addition to $S_{H}$ is because $S_{H}$ of Eq. (\ref{varH}) alone does not
warrant the Schr\"{o}dinger dynamics, because they are physically independent
as already stressed in Sec. \ref{sec:RealValue}. As another obvious example
assume a hypothetical (or Klein-Gordon) dynamics
\begin{equation}
\frac{\partial^{2}}{\partial t^{2}}\psi=\hat{H}\psi. \label{KG}%
\end{equation}
Then it should be required that
\begin{equation}
S_{W}(\mathbf{u,v})=\int_{\mathit{C}}\left(  \mathbf{v\cdot\dot{u}%
}-\left\langle \psi\left\vert \frac{\partial^{2}}{\partial t^{2}}\right\vert
\psi\right\rangle \right)  dt \label{KGKG}%
\end{equation}
even if the same Hamiltonian operator is adopted. Hence the condition $\delta
S_{W}=0$ determines the way of propagation of the flow arising from $\psi$,
while $\delta S_{H}=0$ is \textit{mainly} responsible for energetics.

\subsubsection{Working equations coupling the energy and flux conservations}

Equations (\ref{WWu})-(\ref{WWv}) are nonlinear and are not useful in the
present form. Besides, it is unclear what drives the flow there. We therefore
implant $d\mathbf{u/}dt$ and $d\mathbf{v/}dt$ of Eqs. (\ref{LA21}%
)-(\ref{LA22}) into Eqs. (\ref{WWu})-(\ref{WWv}), respectively, such that%
\begin{equation}
\frac{d\mathbf{u}}{dt}=\frac{\partial}{\partial\mathbf{v}}\left(
\frac{\partial\left\langle \psi\left\vert \hat{H}\right\vert \psi\right\rangle
}{\partial\mathbf{v}}\cdot\mathbf{j}_{\mathbf{u}}-\frac{\partial\left\langle
\psi\left\vert \hat{H}\right\vert \psi\right\rangle }{\partial\mathbf{u}}%
\cdot\mathbf{j}_{\mathbf{v}}\right)  \label{HWu}%
\end{equation}
and%
\begin{equation}
\frac{d\mathbf{v}}{dt}=-\frac{\partial}{\partial\mathbf{u}}\left(
\frac{\partial\left\langle \psi\left\vert \hat{H}\right\vert \psi\right\rangle
}{\partial\mathbf{v}}\cdot\mathbf{j}_{\mathbf{u}}-\frac{\partial\left\langle
\psi\left\vert \hat{H}\right\vert \psi\right\rangle }{\partial\mathbf{u}}%
\cdot\mathbf{j}_{\mathbf{v}}\right)  . \label{HWv}%
\end{equation}
These serve as our working equations, which should be integrated\ together
with Eqs. (\ref{LA21})-(\ref{LA22}).

Incidentally, the factor in Eqs. (\ref{HWu}) and (\ref{HWv}), for instance, is
rewritten as
\begin{align}
&  \frac{\partial\left\langle \psi\left\vert \hat{H}\right\vert \psi
\right\rangle }{\partial\mathbf{v}}\cdot\mathbf{j}_{\mathbf{u}}-\frac
{\partial\left\langle \psi\left\vert \hat{H}\right\vert \psi\right\rangle
}{\partial\mathbf{u}}\cdot\mathbf{j}_{\mathbf{v}}\nonumber\\
&  =\left(
\begin{array}
[c]{cc}%
\partial\left\langle \psi\left\vert \hat{H}\right\vert \psi\right\rangle
/\partial\mathbf{u} & \partial\left\langle \psi\left\vert \hat{H}\right\vert
\psi\right\rangle /\partial\mathbf{v}%
\end{array}
\right)  \left(
\begin{array}
[c]{cc}%
\mathbf{0} & -\mathbf{I}\\
\mathbf{I} & \mathbf{0}%
\end{array}
\right) \nonumber\\
&  \times\left(
\begin{array}
[c]{c}%
\mathbf{j}_{\mathbf{u}}\\
\mathbf{j}_{\mathbf{v}}%
\end{array}
\right)  ,
\end{align}
which is a symplectic inner product between the Hamiltonian derivative vector
and the flux vector, thereby representing a coupling between the Hamilton
dynamics and the flow dynamics for the quantum distribution $\psi$. We thus
have prepared all the necessary quantities to propagate the Schr\"{o}dinger
function in the parameter space.

\subsubsection{Connecting piecewise trajectories}

An exact solution of the Schr\"{o}dinger equation having a proper set of
infinite number of parameters $\psi(\mathbf{u}(t),\mathbf{v}(t))$ should
naturally satisfy the two dynamical equations, Eqs. (\ref{LA21})-(\ref{LA22})
from the Hamilton dynamics and Eqs. (\ref{WWu})-(\ref{WWv}) (or Eqs.
(\ref{HWu})-(\ref{HWv})) from the flow dynamics, which is obvious by construction.

As for our non-exact trial functions, we need to consider a technical aspect
in a little more systematic manner. Suppose we carry out a short time ($\Delta
t$) propagation in Eqs. (\ref{LA21})-(\ref{LA22}), which is formally written
as%
\begin{equation}
\left(
\begin{array}
[c]{c}%
\mathbf{u}(t+\Delta t)\\
\mathbf{v}(t+\Delta t)
\end{array}
\right)  _{H}=\mathbf{F}\left(  t;\Delta t\right)  \left(
\begin{array}
[c]{c}%
\mathbf{u}(t)\\
\mathbf{v}(t)
\end{array}
\right)  _{H}%
\end{equation}
and likewise from Eqs. (\ref{HWu})-(\ref{HWv}) we have%
\begin{equation}
\left(
\begin{array}
[c]{c}%
\mathbf{u}(t+\Delta t)\\
\mathbf{v}(t+\Delta t)
\end{array}
\right)  _{W}=\mathbf{G}\left(  t;\Delta t\right)  \left(
\begin{array}
[c]{c}%
\mathbf{u}(t)\\
\mathbf{v}(t)
\end{array}
\right)  _{W}.
\end{equation}
The suffices $H$ and $W$ indicate the Hamilton dynamics and the flow dynamics,
respectively. Finite time solution of each is to be formally given as
%\onecolumngrid%
\begin{align}
&  \left(
\begin{array}
[c]{c}%
\mathbf{u}(t)\\
\mathbf{v}(t)
\end{array}
\right)  _{H}\nonumber\\
&  =\mathbf{F}\left(  t-\Delta t;\Delta t\right)  \cdots\mathbf{F}\left(
\Delta t;\Delta t\right)  \mathbf{F}\left(  0;\Delta t\right)  \left(
\begin{array}
[c]{c}%
\mathbf{u}(0)\\
\mathbf{v}(0)
\end{array}
\right)
\end{align}
and
\begin{align}
&  \left(
\begin{array}
[c]{c}%
\mathbf{u}(t)\\
\mathbf{v}(t)
\end{array}
\right)  _{W}\nonumber\\
&  =\mathbf{G}\left(  t-\Delta t;\Delta t\right)  \cdots\mathbf{G}\left(
\Delta t;\Delta t\right)  \mathbf{G}\left(  0;\Delta t\right)  \left(
\begin{array}
[c]{c}%
\mathbf{u}(0)\\
\mathbf{v}(0)
\end{array}
\right)  .
\end{align}
If $\psi(\mathbf{u}(t),\mathbf{v}(t))$ was exact globally in the parameter
space, it should hold that $\mathbf{F}\left(  t;\Delta t\right)
=\mathbf{G}\left(  t;\Delta t\right)  $ under the same initial conditions and
\begin{equation}
\left(
\begin{array}
[c]{c}%
\mathbf{u}(t)\\
\mathbf{v}(t)
\end{array}
\right)  _{H}=\left(
\begin{array}
[c]{c}%
\mathbf{u}(t)\\
\mathbf{v}(t)
\end{array}
\right)  _{W}.
\end{equation}
However, in our practical approximations, these two solutions at a finite time
$t$ can deviate from each other, depending on the physical conditions given.
We therefore assume that the simultaneous variational principles $\delta
S_{H}=0$ and $\delta S_{W}=0$ are satisfied only for a very short time
segments $\Delta t$. The finite time solutions are to be obtained by
connecting the short-time segment solutions of $\mathbf{F}$\ and $\mathbf{G}$
in such a way that%
\begin{align}
&  \left(
\begin{array}
[c]{c}%
\mathbf{u}(t)\\
\mathbf{v}(t)
\end{array}
\right) \nonumber\\
&  =\lim_{\Delta t\rightarrow0}\mathbf{G}\left(  t-\Delta t/2;\Delta
t/2\right)  \mathbf{F}\left(  t-\Delta t;\Delta t/2\right)  \cdots\nonumber\\
&  \times\mathbf{G}\left(  \Delta t/2;\Delta t/2\right)  \mathbf{F}\left(
0;\Delta t/2\right)  \left(
\begin{array}
[c]{c}%
\mathbf{u}(0)\\
\mathbf{v}(0)
\end{array}
\right)  . \label{alt}%
\end{align}
Incidentally, we note the higher-order symplectic integrators can be applied
in practical calculations.\cite{Yoshida}

\subsubsection{A condition for the path dynamics to be well-posed}

In order for $d\mathbf{u}/dt$ and $d\mathbf{v}/dt$\ of Eqs. (\ref{LA21}%
)-(\ref{LA22}) and Eqs. (\ref{Wu})-(\ref{Wv}) (in place of Eqs. (\ref{HWu}%
)-(\ref{HWv}) for shorter notation) to be mathematically consistent and for
the time-propagation by means of these sets of ordinary differential equations
to be well-posed, the following condition needs to be fulfilled%
\begin{equation}
\left(
\begin{array}
[c]{cc}%
\partial\left\langle \psi\left\vert \hat{H}\right\vert \psi\right\rangle
/\partial\mathbf{u} & \partial\left\langle \psi\left\vert \hat{H}\right\vert
\psi\right\rangle /\partial\mathbf{v}\\
\partial\left\langle \left.  \psi\right\vert \dot{\psi}\right\rangle
/\partial\mathbf{u} & \partial\left\langle \left.  \psi\right\vert \dot{\psi
}\right\rangle /\partial\mathbf{v}%
\end{array}
\right)  \left(
\begin{array}
[c]{c}%
d\mathbf{u}/dt\\
d\mathbf{v}/dt
\end{array}
\right)  =0 \label{conb}%
\end{equation}
(see Eq. (\ref{fcon}) and its counterpart for energy conservation). This is a
Poisson bracket of $\left\langle \psi\left\vert \hat{H}\right\vert
\psi\right\rangle $ and $\left\langle \left.  \psi\right\vert \dot{\psi
}\right\rangle $ with respect to the variables $\left(  \mathbf{u}%
,\mathbf{v}\right)  $. And further, for Eq. (\ref{conb}) to give nontrivial
solutions for the vector $\left(  d\mathbf{u}/dt,d\mathbf{v}/dt\right)  $, the
relevant determinant should vanish, that is,%
\begin{equation}
\left\vert
\begin{array}
[c]{cc}%
\partial\left\langle \psi\left\vert \hat{H}\right\vert \psi\right\rangle
/\partial\mathbf{u} & \partial\left\langle \psi\left\vert \hat{H}\right\vert
\psi\right\rangle /\partial\mathbf{v}\\
\partial\left\langle \left.  \psi\right\vert \dot{\psi}\right\rangle
/\partial\mathbf{u} & \partial\left\langle \left.  \psi\right\vert \dot{\psi
}\right\rangle /\partial\mathbf{v}%
\end{array}
\right\vert =0. \label{condet}%
\end{equation}

Likewise, the choice of initial conditions $(\mathbf{u}(0),\mathbf{v}(0))$
should practically satisfy
\begin{equation}
\det\left\vert
\begin{array}
[c]{cc}%
\partial\left\langle \psi\left\vert \hat{H}\right\vert \psi\right\rangle
/\partial u_{i} & \partial\left\langle \psi\left\vert \hat{H}\right\vert
\psi\right\rangle /\partial v_{i}\\
\partial\left\langle \left.  \psi\right\vert \dot{\psi}\right\rangle /\partial
u_{i} & \partial\left\langle \left.  \psi\right\vert \dot{\psi}\right\rangle
/\partial v_{i}%
\end{array}
\right\vert =0 \label{vdet}%
\end{equation}
on each pair of the initial components $(u_{i}(0),v_{i}(0))$ prior to the
integration. Otherwise, the parameter setting is to be judged ill-conditioned.
Choice of the linear dependent and/or inconsistent parameters can be avoided
by monitoring Eq. (\ref{condet}).

\subsection{Conditions for the energy conservation alone to give correct
parameter paths\label{sec:parameters}}

Back to the simplest case of the coherent state dynamics, Sec.
\ref{Sec:coherentstate}, we saw that the Hamilton dynamics alone, that is a
pair of Eqs. (\ref{HLq}) and (\ref{HLp}), suffices to propagate the
Schr\"{o}dinger dynamics (approximately). There are mathematical conditions to
ensure that $\delta S_{W}=0$ for the flow dynamics is fulfilled automatically.
Some are discussed below. Otherwise, both $S_{H}$ and $S_{W}$ are definitely
necessary for the quantum Maupertuis-Hamilton variational principle to
determine the energy and flow dynamics.

The flow-dynamics variational functional $S_{W}$ is identically nullified
\begin{align}
S_{W}  &  =\int\left(  \mathbf{v\cdot\dot{u}-}\left\langle \psi(\mathbf{u,v}%
)\left\vert i\hbar\frac{\partial}{\partial t}\right\vert \psi(\mathbf{u,v}%
)\right\rangle \right)  dt\nonumber\\
&  =\int\left(  \mathbf{v\cdot\dot{u}}-i\hbar\mathbf{\dot{u}}\cdot\left\langle
\left.  \psi\right\vert \frac{\partial\psi}{\partial\mathbf{u}}\right\rangle
-i\hbar\mathbf{\dot{v}\cdot}\left\langle \left.  \psi\right\vert
\frac{\partial\psi}{\partial\mathbf{v}}\right\rangle \right)  dt\nonumber\\
&  =0, \label{zeroSW}%
\end{align}
when
\begin{equation}
-\mathbf{j}_{u}=\mathbf{v} \label{parav}%
\end{equation}
and
\begin{equation}
-\mathbf{j}_{\mathbf{v}}=\mathbf{0} \label{parau}%
\end{equation}
are simultaneously fulfilled (see Eqs. (\ref{Ju}) and (\ref{Jv})). Such pairs
of the parameters, referred to as the dynamical pair, reduce the entire
variational functional to%
\begin{align}
&  \delta\int\left\langle \psi(\mathbf{u,v})\left\vert \left(  i\hbar
\frac{\partial}{\partial t}-\hat{H}\right)  \right\vert \psi(\mathbf{u,v}%
)\right\rangle dt\nonumber\\
&  =\delta\int\left(  \mathbf{v\cdot\dot{u}}-\left\langle \psi(\mathbf{u,v}%
)\left\vert \hat{H}\right\vert \psi(\mathbf{u,v})\right\rangle \right)  dt=0,
\label{reduction}%
\end{align}
which is exactly the same as $\delta S_{H}=0$ of Eq. (\ref{varH}). Thus we are
left only with the ODE of the type of the Hamilton canonical equations, Eqs.
(\ref{LA21})-(\ref{LA22}). Further, it is readily proved that both Eqs.
(\ref{Wu}) and (\ref{Wv}) are accordingly satisfied.
\cite{takatsuka2020maupertuis}

In the case of the coherent state in Sec. \ref{Sec:coherentstate}, we have
already had Eq. (\ref{jq}) and (\ref{jp}), which correspond to Eq.
(\ref{parav}) and (\ref{parau}), respectively. Incidentally, for a little
skewed functional form of the coherent state as
\begin{align}
\psi\left(  q,t\right)   &  =N(1+c\left(  q-q_{0}(t)\right)  )\exp\left(
-\alpha(t)\left(  q-q_{0}(t)\right)  ^{2}\right) \nonumber\\
&  \times\exp\left(  \frac{i}{\hbar}p_{0}(t)\left(  q-q_{0}(t)\right)
\right)  ,
\end{align}
$j_{v}$ is no longer zero, and hence $(q_{0},p_{0})$ is no longer a dynamical
pair either. To determine these parameters we need the help of $\delta
S_{W}=0$. For the same reason, $\left(  q_{01}(t),p_{01}(t)\right)  $ and
$\left(  q_{02}(t),p_{02}(t)\right)  $ in the following linear combination of
one-particle one-dimensional Gaussians
\begin{align}
&  \psi\left(  q,t\right)  =N\sum_{j\geq2}c_{j}(t)\nonumber\\
&  \times\exp\left(  -\alpha_{j}(t)\left(  q-q_{0j}(t)\right)  ^{2}+\frac
{i}{\hbar}p_{0j}(t)\left(  q-q_{0j}(t)\right)  \right)
\end{align}
lose the property of the dynamical pair when the two packets begin to mutually
interfere. In such a higher quantum situation, the condition $\delta S_{H}=0$
alone is not sufficient.

We note that the condition of Eq. (\ref{parav}) and (\ref{parau}) is not
comprehensive to list all the possible conditions under which $\delta S_{W}=0$
is not demanded.

\subsection{Integral invariance for the quantum reduced action in parameter
space\label{sec:invariance}}

We note that $\mathbf{v\cdot\dot{u}}$ has been introduced to the variational
functionals in order to find the stationary paths and invariances in the
parameter spaces in analogy to classical mechanics but not to induce the
Legendre transformation. We nevertheless identify $\mathbf{v\cdot\dot{u}}$ as
a quantity related to an absolute invariance in the parameter space
\cite{Arnold} as follows. Under an assumption that both of the two sets of
Eqs. (\ref{LA21})-(\ref{LA22}) and Eqs. (\ref{Wu})-(\ref{Wv}) are
simultaneously satisfied, it is straightforward to see that the divergence of
the vector field $\left(  d\mathbf{u}/dt,d\mathbf{v}/dt\right)  $ be zero
\cite{Arnold}, that is,%
\begin{equation}
\frac{\partial}{\partial\mathbf{u}}\cdot\frac{d\mathbf{u}}{dt}+\frac{\partial
}{\partial\mathbf{v}}\cdot\frac{d\mathbf{v}}{dt}=0,
\end{equation}
which claims that $\left(  \mathbf{u}(t),\mathbf{v}(t)\right)  $ forms an
incompressible flow in parameter space. This is similar to incompressible
phase flow in classical dynamics, in which the integral invariances are
captured.\cite{Arnold} Similarly, we may take a continuous area $\Omega$
encircled by its boundary $\partial\Omega$, which is to be carried by the flow
lines of $\left(  \mathbf{u}(t),\mathbf{v}(t)\right)  $. Then we have an
integral invariance, along with the Stokes theorem%

\begin{align}
\int_{\partial\Omega}\mathbf{v\cdot\dot{u}}dt  &  =\int\int_{\Omega
}d\mathbf{v\wedge}d\mathbf{u}\nonumber\\
&  \mathbf{=}\mathbf{-}\int\int_{\Omega}d\mathbf{u\wedge}d\mathbf{v=-}%
\int_{\partial\Omega}\mathbf{u\cdot\dot{v}}dt, \label{wedge}%
\end{align}
This quantity is conserved for an area $\Omega$ to be carried along $\left(
\mathbf{u}(t),\mathbf{v}(t)\right)  $ by the unique and incompressible
flow.\cite{Arnold} This is one of the geometrical characteristics of
Schr\"{o}dinger dynamics.

\subsection{Summary of this section: Variational principle in parameter
space\label{sec:conclusion}}

We have studied the time-dependent variational principle behind the
Schr\"{o}dinger dynamics by splitting
\begin{equation}
\delta\int\left\langle \psi\left\vert \left(  i\hbar\frac{\partial}{\partial
t}-\hat{H}\right)  \right\vert \psi\right\rangle dt=0 \label{standard}%
\end{equation}
into the set of two coupled Maupertuis-Hamilton dynamics
\begin{equation}
\delta S_{H}=\delta\int\left(  \mathbf{v\cdot\dot{u}}-\left\langle
\psi\left\vert \hat{H}\right\vert \psi\right\rangle \right)  dt=0 \label{EEE}%
\end{equation}
and%
\begin{equation}
\delta S_{W}=\delta\int\left(  \mathbf{v\cdot\dot{u}}-\left\langle
\psi\left\vert i\hbar\frac{\partial}{\partial t}\right\vert \psi\right\rangle
\right)  dt=0. \label{WWW}%
\end{equation}
These two variational principles represent different physics in themselves. It
seems miracle to the present author that Schr\"{o}dinger abruptly combined
them into a single partial differential equation. Equation (\ref{EEE})
corresponds to the classical Hamilton least action principle, while the latter
is responsible for the flow dynamics. The conservation laws emerge not only
for energy but for the so-called \textquotedblleft
probability\textquotedblright\ current. By eliminating the term%
\begin{equation}
\delta\int\mathbf{v\cdot\dot{u}}dt
\end{equation}
from Eqs. (\ref{EEE})-(\ref{WWW}) we attain Eq. (\ref{standard}) after all.

We have to go back to Sec. \ref{sec:RealValue} to see the physical basis on
how the energy functional $\left\langle \psi\left\vert \hat{H}\right\vert
\psi\right\rangle $ and probability flow functional $\left\langle
\psi\left\vert i\hbar\frac{\partial}{\partial t}\right\vert \psi\right\rangle
$ should have those functional forms. Recall that we formulated there the
minimal Schr\"{o}dinger equation from scratch by imposing the flux
conservation and the energy conservation in addition to the space-time
translational symmetry.

\section{Roles of the classical paths beyond the semiclassical
mechanics\label{sec:ADFpath}}

We further proceed to analyze how the Schr\"{o}dinger function is to be
practically propagated in the present configuration space.

\subsection{Introductory remarks: Deep into the Schr\"{o}dinger dynamics}

It is well known that quantum mechanics converges (not absolute convergence
though) to classical mechanics as $\hbar\rightarrow0$. On the other side of
large $\hbar$, the Schr\"{o}dinger functions bear the
Huygens-like\cite{Goussev-Huygens} (wave-like) properties like diffraction,
bifurcation, and so on. To comprehend this strange property of the
Schr\"{o}dinger dynamics, it is natural to compare the Schr\"{o}dinger
equation with the Hamilton-Jacobi equation, which has a wave-like solution
over an ensemble of classical trajectories. Another standard method of
asymptotic analysis is due to Brillouin, Wentzel and Kramers (WKB) and
Jeffreys in which $W$ in%
\begin{equation}
\psi=A\exp\left(  \frac{i}{\hbar}W\right)
\end{equation}
is expanded in powers of $\hbar$.\cite{Schiff, Messiah} The hierarchical
transport equations by Maslov provides with another fundamental
method.\cite{Maslov} Also, a variety of theories of similar semiclassical
mechanics have been
proposed.\cite{U-Miller1,U-Miller2,berry1972semiclassical,Smilansky,ChildBook91,Lasser,de2016principles,Brack}
In these expansions, while the lowest order term exhibits classical mechanics
immediately, it is not straightforward to climb the power of $\hbar$ to the
high orders. It is not therefore easy to track how the classical nature in the
Schr\"{o}dinger function is gradually modulated and ends up with the full
quantum nature after all. The genuine quantum nature must be identified at
some stage on the way from classical to quantum feature.

In the operator algebra of quantum mechanics, the transition from classical to
quantum mechanics is clearly proclaimed by replacement of scalar quantities to
operators such as $E\rightarrow\hat{H}=i\hbar\frac{\partial}{\partial t},$
$p\rightarrow\hat{p}=-i\hbar\vec{\nabla}$, along with changing the Poisson
bracket to the commutator\cite{dirac1981principles} and so on. No ambiguity
exists in this type of quantization. However, we do want to understand what is
the continuous and/or discontinuous relationship between classical and quantum dynamics.

We therefore analyze the space-time propagation of the Schr\"{o}dinger
function from the viewpoint of path dynamics in the following three steps.
First, by factorizing the Hamilton-Jacobi component out of a Schr\"{o}dinger
equation, we proceed to describe semiclassical mechanics in our
way.\cite{Paper-I} There are two stages in the semiclassical mechanics, in
both stages of which the semiclassical Schr\"{o}dinger function is propagated
along classical paths.\cite{Paper-II} And beyond the semiclassical stage, we
see the successive branching of classical paths, thereby bringing about the
Huygens-like (wave-like) property into the
dynamics.\cite{takatsuka2023schrodinger} As the quantum nature becomes deeper,
those branching paths are seen to dissolve into the sea of wave-like nature of
the Schr\"{o}dinger function.

\subsection{Separation of the primitive classical dynamics from the
Schr\"{o}dinger function\label{sec:ADF}}

Let us begin with the Malov-type representation of a Schr\"{o}dinger
function.\cite{Maslov} For a short time interval $\left[  t,t+\Delta t\right]
$, the total Schr\"{o}dinger function can be written as%

\begin{equation}
\psi\left(  q,t\right)  =F(q,t)\exp\left(  \frac{i}{\hbar}S_{cl}\left(
q,t\right)  \right)  , \label{ADF}%
\end{equation}
on a\ multidimensional coordinate $q$ in configuration space, with $S_{cl}$
locally satisfying the classical Hamilton-Jacobi (HJ) equation%
\begin{equation}
\frac{\partial S_{cl}}{\partial t}+\frac{1}{2m}\left(  \nabla S_{cl}\right)
^{2}+V=0, \label{HJ}%
\end{equation}
which defines relevant classical trajectories during the interval, with $m$
being the mass.\cite{Atsuko,Atsuko2,Takatsuka2001,Paper-I,Paper-II} We assume
that the complex amplitude function $F(q,t)$ is compact enough in comparison
with the range of a potential function $V(q)$ under study. Thus the primitive
classical component has been factored out from $\psi\left(  q,t\right)  $,
leaving the quantum component $F(q,t)$ behind. We refer to $F(q,t)$ as Action
Decomposed Function (ADF),\cite{Atsuko} which is found to satisfy a liner
dynamical equation%

\begin{equation}
\frac{\partial F(q,t)}{\partial t}=\frac{1}{m}\left(  -p\cdot\nabla-\frac
{1}{2}(\nabla\cdot p)\right)  F(q,t)+\frac{i\hbar}{2m}\nabla^{2}F(q,t),
\label{eom}%
\end{equation}
where $p$ is a classical momentum at $\left(  q,t\right)  $ as $p=\nabla
S_{cl}\left(  q,t\right)  $. It is natural to expect that the simple form of
Eq. (\ref{ADF}) becomes \ less valid in a long-time and wide-space dynamics,
where $\psi\left(  q,t\right)  $ is to experience a large variation of the
qualitative nature of the potential function.

As seen in Eq. (\ref{eom}), the classical momentum $p$ is left behind in the
dynamics of $F(q,t)$. The reason for this comes from the property of the
Laplacian in the Hamiltonian such that%

\begin{align}
&  \nabla^{2}\left[  F(q,t)\exp\left(  \frac{i}{\hbar}S_{cl}\left(
q,t\right)  \right)  \right] \nonumber\\
&  =[F\left(  \frac{i}{\hbar}\right)  ^{2}\left(  \nabla S_{cl}\left(
q,t\right)  \right)  ^{2}\label{Lcl}\\
&  +\frac{i}{\hbar}F\nabla\cdot\nabla S_{cl}\left(  q,t\right)  +2\frac
{i}{\hbar}\left(  \nabla S_{cl}\left(  q,t\right)  \right)  \cdot\nabla
F\label{Lsemi}\\
&  +\nabla^{2}F(q,t)]\exp\left(  \frac{i}{\hbar}S_{cl}\left(  q,t\right)
\right)  . \label{Lfq}%
\end{align}
The term of Eq. (\ref{Lcl}) is absorbed into Eq. (\ref{HJ}). The last term Eq.
(\ref{Lfq}) is responsible for the pure quantum diffusion, and the term of Eq.
(\ref{Lsemi}) couples the classical and pure quantum terms as Eq. (\ref{eom})
and serves as a semiclassical interaction.

Notice the difference of the Maslov-type representation Eq. (\ref{ADF}) from
the Bohm one\cite{Bohm-text,Wyatttext,sanz2013trajectory}%
\begin{equation}
\psi\left(  q,t\right)  =R(q,t)\exp\left(  \frac{i}{\hbar}S_{B}\left(
q,t\right)  \right)  , \label{Bohm}%
\end{equation}
where both $R(q,t)$ and $S_{B}\left(  q,t\right)  $ are required to be
real-valued functions, and the so-called quantum potential $-\frac{\hbar}%
{2m}\frac{\nabla^{2}R}{R}$ enters the dynamics for $S_{B}$. In contrast, Eq.
(\ref{eom}) for $F(q,t)$ is linear, but as a price\ it is complex in general.
We use the mass weighted coordinates throughout, in which all the masses ($m$)
is scaled to unity $m=1$, and thereby $p$ turns numerically equivalent to the
corresponding velocity $v$. Therefore, Eq. (\ref{eom}) is rewritten in a
simpler form as%
\begin{equation}
\frac{D}{Dt}F(q,t)=\left[  -\frac{1}{2m}(\nabla\cdot p)+\frac{i\hbar}%
{2m}\nabla^{2}\right]  F(q,t), \label{eom2}%
\end{equation}
where the Lagrange picture in fluid dynamics is adopted by defining the
derivative as $\frac{D}{Dt}=\frac{\partial}{\partial t}+v\cdot\nabla$ (since
$m=1$). Thus, $F(q-q_{cl}\left(  t\right)  ,t)$ is to be carried along a
classical trajectory $q_{cl}\left(  t\right)  $ serving as a fluid line. Then
the Trotter decomposition\cite{Schulman} valid for a very short time--interval
$\Delta t$ gives%
\begin{align}
&  F(q-q_{cl}\left(  t+\Delta t\right)  ,t+\Delta t)\nonumber\\
&  \simeq\exp\left[  \frac{i\hbar}{2m}\Delta t\nabla^{2}\right]  \exp\left[
-\frac{1}{2m}(\nabla\cdot p)\Delta t\right]  F(q-q_{cl}\left(  t\right)
,t)\nonumber\\
&  \simeq\exp\left[  -\frac{1}{2m}(\nabla\cdot p)\Delta t\right]  \exp\left[
\frac{i\hbar}{2m}\Delta t\nabla^{2}\right]  F(q-q_{cl}\left(  t\right)  ,t)
\label{trot}%
\end{align}
and represents the quantum effects in a product between the momentum gradient
($\exp\left[  -\frac{1}{2m}(\nabla\cdot p)\Delta t\right]  $) and the quantum
diffusion ($\exp\left[  \frac{i\hbar}{2m}\Delta t\nabla^{2}\right]  $).

\begin{comment}
To summarize the roles of the Laplacian played in the above
transformation we lay out the action of the Laplacian as
\begin{align}
&  \nabla^{2}\left[  F(q,t)\exp\left(  \frac{i}{\hbar}S_{cl}\left(
q,t\right)  \right)  \right] \nonumber\\
&  =[F\left(  \frac{i}{\hbar}\right)  ^{2}\left(  \nabla S_{cl}\left(
q,t\right)  \right)  ^{2}\label{Lcl}\\
&  +\frac{i}{\hbar}F\nabla\cdot\nabla S_{cl}\left(  q,t\right)  +2\frac
{i}{\hbar}\left(  \nabla S_{cl}\left(  q,t\right)  \right)  \cdot\nabla
F\label{Lsemi}\\
&  +\nabla^{2}F(q,t)]\exp\left(  \frac{i}{\hbar}S_{cl}\left(  q,t\right)
\right)  . \label{Lfq}%
\end{align}
\newline One can readily identify in the right hand side that the first term,
line Eq. (\ref{Lcl}), reproduces the classical kinetic energy, the second and
third terms in line of Eq. (\ref{Lsemi}), give rise to the momentum gradient
terms in Eq. (\ref{trot}), and the fourth one, in line of Eq. (\ref{Lfq}),
represents the quantum diffusion dynamics with an imaginary diffusion constant
$i\hbar\nabla^{2}F/2$. The hierarchical structure of the Schr\"{o}dinger
equation is thus apparent from the classical term of Eq. (\ref{Lcl}) to the
pure quantum term of Eq. (\ref{Lfq}), with the term of Eq. (\ref{Lsemi})
linking the classical and pure quantum.
\end{comment}

\subsection{A semiclassical view of the Schr\"{o}dinger function as quantum
distribution function \label{sec:Rescaling}}

To proceed, we first integrate the momentum gradient propagator\ $\exp\left[
-\frac{1}{2m}(\nabla\cdot p)\Delta t\right]  $ such that \cite{Paper-I}%
\begin{align}
&  \left.  F_{sc}(q-q\left(  t+\Delta t\right)  ,t+\Delta t)\right\vert
_{q=q_{cl}\left(  t+\Delta t\right)  }\nonumber\\
&  \simeq\exp\left[  -\frac{1}{2m}(\nabla\cdot p)\Delta t\right]
F_{sc}(q-q_{cl}\left(  t\right)  ,t)\nonumber\\
&  =\left.  \left(  \frac{\sigma\left(  t\right)  }{\sigma\left(  t+\Delta
t\right)  }\right)  ^{1/2}F_{sc}(q-q_{cl}\left(  t\right)  ,t)\right\vert
_{q=q_{cl}\left(  t\right)  }, \label{MG}%
\end{align}
where the deviation determinant $\sigma\left(  t\right)  $ is expressed as%
\begin{equation}
\sigma\left(  t\right)  =\prod_{i=1}^{N}\wedge\left(  q_{cl}^{i}%
(t)-q_{cl}(t)\right)  , \label{mg2}%
\end{equation}
which is an $N$-dimensional orientable\ tiny volume surrounding the point
$q_{cl}\left(  t\right)  $ in configuration space, with $q_{cl}^{i}\left(
t\right)  $ ($i=1,\cdots,N$) being the configuration space point of an $i$th
classical trajectories \textit{nearby} the reference path $q_{cl}\left(
t\right)  $. Equation (\ref{MG}) is equivalent to
\begin{align}
&  \left.  \sigma\left(  t+\Delta t\right)  ^{1/2}F_{sc}(q-q_{cl}\left(
t+\Delta t\right)  ,t+\Delta t)\right\vert _{q=q_{cl}\left(  t+\Delta
t\right)  }\nonumber\\
&  =\left.  \sigma\left(  t\right)  ^{1/2}F_{sc}(q-q_{cl}\left(  t\right)
,t)\right\vert _{q=q_{cl}\left(  t\right)  }, \label{ADF2}%
\end{align}
The Planck constant does not yet appear in this level.

Equation (\ref{ADF2}) represents a semiclassical conservation law. By squaring
to
\begin{align}
&  \sigma\left(  t+\Delta t\right)  \left\vert F_{sc}(q-q_{cl}\left(  t+\Delta
t\right)  ,t+\Delta t)\right\vert ^{2}\nonumber\\
&  =\sigma\left(  t\right)  \left\vert F_{sc}(q-q_{cl}\left(  t\right)
,t)\right\vert ^{2}, \label{F0}%
\end{align}
which is valid up to the normalization, we have%

\begin{align}
&  \int dq\sigma\left(  t+\Delta t\right)  \left\vert F_{sc}(q-q_{cl}\left(
t+\Delta t\right)  ,t+\Delta t)\right\vert ^{2}\nonumber\\
&  =\int dq\sigma\left(  t\right)  \left\vert F_{sc}(q-q_{cl}\left(  t\right)
,t)\right\vert ^{2}, \label{F1}%
\end{align}
which is further rewritten as%

\begin{align}
&  \int dq_{t+\Delta t}\left\vert F_{sc}(q-q_{cl}\left(  t+\Delta t\right)
,t+\Delta t)\right\vert ^{2}\nonumber\\
&  =\int dq_{t}\left\vert F_{sc}(q-q_{cl}\left(  t\right)  ,t)\right\vert
^{2}, \label{F2}%
\end{align}
where $dq_{t}$ and $dq_{t+\Delta t}$ are the time-and-space dependent volume
elements for the integration. Comparison of Eqs. (\ref{F1}) and (\ref{F2})
leads to
\begin{equation}
dq=\frac{dq_{t}}{\sigma\left(  t\right)  }=\frac{dq_{t+\Delta t}}%
{\sigma\left(  t+\Delta t\right)  } \label{F3}%
\end{equation}

To clarify further the physical meaning of $F_{sc}$, rewriting Eq.
(\ref{ADF2}) in a symmetric manner along with the square root\ of the volume
element\cite{Takatsuka2001} for semiclassical integrals (or the half-density)%
\begin{equation}
dq_{t}=dq_{t}^{1/2}dq_{t}^{1/2\ast}=\left\vert dq_{t}\right\vert ,
\end{equation}
we rewrite Eq. (\ref{F0}) in a compact form up to the phase
as\cite{Takatsuka2001}%

\begin{equation}
F_{sc}\left(  q{_{t+\Delta t},t}+\Delta t\right)  dq_{t+\Delta t}^{1/2}%
=F_{sc}\left(  q{_{t},t}\right)  dq_{t}^{1/2}, \label{15}%
\end{equation}
where the square root of $dq_{t}$ is defined by%
\begin{equation}
dq_{t}^{1/2}\equiv\exp\left[  {{\frac{{i\pi}M\left(  q_{0},q_{t}\right)  }{2}%
}}\right]  \left\vert dq_{t}\right\vert ^{1/2},
\end{equation}
with the Maslov index $M\left(  q_{0},q_{t}\right)  $ being the Maslov index
counting the number of the change of sign of $\sigma\left(  t\right)  $ of Eq.
(\ref{mg2}) up to the degeneracy.\cite{Paper-I,Paper-II}

The physical meaning of the above conservation law, or the rescaling rule of
the length in $q$-space is described as follows: Suppose a time propagation of
a phase-space volume element ($\Delta q_{t}\Delta p_{t}$) in classical
dynamics, which is conserved by the Liouville theorem, that is, $\Delta
q_{t+\Delta t}\Delta p_{t+\Delta t}=\Delta q_{t}\Delta p_{t}$ by setting a
phase space distribution function $\Gamma(q,p,t)=1$ within the area $\Delta
q_{t}\Delta p_{t}$ and $\Gamma(q,p,t)=0$ otherwise. Reading $\Delta
q_{t}=dq_{t}$ as a tiny volume element in configuration space (for $t+\Delta
t$ as well), and comparing this conservation law with another conservation law
in semiclassics, Eq. (\ref{15}), we see $F_{sc}\left(  q{_{t},t}\right)  ^{2}$
in $F_{sc}\left(  q{_{t},t}\right)  ^{2}dq_{t}$ serve as a classical
distribution density function (without respect to the Maslov phase).
Therefore, $F_{sc}\left(  q{_{t},t}\right)  dq_{t}^{1/2}$ can be regarded as a
distribution covering the tiny variable space $dq_{t}^{1/2}$, and $F_{sc}$ and
$\psi$ as well can be regarded as \textquotedblleft a quantum (coherent)
distribution function\textquotedblright\ per $dq_{t}^{1/2}$.

\subsection{Further quantum dynamics manifesting in a complex Gaussian
Schr\"{o}dinger function \label{sec:hidden}}

Beyond the above semiclassics, we next proceed to the deeper quantum level,
which is the quantum diffusion $\exp\left[  \frac{i\hbar}{2m}\Delta
t\nabla^{2}\right]  $ in Eq. (\ref{trot}). An integral representation of this
operator is\cite{Schulman}
\begin{align}
&  \left\langle q\left\vert \hat{K}_{dif}\right\vert f\right\rangle
\nonumber\\
&  =\left(  \frac{m}{2\pi i\hbar\Delta t}\right)  ^{1/2}\int_{-\infty}%
^{\infty}dy\exp\left[  \frac{im(q-y)^{2}}{2\Delta t}\right]  f\left(
y\right)  . \label{kernel}%
\end{align}
Choosing $f\left(  y\right)  $ to be a complex Gaussian function, Eq.
(\ref{kernel}) proceeds analytically. The Gaussian function thus chosen should
reflect the stochastic motion of a one-world path considered in Sec.
\ref{sec:Stochastic}. Recall that the kernel in Eq. (\ref{kernel}) is
mathematically equivalent to that of the Wiener process if we regard
$im/2\hbar$ as a diffusion constant $-1/4D$, and the normalization constant
$\left(  m/2\pi i\hbar\Delta t\right)  ^{1/2}$ as $\left(  1/4\pi D\Delta
t\right)  ^{1/2}$. Therefore $\hat{K}_{dif}$ in Eq. (\ref{kernel}) reflects
the hidden Brownian motion of the diffusion constant $D=-i2m/\hbar$.

\subsubsection{Semiclassics for the Gaussian}

We resume the dynamics for the action decomposed function $F(q,t)$ by assuming
a Gaussian function of the compact form%

\begin{equation}
G(q-q_{cl}\left(  t\right)  ,t)=g\left(  t\right)  \exp\left[  -\gamma\left(
t\right)  (q-q_{cl}\left(  t\right)  )^{2}\right]  , \label{GQ}%
\end{equation}
where $g\left(  t\right)  $ is a normalization constant. $g\left(  t\right)  $
is important, since the semiclassical mechanics considered above is about the
rescaling of the configuration space.

A semiclassics for the Gaussian function was extensively studied by Heller,
who assumed that the potential function covered by the Gaussian is locally
approximated by a quadratic form.\cite{HellerTG,Heller18} Here we are taking a
different approach. The momentum gradient part can be formally treated for the
Gaussian function too as%

\begin{align}
&  G^{mg}\left(  q-q_{cl}\left(  t+\Delta t\right)  ,t+\Delta t\right)
)\nonumber\\
&  =g^{mg}\left(  t+\Delta t\right)  \exp(-\left[  \frac{\sigma\left(
t\right)  }{\sigma\left(  t+\Delta t\right)  }(q-q_{cl}\left(  t+\Delta
t\right)  )\right]  ^{\dag}\nonumber\\
&  \times\left(  \gamma^{mg}\left(  t\right)  \right)  \left[  \frac
{\sigma\left(  t\right)  }{\sigma\left(  t+\Delta t\right)  }(q-q_{cl}\left(
t+\Delta t\right)  )\right]  ), \label{proj}%
\end{align}
where $g^{mg}\left(  t+\Delta t\right)  $ renormalizes the function. The
result is simpler written as%
\begin{align}
&  G^{mg}\left(  q-q_{cl}\left(  t+\Delta t\right)  ,t+\Delta t\right)
)=g^{mg}\left(  t\right)  \left(  \frac{\sigma\left(  t\right)  }%
{\sigma\left(  t+\Delta t\right)  }\right)  ^{1/2}\nonumber\\
&  \times\exp\left[  -\left(  \gamma^{mg}\left(  t+\Delta t\right)  \right)
(q-q_{cl}\left(  t+\Delta t\right)  )^{2}\right]  , \label{MG2}%
\end{align}
with
\begin{equation}
\gamma^{mg}\left(  t+\Delta t\right)  =\gamma\left(  t\right)  \left(
\frac{\sigma\left(  t\right)  }{\sigma\left(  t+\Delta t\right)  }\right)
^{2} \label{NG8}%
\end{equation}
and
\begin{equation}
g^{mg}\left(  t+\Delta t\right)  =g\left(  t\right)  \left(  \frac
{\sigma\left(  t\right)  }{\sigma\left(  t+\Delta t\right)  }\right)  ^{1/2},
\label{NG7}%
\end{equation}
where suffix $mg$ on $\gamma$ and $g$ indicates that the effects of the
momentum gradient \textit{only} have been taken into account in this stage.

\subsubsection{Modulation of the complex Gaussian exponent by the quantum
diffusion}

\paragraph{Dynamics of the exponents}

Further a direct application of (\ref{kernel}) on Eq. (\ref{MG2}) gives%
\begin{align}
&  G(q-q_{cl}\left(  t+\Delta t\right)  ,t+\Delta t)\nonumber\\
&  =\int dyK_{dif}\left(  q,y,\Delta t\right)  G^{mg}(y-q_{cl}\left(  t+\Delta
t\right)  ,t+\Delta t)\nonumber\\
&  =g\left(  t\right)  \left(  \frac{\sigma\left(  t\right)  }{\sigma\left(
t+\Delta t\right)  }\frac{1/\gamma^{mg}}{1/\gamma^{mg}+i/b}\right)
^{1/2}\nonumber\\
&  \times\exp\left[  -\frac{1}{1/\gamma^{mg}+i/b}\left(  q-q_{cl}\left(
t+\Delta t\right)  \right)  ^{2}\right]  \label{GG}%
\end{align}
with $b=m/2\hbar\Delta t.$

To make the expressions better visualized, we write $G$ with the following form%

\begin{equation}
G(q-q_{cl}\left(  t\right)  ,t)=g\left(  t\right)  \exp\left[  -\frac
{1}{c\left(  t\right)  +id\left(  t\right)  }(q-q_{cl}\left(  t\right)
)^{2}\right]  , \label{CD1}%
\end{equation}
with
\begin{equation}
g\left(  t\right)  =\left(  \frac{2}{\pi}\right)  ^{1/4}\left[  c\left(
t\right)  \right]  ^{1/4}\left[  c\left(  t\right)  +id\left(  t\right)
\right]  ^{-1/2}, \label{gnorm}%
\end{equation}
making the correspondence with Eq. (\ref{GQ}) with $\gamma\left(  t\right)
^{-1}=c\left(  t\right)  +id\left(  t\right)  $, where both $c\left(
t\right)  $ and $d\left(  t\right)  $ are real-valued. It is found that the
parameters $c\left(  t\right)  $ and $d\left(  t\right)  $ satisfy the
following dynamical equations, respectively,\cite{Paper-II}%

\begin{align}
c\left(  t+\Delta t\right)   &  =c\left(  t\right)  \left(  \frac
{\sigma\left(  t+\Delta t\right)  }{\sigma\left(  t\right)  }\right)
^{2}\text{ \ or \ }\nonumber\\
\dot{c}\left(  t\right)   &  =2\frac{\dot{\sigma}\left(  t\right)  }%
{\sigma\left(  t\right)  }c\left(  t\right)  \label{CN95}%
\end{align}
and%
\begin{align}
d\left(  t+\Delta t\right)   &  =d\left(  t\right)  \left(  \frac
{\sigma\left(  t+\Delta t\right)  }{\sigma\left(  t\right)  }\right)
^{2}+\frac{2\hbar\Delta t}{m}\text{ \ or \ }\nonumber\\
\dot{d}\left(  t\right)   &  =2\frac{\dot{\sigma}\left(  t\right)  }%
{\sigma\left(  t\right)  }d\left(  t\right)  +\frac{2\hbar}{m}. \label{CN94}%
\end{align}
Therefore, $c\left(  t\right)  $ [the real part of $1/\gamma\left(  t\right)
$] is responsible only for the semiclassical flow, and $d\left(  t\right)  $
[the imaginary part of $1/\gamma\left(  t\right)  $] is responsible for
quantum diffusion along with its coupling with the semiclassical flow.

The Planck constant finally appears in this stage but only in the dynamics of
$d\left(  t\right)  $. The effects of the momentum gradient and the quantum
diffusion are mixed up when $c\left(  t\right)  $ and $d\left(  t\right)  $
appear together in $\gamma\left(  t+\Delta t\right)  $. We refer to the
present normalized-Gaussian representation of ADF as ADF-NVG in what follows.

\subsubsection{Experimentally uncontrollable nature of the complex Gaussian}

The physical implication behind the complex exponent $\left(  c\left(
t\right)  ,d\left(  t\right)  \right)  $ is as follows. Defining a quantity
$\zeta(t)$ as%
\begin{equation}
\zeta(t)=d(t)/\sigma\left(  t\right)  , \label{zeta}%
\end{equation}
we find the following Wronskian relation\cite{Paper-II} taken out of Eq.
(\ref{CN94})
\begin{equation}
\sigma\left(  t\right)  \dot{\zeta}(t)-\dot{\sigma}\left(  t\right)
\zeta(t)=\left\vert
\begin{array}
[c]{cc}%
\sigma\left(  t\right)  & \dot{\sigma}\left(  t\right) \\
\zeta(t) & \dot{\zeta}(t)
\end{array}
\right\vert =\frac{2\hbar}{m}. \label{Wronskian}%
\end{equation}
Therefore a vector $(\sigma\left(  t\right)  ,\zeta(t))$ is basically rotating
on $\sigma-\zeta$ plane with a relevant \textquotedblleft angular
momentum\textquotedblright, which is directed vertical to the plane and is
conserved. This Wronskian relation is valid for the \textquotedblleft
rotational motion\textquotedblright\ in the $\sigma-\zeta$ space, which is
realized for vibration and libration. This is not the case for the rotational
motion of a pendulum and free particle motion, because $\sigma\left(
t\right)  $ does not oscillate.

We thus have faced an experimentally uncontrollable parameter $d(t)$ emerging
from the quantum diffusion, and the Planck constant begins to play its role in
this stage. Further, there is no way to choose the initial exponent $1/(c+id)$
experimentally. Therefore, even if one can track the center of the single
Gaussian in terms of the classical paths, the Gaussian width is of nature far
from classical dynamics.

\subsubsection{Limitation and breakdown of the Gaussian average}

In an analogy to the classical stochastic dynamics, the simple Gaussian itself
suggests the presence of a stochastic motion around the center (average) path.
And, so far, we have assumed that the classical motion of the Gaussian center
can well average over the fast stochastic quantum dynamics. However, it is
obvious that the single Gaussian representation is a bad approximation in case
where the potential range is shorter than or equal to the width of the
Gaussian, particularly in those systems where a Schr\"{o}dinger function is
expected to undergo bifurcation (branching). In such a case, we actually need
to explicitly track the individual stochastic paths, since those possibly fast
stochastic paths can trigger to deviate the average motion of the Gaussian. We
therefore proceed to this aspect.

\subsection{Splitting of a single Gaussian and classical
paths\label{sec:splitting}}

As the genuine quantum nature is deepened, the notion of classical dynamics is
getting lost gradually (continuously and not abruptly), and eventually we find
that classical paths have to split into many pieces and each dissolves in to a
distribution function deformed from the simple Gaussian. We below consider how
such a Gaussian deformation and branching can be represented and be tracked
through. \ 

\subsubsection{The Weierstrass transform}

Consider a case in which a Gaussian wavepacket sits on a potential function,
the range of spatial variation of which is shorter than the width of the
Gaussian. Then the different parts within the extent of the Gaussian are
supposed to be driven by the different force fields. In such a case it is
necessary to re-express the Gaussian in a set of smaller (short-ranged)
Schr\"{o}dinger functions. Among the many possible ways for the
re-representation mathematically, we here adopt a method which rigorously
allows to split a parent Gaussian to a continuous set of daughter
Gaussians.\cite{takatsuka2023schrodinger} (We note that the aim of this study
is not to seek for a convenient numerical approximation.)

We resume with a Gaussian wavepacket of the Maslov type or the coherent-state
type at a time $t$ of the form%
\begin{equation}
G(q-q_{0}\left(  t\right)  ,t)=\exp\left(  -\gamma_{1}\left(  q-q_{0}\right)
^{2}+\frac{i}{\hbar}p_{0}\left(  q-q_{0}\right)  \right)  , \label{Gauss00}%
\end{equation}
where $\left(  q_{0},p_{0}\right)  $ is the center, and the exponent
$\gamma_{1}$ is complex as $\gamma_{1}\equiv\gamma_{1}^{r}+i\gamma_{1}^{c},$
where $\gamma_{1}^{r}$ ($>0$) and $\gamma_{1}^{c}$ are the real and imaginary
parts, respectively. We thus have two phase terms in the exponent of Eq.
(\ref{Gauss00}), which are%
\begin{equation}
\exp\left(  \frac{i}{\hbar}\left[  p_{0}\left(  q-q_{0}\right)  -\hbar
\gamma_{1}^{c}\left(  q-q_{0}\right)  ^{2}\right]  \right)  . \label{phase2}%
\end{equation}
Aside from these phase parts, we apply the following integral transform only
for the real part $\exp\left(  -\gamma_{1}^{r}\left(  q-q_{0}\right)
^{2}\right)  $ in such a way that
\begin{align}
&  \exp\left(  -\gamma_{1}^{r}\left(  q-q_{0}\right)  ^{2}\right) \nonumber\\
&  =\int d\Omega\exp\left(  -\gamma_{2}^{r}\left(  q-q_{0}-\Omega\right)
^{2}\right)  \exp\left(  -A\Omega^{2}\right) \nonumber\\
&  =\sqrt{\frac{\pi}{A+\gamma_{2}^{r}}}\exp\left[  -\left(  \gamma_{2}%
^{r}-\frac{\left(  \gamma_{2}^{r}\right)  ^{2}}{A+\gamma_{2}^{r}}\right)
\left(  q-q_{0}\right)  ^{2}\right]  \label{transreal}%
\end{align}
under the condition $A>0$. This is essentially the Weierstrass transform
(W-transform for short). The exponents in Eq. (\ref{transreal}) can be chosen as%

\begin{equation}
\gamma_{1}^{r}=\gamma_{2}^{r}-\frac{\left(  \gamma_{2}^{r}\right)  ^{2}%
}{A+\gamma_{2}^{r}} \label{A3}%
\end{equation}
or in more compact form
\begin{equation}
\frac{1}{\gamma_{2}^{r}}\equiv\frac{1}{\gamma_{1}^{r}}-\frac{1}{A}.
\label{omg-int5}%
\end{equation}
Thus the single Gaussian at the left hand side of Eq. (\ref{transreal}) is
transformed to an continuous set of other Gaussians of a common exponent
$\gamma_{2}^{r}$, each being centered at $q=q_{0}+\Omega$. We refer to these
Gaussians containing $\Omega$ as $\Omega$-Gaussian. Since $A>0$, $\gamma
_{2}^{r}>\gamma_{1}^{r}$ in Eq. (\ref{omg-int5}), and the newly born Gaussians
are sharper and narrower. ($\gamma_{2}^{r}$ may be relaxed to a smaller value
in the subsequent dynamics.) The magnitude of $\gamma_{2}^{r}$ can be chosen
such that the potential function can fit well in the potential range.

Equation (\ref{Gauss00}) is then rewritten as%
\begin{align}
&  \exp\left(  -\gamma_{1}\left(  q-q_{0}\right)  ^{2}+\frac{i}{\hbar}%
p_{0}\left(  q-q_{0}\right)  \right) \nonumber\\
&  =\left(  \frac{A+\gamma_{2}^{r}}{\pi}\right)  ^{1/2}\int d\Omega\exp\left(
-A\Omega^{2}\right)  \exp\left(  \frac{i}{\hbar}p_{0}\Omega-i\gamma_{1}%
^{c}\Omega^{2}\right) \nonumber\\
&  \times\exp\left(  -\gamma_{2}\left(  q-q_{\Omega}\right)  ^{2}+\frac
{i}{\hbar}p_{\Omega}\left(  q-q_{\Omega}\right)  \right)  \label{gamm}%
\end{align}
with the new centers of the split Gaussians%
\begin{equation}
q_{\Omega}\equiv q_{0}+\Omega, \label{posshift}%
\end{equation}
and%
\begin{equation}
p_{\Omega}\equiv p_{0}-2\hbar\gamma_{1}^{c}\Omega, \label{momshift}%
\end{equation}
with%
\begin{equation}
\gamma_{2}\equiv\gamma_{2}^{r}+i\gamma_{1}^{c}. \label{gam12}%
\end{equation}
(Notice $\gamma_{2}^{c}\equiv\gamma_{1}^{c}$). Thus the center of each
daughter Gaussian is located at ($q_{\Omega},p_{\Omega}$) ready for the next
propagation. All these split Gaussians are kept coherently connected to each
other through $\int d\Omega$ in Eq. (\ref{gamm}). The role of the imaginary
part of the exponent $\gamma_{1}^{c}$ on $p_{\Omega}$ (see Eq. (\ref{momshift}%
)) is quite critical.

We here detour back to Eq. (\ref{v-density}) for the local velocity, which is
rewritten for\ the local momentum as%

\begin{equation}
p^{\text{local}}(q,t)=\hbar\operatorname{Im}\frac{\vec{\nabla}\psi}{\psi}.
\end{equation}
For the mother Gaussian in Eq. (\ref{gamm})%
\begin{equation}
\exp\left(  -\gamma_{1}\left(  q-q_{0}\right)  ^{2}+\frac{i}{\hbar}%
p_{0}\left(  q-q_{0}\right)  \right)  ,
\end{equation}
the local momentum $p^{\text{local}}(q,t)$ is
\begin{equation}
-2h\left(  \operatorname{Im}\gamma_{1}\right)  \left(  q-q_{0}\right)  +p_{0}
\label{newmom}%
\end{equation}
and the daughter Gaussians shift their central position according to Eq.
(\ref{posshift}), or $q_{0}=q_{\Omega}-\Omega,$ which also shifts the local
momentum of Eq. (\ref{newmom}) as%

\begin{equation}
-2h\gamma_{1}^{c}\left(  q-q_{\Omega}\right)  )-2h\gamma_{1}^{c}\Omega+p_{0}.
\end{equation}
($\operatorname{Im}\gamma_{1}=\gamma_{1}^{c}$). Thus each daughter Gaussian
has a shifted momentum $p_{\Omega}$ of (\ref{momshift}) at its central
position $q_{\Omega}$ (\ref{posshift}).

Incidentally, one may want to discretize\ the continuous transformation such
Eq. (\ref{gamm}) with a quadrature such that%
\begin{align}
&  \int d\Omega\exp\left(  -A\Omega^{2}\right)  f\left(  \Omega\right)
\nonumber\\
&  \simeq\sum_{k}w_{k}\exp\left(  -A\Omega_{k}^{2}\right)  f\left(  \Omega
_{k}\right)  , \label{expand}%
\end{align}
where $\Omega_{k}$ indicates the quadrature point with their weighting factor
$w_{k}$.

\subsubsection{Recursive processes of ADF-NVG dynamics and proliferation of
daughter packets \label{sec:OMG-GAUSS}}

We next repeat the W-transform as a branched packet passes through another
deep quantum region. From an initial single squeezed Gaussian at time $t_{0}$%
\begin{align}
&  G(q-q_{0}\left(  t_{0}\right)  ,t_{0})\nonumber\\
&  =n_{\gamma_{0}}(t_{0})\exp\left[  -\gamma_{0}(t_{0})(q-q_{0}(t_{0}%
))^{2}+\frac{i}{\hbar}p_{0}(t_{0})\left(  q-q_{0}(t_{0})\right)  \right]  ,
\label{initialG}%
\end{align}
where $n_{\gamma}$ is the normalization factor, let us propagate it from
$t_{0}$ to $t_{1}=t_{0}+\Delta t$ (with $\Delta t$ being sufficiently small,
if necessary. Yet, it can be practically elongated to a finite length
depending on how smooth the potential function is). Suppose an ADF-NVG is
carried out from $t_{0}$ to $t_{1}$ without splitting such that
\begin{align}
&  \exp\left[  -\frac{i}{\hbar}\hat{H}(t_{1}-t_{0})\right] \nonumber\\
&  \times\exp\left[  -\gamma_{0}(t_{0})(q-q_{0}(t_{0}))^{2}+\frac{i}{\hbar
}p_{0}(t_{0})\left(  q-q_{0}(t_{0})\right)  \right] \nonumber\\
&  =\frac{n_{\gamma_{0}}(t_{1})}{n_{\gamma_{0}}(t_{0})}\exp\left[  -\gamma
_{0}(t_{1})(q-q_{0}(t_{1}))^{2}+\frac{i}{\hbar}p_{0}(t_{1})\left(
q-q_{0}(t_{1})\right)  \right] \nonumber\\
&  \equiv G(q-q_{0}\left(  t_{1}\right)  ,t_{1}). \label{Gprop}%
\end{align}
Then we perform a W-transform at time $t_{1}$ as%
\begin{align}
&  \exp[-\gamma_{0}(t_{1})\left(  q-q_{0}(t_{1})\right)  ^{2}+\frac{i}{\hbar
}p_{0}\left(  t_{1}\right)  \left(  q-q_{0}\left(  t_{1}\right)  \right)
]\nonumber\\
&  =\left(  \frac{A_{1}+\gamma_{1}^{r}(t_{1})}{\pi}\right)  ^{1/2}\nonumber\\
&  \times\int d\Omega_{1}\exp\left(  -A_{1}\Omega_{1}^{2}\right)  \exp\left(
\frac{i}{\hbar}p_{0}(t_{1})\Omega_{1}-i\gamma_{0}^{c}(t_{1})\Omega_{1}%
^{2}\right) \nonumber\\
&  \times\exp\left[  -\gamma_{1}(t_{1})\left(  q-q_{\Omega_{1}}(t_{1})\right)
^{2}+\frac{i}{\hbar}p_{\Omega_{1}}(t_{1})\left(  q-q_{\Omega_{1}}%
(t_{1})\right)  \right]
\end{align}
with an appropriate choice of the exponent $\gamma_{1}$. Following the above
rule, $A_{1}$ is to be determined by
\begin{equation}
\frac{1}{\gamma_{1}^{r}(t_{1})}=\frac{1}{\gamma_{0}^{r}(t_{1})}-\frac{1}%
{A_{1}}, \label{Agamma}%
\end{equation}
and%
\begin{equation}
\gamma_{1}^{c}(t_{1})=\gamma_{0}^{c}(t_{1}), \label{gam01}%
\end{equation}
with%
\begin{equation}
q_{\Omega_{1}}(t_{1})=q_{0}(t_{1})+\Omega_{1} \label{qmod}%
\end{equation}
and%
\begin{equation}
p_{\Omega_{1}}(t_{1})=p_{0}(t_{1})-2\hbar\gamma_{1}^{c}(t_{1})\Omega_{1}.
\label{pmod}%
\end{equation}

We can further propagate the $\Omega$-Gaussians from $t_{1}$ to $t_{2}%
=t_{1}+\Delta t$ such that%

\begin{align}
&  \exp\left[  -\frac{i}{\hbar}\hat{H}(t_{2}-t_{1})\right] \nonumber\\
&  \times\exp[-\gamma_{0}(t_{1})\left(  q-q_{0}(t_{1})\right)  ^{2}+\frac
{i}{\hbar}p_{0}\left(  t_{1}\right)  \left(  q-q_{0}\left(  t_{1}\right)
\right)  ]\nonumber\\
&  =\left(  \frac{A_{1}+\gamma_{1}^{r}(t_{1})}{\pi}\right)  ^{1/2}\nonumber\\
&  \times\int d\Omega_{1}\exp\left(  -A_{1}\Omega_{1}^{2}\right)  \exp\left(
\frac{i}{\hbar}p_{0}(t_{1})\Omega_{1}-i\gamma_{1}^{c}(t_{1})\Omega_{1}%
^{2}\right) \nonumber\\
&  \times\exp\left[  -\frac{i}{\hbar}\hat{H}(t_{2}-t_{1})\right] \nonumber\\
&  \times\exp\left[  -\gamma_{1}(t_{1})\left(  q-q_{\Omega_{1}}(t_{1})\right)
^{2}+\frac{i}{\hbar}p_{\Omega_{1}}(t_{1})\left(  q-q_{\Omega_{1}}%
(t_{1})\right)  \right]  . \label{omg-int6}%
\end{align}
Note that each of the $\Omega$-Gaussians can be propagated independently,
since ADF-NVG dynamics is linear. We thus can rewrite the right hand side of
Eq. (\ref{omg-int6}) with use of Eq. (\ref{Gprop}) as%
\begin{align}
&  \left(  \frac{A_{1}+\gamma_{1}^{r}(t_{1})}{\pi}\right)  ^{1/2}\nonumber\\
&  \int d\Omega_{1}\exp\left(  -A_{1}\Omega_{1}^{2}\right)  \exp\left(
\frac{i}{\hbar}p_{0}\Omega_{1}-i\gamma_{0}^{c}\Omega_{1}^{2}\right)
\frac{n_{\gamma_{1}\Omega_{1}}(t_{2})}{n_{\gamma_{1}\Omega_{1}}(t_{1}%
)}\nonumber\\
&  \times\exp\left[  -\gamma_{1}(t_{2})\left(  q-q_{\Omega_{1}}(t_{2}\right)
^{2}+\frac{i}{\hbar}p_{\Omega_{1}}\left(  t_{2}\right)  \left(  q-q_{\Omega
_{1}}(t_{2}\right)  )\right]  , \label{omg-int7}%
\end{align}
where $n_{\gamma_{1}\Omega_{1}}(t)$ is the normalization constant for the
coherent state of%
\begin{equation}
\exp\left[  -\gamma_{1}(t)(q-q_{\Omega_{1}}(t))^{2}+\frac{i}{\hbar}%
p_{\Omega_{1}}(t)\left(  q-q_{\Omega_{1}}(t)\right)  \right]  .\nonumber
\end{equation}
In the ADF-NVG dynamics, its Gaussian center $\left(  q_{\Omega_{1}}%
(t_{1}),p_{\Omega_{1}}(t_{1})\right)  $ proceeds classically as%
\begin{equation}
\left(  q_{\Omega_{1}}(t_{1}),p_{\Omega_{1}}(t_{1})\right)  \text{
}\underrightarrow{\text{ \ classical \ }}\text{ }(q_{\Omega_{1}}\left(
t_{2}\right)  ,p_{\Omega_{1}}\left(  t_{2}\right)  ) \label{Qpath}%
\end{equation}
during this time interval. And of course the Gaussians at $(q_{\Omega_{1}%
}\left(  t_{2}\right)  ,p_{\Omega_{1}}\left(  t_{2}\right)  )$ are subject to
another W-transform, if necessary.

\subsection{Path integrals over many coherently branching paths
\label{sec.Pathint}}

Multiple applications of the W-transform can be performed to proliferate the
$\Omega$-Gaussians of the next generations with different exponents. As an
example, the process is repeated from time $t_{0}$ to $t_{4}$ via $t_{1}$,
$t_{2}$, and $t_{3}$, staring from the initial squeezed Gaussian of Eq.
(\ref{initialG}): It schematically looks%

\begin{align}
&  \exp\left[  \frac{i}{\hbar}\hat{H}\left(  t_{4}-t_{3}\right)  \right]
\exp\left[  \frac{i}{\hbar}\hat{H}\left(  t_{3}-t_{2}\right)  \right]
\exp\left[  \frac{i}{\hbar}\hat{H}\left(  t_{2}-t_{1}\right)  \right]
\nonumber\\
&  \times\exp\left[  \frac{i}{\hbar}\hat{H}\left(  t_{1}-t_{0}\right)
\right]  G(q-q_{0}\left(  t_{0}\right)  ,t_{0})\nonumber\\
&  =\left(  \frac{A_{1}+\gamma_{1}^{r}(t_{1})}{\pi}\right)  ^{1/2}\int
d\Omega_{1}\nonumber\\
&  \times\exp\left(  -A_{1}\Omega_{1}^{2}\right)  \frac{n_{\gamma_{1}%
\Omega_{1}}(t_{2})}{n_{\gamma_{1}\Omega_{1}}(t_{1})}\exp\left(  \frac{i}%
{\hbar}p_{\Omega_{0}}(t_{1})\Omega_{1}-i\gamma_{1}^{c}(t_{1})\Omega_{1}%
^{2}\right) \nonumber\\
&  \times\left(  \frac{A_{2}+\gamma_{2}^{r}(t_{2})}{\pi}\right)  ^{1/2}\int
d\Omega_{2}\nonumber\\
&  \times\exp\left(  -A_{2}\Omega_{2}^{2}\right)  \frac{n_{\gamma_{2}%
\Omega_{2}}(t_{3})}{n_{\gamma_{2}\Omega_{2}}(t_{2})}\exp\left(  \frac{i}%
{\hbar}p_{\Omega_{1}}(t_{2})\Omega_{2}-i\gamma_{2}^{c}(t_{2})\Omega_{2}%
^{2}\right) \nonumber\\
&  \times\left(  \frac{A_{3}+\gamma_{3}^{r}(t_{3})}{\pi}\right)  ^{1/2}\int
d\Omega_{3}\nonumber\\
&  \times\exp\left(  -A_{3}\Omega_{3}^{2}\right)  \frac{n_{\gamma_{3}%
\Omega_{3}}(t_{4})}{n_{\gamma_{3}\Omega_{3}}(t_{3})}\exp\left(  \frac{i}%
{\hbar}p_{\Omega_{2}}(t_{3})\Omega_{3}-i\gamma_{3}^{c}(t_{3})\Omega_{3}%
^{2}\right) \nonumber\\
&  \times\exp[-\gamma_{3}(t_{4})\left(  q-q_{\Omega_{123}}(t_{4})\right)
^{2}+\frac{i}{\hbar}p_{\Omega_{123}}(t_{4})\left(  q-q_{\Omega_{123}}%
(t_{4})\right)  ], \label{OMG23}%
\end{align}
where $q_{\Omega_{123}}(t_{4})$, for instance, indicates the history of a
path, which is one of the paths that have experienced the transform with the
parameters $\Omega_{1}$, $\Omega_{2}$, and $\Omega_{3}$. The other notations
are rather straightforward. For instance,%
\begin{equation}
\left(  q_{\Omega_{123}}(t_{3}),p_{\Omega_{123}}(t_{3})\right)  \text{
}\underrightarrow{\text{ \ classical \ }}\text{ }(q_{\Omega_{123}}\left(
t_{4}\right)  ,p_{\Omega_{123}}\left(  t_{4}\right)  ).
\end{equation}
The parameters in Eq. (\ref{OMG23}) are given in a recursive manner that%

\begin{equation}
q_{\Omega_{j}}(t_{j})=q_{j-1}(t_{j})+\Omega_{j}, \label{pathq}%
\end{equation}
and%
\begin{equation}
p_{\Omega_{j}}(t_{j})=p_{j-1}(t_{j})-2\hbar\gamma_{0}^{c}(t_{j})\Omega_{j}
\label{pathp}%
\end{equation}
and the Gaussian exponents are given as%
\begin{equation}
\gamma_{j}(t)=\gamma_{j}^{r}(t)+i\gamma_{j}^{c}(t)=\frac{1}{c_{j}%
(t)+id_{j}(t)}%
\end{equation}
way back to Eq. (\ref{CD1}) in the ADF-NVG scheme. We assume that at each
transform at time $t_{j}$, the equality like Eq. (\ref{Agamma}) holds as%
\begin{equation}
\frac{1}{\gamma_{j}^{r}(t_{j})}=\frac{1}{\gamma_{j-1}^{r}(t_{j})}-\frac
{1}{A_{j}}%
\end{equation}
and
\begin{equation}
\gamma_{j}^{c}(t_{j})=\gamma_{j-1}^{c}(t_{j})
\end{equation}
for $j=1,2,\cdots$. The successive W-transforms are tracked schematically
through the nonclassical branching paths as%

\begin{equation}
\left(
\begin{array}
[c]{c}%
q_{0}(t_{0})\\
p_{0}(t_{0})\\
S_{0}\\
0\\
\gamma_{0}(t_{0})\\
0
\end{array}
\right)  \rightarrow\left(
\begin{array}
[c]{c}%
q_{\Omega_{1}}(t_{1})\\
p_{\Omega_{1}}(t_{1})\\
S_{1}\\
\Omega_{1}\\
\gamma_{_{1}}(t_{1})\\
A_{1}%
\end{array}
\right)  \rightarrow\left(
\begin{array}
[c]{c}%
q_{\Omega_{12}}(t_{2})\\
p_{\Omega_{12}}(t_{2})\\
S_{12}\\
\Omega_{2}\\
\gamma_{2}(t_{2})\\
A_{2}%
\end{array}
\right)  \rightarrow(t_{3})\rightarrow\cdots, \label{O-path}%
\end{equation}
where $S_{i}$ denote the phases taken along the relevant paths.

Thus the paths of the centers of the proliferated Gaussian form a fractal
feature and/or successive pitchfork bifurcations. However, those paths can
merge and separate on the way to a goal in arbitrarily many times, and
therefore an initial point and a final end point are not uniquely tied in
one-to-one correspondence. The ensemble of those bifurcating wavepackets, each
as a part of a quantum distribution function within the coherent sum, can now
pass through the two slits and meet one another to coherently interfere afterward.

\subsection{ Note on Bohr's instantaneous collapse and semiclassical
divergence of a Schr\"{o}dinger function \label{sec:mesoscopic}}

Before concluding the present analysis on the Schr\"{o}dinger dynamics, we
study about the \textquotedblleft instantaneous\textquotedblright\ collapse of
a Schr\"{o}dinger function to a point at the measuring board in the double
slit experiment. This concept, not an experimental fact and among the most
intense controversy, sits in the heart of the Copenhagen interpretation of
quantum
mechanics.\cite{auletta2001foundations,home2007einstein,whitaker2012new,freire2022oxford}
The most essential question is whether the Schr\"{o}dinger functions actually
represent a physical substance. If so, such an instantaneous motion breaks the
theory of relativity and locality, as Einstein allegedly made a strong objection.

We in this article regard that the Schr\"{o}dinger function should be a
real-valued distribution function of two component vector as explicitly
formulated in Sec. \ref{sec:RealValue} and that it does not represent any
physical wave. Besides, we have figured out the one-world quantum path in Sec.
\ref{sec:Stochastic}. Therefore, we deem that such an instantaneous collapse
of a Schr\"{o}dinger function is not a realistic physical process. It is then
reasonable to conceive that each spot on the measurement board is marked by a
particle running on a one-world path studied in Sec. \ref{sec:Stochastic}. The
sum of all the one\symbol{94}world paths are supposed to reproduce the entire
Schr\"{o}dinger dynamics through the propagation of the density $\rho(q,t)$
and the Schr\"{o}dinger vector $\bar{\psi}(q,t)$ as the factorization of
$\rho(q,t)$.

However, since our Schr\"{o}dinger dynamics does not take account of any
measurement process, we need to comprehend what is expected to happen in the
relevant Schr\"{o}dinger functions.

\subsubsection{Phase destruction and flux termination}

First, to emphasize the linearity of the Schr\"{o}dinger functions, let us
assume in the double slit experiment that the measuring board is removed so
that the pieces of the Schr\"{o}dinger functions after the passage of two
slits can propagate freely. We consider two Schr\"{o}dinger packets, each
having passed through a different slit. Suppose that the total Schr\"{o}dinger
function is written as a superposition of each which is represented in the
Bohm form as
\begin{equation}
\psi(q,t)=R_{1}(q,t)\exp\left(  \frac{i}{\hbar}S_{1}(q,t)\right)
+R_{2}(q,t)\exp\left(  \frac{i}{\hbar}S_{2}(q,t)\right)  . \label{wa}%
\end{equation}
Since the associated quantum flux (current of the probability density) is
given\cite{Schiff,Hanasaki-flux} as%
\begin{equation}
\vec{j}=\frac{i}{2m}\left[  \psi^{\ast}(t)\vec{\nabla}\psi(t)-\left(
\vec{\nabla}\psi^{\ast}(t)\right)  \psi(t)\right]  , \label{fluxSchiff}%
\end{equation}
the total flux arising from $\psi(q,t)$ at any $q$ and $t$ is written down as
\begin{align}
-m\vec{j}  &  =R_{1}(q,t)^{2}\ \vec{\nabla}S_{1}+R_{2}(q,t)^{2}\ \vec{\nabla
}S_{2}\nonumber\\
&  +\hbar\left[  R_{2}(q,t)\vec{\nabla}R_{1}(q,t)-R_{1}(q,t)\vec{\nabla}%
R_{2}(q,t)\right] \nonumber\\
&  \times\sin\left(  \frac{1}{\hbar}\left(  S_{1}-S_{2}\right)  \right)
\nonumber\\
&  +R_{1}(q,t)R_{2}(q,t)\left[  \vec{\nabla}S_{1}+\vec{\nabla}S_{2}\right]
\nonumber\\
&  \times\cos\left(  \frac{1}{\hbar}\left(  S_{1}-S_{2}\right)  \right)  ,
\label{intf}%
\end{align}
which obviously indicates that when $R_{1}$ and $R_{2}$ are well separated
from each other to an extent satisfying%
\begin{equation}
R_{2}(q,t)\vec{\nabla}R_{1}(q,t)-R_{1}(q,t)\vec{\nabla}R_{2}(q,t)=0
\label{RR1}%
\end{equation}
and%
\begin{equation}
R_{1}(q,t)R_{2}(q,t)=0, \label{RR2}%
\end{equation}
it results that%
\begin{equation}
-m\vec{j}=R_{1}(q,t)^{2}\ \vec{\nabla}S_{1}+R_{2}(q,t)^{2}\ \vec{\nabla}S_{2}.
\label{fluxmixed}%
\end{equation}
Therefore the total flux is simply the sum of those of each one, as though
nothing had happened even after their once-overlapping.

Meanwhile, the interference pattern in the double-slit experiment can arise
when we perform a measurement with the sensing board, which destroys the phase
of the Schr\"{o}dinger function. An interference pattern actually appears only
when the board is set at a place where the overlap is materialized in Eq.
(\ref{intf}). Otherwise the packets simply come across and leave from each
other as though they were mutually transparent.

Then let us assume that a Schr\"{o}dinger function $\psi(q,t)$ is being
captured on an imaging screen where the dynamical phase of $\psi(q,t)$ is
destroyed. Phase randomization is surely one of the origins of
classicalization of quantum dynamics.\cite{Yaffe} Write a Schr\"{o}dinger
function $\psi(q,t)$ here again in a form a la Bohm, with $R$ and $S_{B}$ the
amplitude and phase functions. Then the associated quantum flux of Eq.
(\ref{fluxSchiff})is reduced in a rather classical form to%

\begin{equation}
\vec{j}(q,t)=\left\vert R(q,t)\right\vert ^{2}\frac{1}{m}\nabla S_{B}(q,t),
\label{flux3}%
\end{equation}
where $m$ indicates again the mass of a particle under study. The term $\nabla
S_{B}(q,t)/m$ is a quantum analog of the classical velocity represented by the
Schr\"{o}dinger function. If a destructive interference diminishes the phase
$iS_{B}(q,t)/\hbar$, then $\vec{j}(q,t)$ is also reduced to zero. The quantum
mechanical equation of continuity freezes the density $\rho(q,t)$ by
\begin{equation}
\frac{d}{dt}\left\vert \psi(t)\right\vert ^{2}=-\vec{\nabla}\cdot\vec{j}%
\simeq0 \label{flux}%
\end{equation}
after $t\sim t_{x},$ which is the phase destruction time. Thus $\left\vert
\psi(q(t_{x}),t_{x})\right\vert ^{2}$ should be left as a static quantity
after $t_{x}$. In our Gaussian dynamics of ADF of Eq. (\ref{CD1}), $S_{B}$ is
essentially%
\begin{equation}
\frac{1}{\hbar}S_{B}(q,t)\simeq\frac{1}{\hbar}S_{cl}(q,t)-\frac{d(t)}%
{c^{2}(t)+d^{2}(t)}(q-q_{cl}(t))^{2} \label{Bohm-ADF}%
\end{equation}
and therefore $\nabla S_{B}(q,t)=0$ can be realized not only when $\nabla
S_{cl}(q,t)=p_{cl}(t)=0$ but in case where the stationarity holding $d(t)=0$
is simultaneously realized. Therefore, the turning-point condition
$p_{cl}(t)=0$ and $c(t)=0$, at which the semiclassical Schr\"{o}dinger
function such as Eq. (\ref{MG}) is known to diverge, is not enough to
eliminate the flux. To distinguish the capture of a particle and the possible
turning-back at the measuring board, the condition $p_{cl}(t)=0$ and $c(t)=0$
is insufficient. We consider this aspect next.

\subsubsection{Divergence of the semiclassical Schr\"{o}dinger function and
quantum smoothing at caustics}

Suppose that one piece of the Gaussians reaches a measurement panel of the
double slit experiment. The analysis so far made suggests that there is no
mathematical mechanism that makes the Gaussians packet collapse to a spot as a
delta function even at caustics and/or turning points. We therefore need to
study rather precisely the semiclassical divergence of the Schr\"{o}dinger
function at caustics, focal points, and turning points to confirm how
different it is from the \textquotedblleft instantaneous
collapse\textquotedblright.

Equation (\ref{ADF2}) shows that a semiclassical wavepacket diverges at points
of $\sigma\left(  t\right)  =0$. These singularities commonly appear in the
asymptotic relation in between classical and quantum mechanics. In the present
context, such divergence is to be smoothed to a finite function by the
diffusion term.\cite{Paper-II} In the Gaussian dynamics with the momentum
gradient only, we may set $d\left(  t\right)  =0$ at the outset in Eq.
(\ref{CD1}), and the divergence condition is essentially equivalent to
$c\left(  t\right)  =0$, since Eq. (\ref{CN95}) results in \
\begin{equation}
c\left(  t\right)  =\frac{c\left(  0\right)  }{\sigma\left(  0\right)  ^{2}%
}\sigma\left(  t\right)  ^{2}. \label{ctct}%
\end{equation}
Yet, the Wronskian relation, Eq. (\ref{Wronskian}) suggests that the quantum
diffusion should prevent those singularities. More precisely, the following
mathematical mechanism works: Let $t^{\ast}$ be the time of a caustic point is
realized, that is, $\sigma\left(  t^{\ast}\right)  =0$ and $c\left(  t^{\ast
}\right)  =0$. Then we may expand $\sigma\left(  t\right)  $ around the time
$t^{\ast}$ as%
\begin{equation}
\sigma\left(  t^{\ast}+\varepsilon\right)  =\sigma_{1}\varepsilon+\sigma
_{2}\varepsilon^{2}+\sigma_{3}\varepsilon^{3}+\cdots
\end{equation}
Then Eq. (\ref{ctct}) gives the leading term of $c\left(  t^{\ast}%
+\varepsilon\right)  $ as
\begin{equation}
c\left(  t^{\ast}+\varepsilon\right)  \simeq\frac{c\left(  0\right)  }%
{\sigma\left(  0\right)  ^{2}}\sigma_{1}^{2}\varepsilon^{2}.
\end{equation}
By defining $d\left(  t\right)  =\sigma\left(  t\right)  ^{2}f\left(
t\right)  $, we find%
\begin{equation}
\text{ \ \ }\dot{f}\left(  t\right)  =\frac{2\hbar}{m}\sigma\left(  t\right)
^{-2}\text{ \ \ \ and \ \ \ }f\left(  t^{\ast}+\varepsilon\right)
\simeq-\frac{2\hbar}{m\sigma_{1}^{2}}\frac{1}{\varepsilon}. \label{CN162}%
\end{equation}
and therefore have
\begin{equation}
d\left(  t^{\ast}+\varepsilon\right)  \simeq-\frac{2\hbar}{m}\varepsilon.
\label{d2}%
\end{equation}
The real part of the exponent of the Gaussian of Eq. (\ref{CD1})\ accordingly
behave as%
\begin{equation}
\operatorname{Re}\frac{1}{c\left(  t^{\ast}+\varepsilon\right)  +id\left(
t^{\ast}+\varepsilon\right)  }\rightarrow\frac{c\left(  0\right)  }%
{\sigma\left(  0\right)  ^{2}}\sigma_{1}^{2}\left(  \frac{m}{2\hbar}\right)
^{2}\text{ \ \ as\ }\varepsilon\rightarrow0, \label{remove}%
\end{equation}
which is finite, where $c\left(  0\right)  $ and $\sigma\left(  0\right)  $
are their initial values at $t=0$. Thus, it turns out that the Gaussian does
not collapse to a $\delta$-function, as long as it is subject to the full
quantum dynamics. It is smoothed from a delta-function to a sharp Gaussian.

Note also that the colliding of a particle at a turning point makes the
Gaussian very sharp only to the direction normal to the measuring board. In
the other directions parallel to the board, such a semiclassical divergence
does not take place.

\subsubsection{Possible finite-time shrink of the Gaussian packet by acquiring
an effective mass}

The Tonomura experiment\cite{Tonomura1989} of \textquotedblleft double-slit
experiment\textquotedblright\ is set such that when an electron comes across
the measuring panel or a set of apparatus like the Charge-Coupled Device (CCD)
sensors, which emit photons by electron impact. Collision of electrons with
the fluorescent molecules is essentially a matter of quantum theory of
electron scattering by molecules.\cite{Takatsuka1984,McKoy} This is simply a
quantum process of electron attachment photoemission.

When a Gaussian\ happens to be captured by a sensor, it is imagined to happen
\begin{equation}
\sigma\left(  t\right)  \rightarrow0,
\end{equation}
due to $p_{cl}(t)=0$, but as discussed above the instantaneous collapse of the
semiclassical wavepacket cannot happen as a matter of fact. Instead, the
capture of electron by the CCD molecules can be viewed as a process that the
electron acquires a huge effective mass of the set of sensing molecules.
Symbolically a process
\begin{equation}
m\rightarrow\text{huge,} \label{mass}%
\end{equation}
is realized in some short yet finite time. This is somewhat like the notion in
\textit{heavy electron} theory of electron
(super)conductivity.\cite{steglich1979superconductivity,fisk1986heavy,yang2012emergent}
Then, with the huge effective mass, it follows that $\zeta(t)\rightarrow0$ in
Eq. (\ref{Wronskian}) and consequently $d(t)\rightarrow0$ in Eq. (\ref{zeta}).
[The behavior of this $d(t)\rightarrow0$ due to $m\rightarrow\infty$ should be
distinguished from $d(t)\rightarrow0$ due to Eq. (\ref{d2}). $d(t)\rightarrow
0$ is realized independently of $c(t)\rightarrow0.$] Hence $d(t)\rightarrow0$
and $c(t)\rightarrow0$ works to freeze the quantum phase in Eq.
(\ref{Bohm-ADF}) at the absorbing place. Thus the quantum flux is terminated
not only in the direction normal to the measurement panel but in all the
directions in Eq. (\ref{flux}), which fixes $\left\vert \psi(t)\right\vert
^{2}$ at classical measurement point. Therefore, the real part of the Gaussian
exponent is expected to behave as%
\begin{equation}
\frac{c(t)}{c(t)^{2}+d(t)^{2}}\rightarrow\text{huge,} \label{mass2}%
\end{equation}
which is confirmed in Eq. (\ref{remove}) by putting $m\rightarrow\infty$ .
Then the \textquotedblleft mesoscopic\textquotedblright\ Gaussian should look
\begin{equation}
\left\vert G_{\gamma}(q-q\left(  t\right)  ,t)\right\vert ^{2}\rightarrow
\left\vert N^{\prime}(t_{x})\right\vert ^{2}\triangle(q-q(t_{x}),t_{x}),
\label{schrink}%
\end{equation}
where $\triangle(x)$ is a function narrowly-localized in vicinity of $x$. Only
the molecule(s) within the range $\triangle(x)$ are supposed to emit light,
leaving a spot. This describes, only symbolically, a finite-time
shrink-in-space of the Schr\"{o}dinger\ distribution function.

\subsection{Summary of this section: Path branching in deep quantum dynamics}

We have studied the mechanism of time-propagation of the Schr\"{o}dinger
function from the view point of a relationship between the quantum
distribution function and classical paths and their branching. First we have
separated the classical component that satisfies the Hamilton-Jacobi equation
out of a Schr\"{o}dinger function and factored what we call the action
decomposed function (ADF). With ADF we revisited semiclassical mechanics, in
which the rescaling of the configuration space along with the emergence of
Maslov phase are highlighted. To proceed further in analytic manner, we chose
a complex exponent Gaussian function as an ADF. Although the Gaussian center
runs along a classical path, its complex exponents undergo a very
characteristic dynamics under the coupling of the configuration-space
rescaling and the quantum diffusion. Those hidden dynamics dominate the
dynamics of the Gaussian exponents.

The quantum diffusion has already brought a Gaussian deep into the quantum
realm. However, the single Gaussian representation of the Schr\"{o}dinger
function is far from sufficient to reproduce the wave-like behavior of the
coherent distribution function, which are bifurcation, diffraction, and those
that arise from the Huygens principle in quantum
mechanics.\cite{Goussev-Huygens} We hence introduce a mathematical mechanism
to transform a Gaussian into a continuous set of daughter Gaussians\ having
different exponents. With this transform a Gaussian can split to smaller
Gaussians, and each local path is branched to infinitely many pieces. A
Schr\"{o}dinger function is thus time-propagated by a series of branching
paths, each carrying a piece of complex Gaussian. Thus, as the quantum nature
of the system becomes deeper, the paths dissolves into the quantum sea as
finer pieces. The manner of the entire propagation can be represented in the
form of path integrals over those branching pieces.

\section{Concluding remarks \label{sec:Conc}}

We have studied the nature of the Schr\"{o}dinger dynamics and Schr\"{o}dinger
functions: First in Sec. \ref{sec:RealValue}, we factorized the distribution
function of particles $\rho(q,t)$ to a two component real-valued vector (the
Schr\"{o}dinger vector), and imposing the space-time translational symmetry
and the equation of continuity (flux conservation) under a given Hamiltonian.
Then the variational condition applied to the equation of continuity chooses
the Schr\"{o}dinger vector as a real-valued function that gives the most
robust (probable) distribution among those satisfying the equation continuity.
The real-valued Schr\"{o}dinger vector is readily transformed to the ordinary
complex-valued counterpart. The physical meaning of the Schr\"{o}dinger
function is therefore regarded as a distribution function by definition. It
was also found that the Schr\"{o}dinger equation is internally consistent. The
present reconstruction of the Schr\"{o}dinger equation covers only the minimal
arena of non-relativistic quantum theory. Yet, we believe we have come to the
basic understanding and interpretation of the Schr\"{o}dinger dynamics.

It is well known that the Schr\"{o}dinger function does not describe the
dynamics of the individual single incident (event) each, but it represents a
coherent superposition of an infinite (and inseparable) ensemble of possible
physical events. In the double-slit experiment electrons are launched
one-by-one and the accumulated spots made by each electron form a fringe
pattern on a measurement board. The fringe pattern is predicted by the
Schr\"{o}dinger function (the absolute square thereof), while we have
formulated in Sec. \ref{sec:Stochastic} a quantum path named one-world
stochastic path that each individual electron is supposed to run on. Both the
Schr\"{o}dinger function and one-world paths are extremely sensitive to the
external perturbation. The one-world quantum paths (indifferentiable almost
everywhere) are taken out of the Feynman-Kac formalism and the Schr\"{o}dinger
equation. The stochastic motion seems to be essential and inherent in quantum
dynamics. In fact, it has been shown that the scaling law with an appropriate
diffusion constant naturally leads to the Bohr quantization condition for
hydrogen atom. We also have shown that the one-world paths loses the quantum
nature and the quantum canonical equations of motion for them are reduced to
the Hamilton (classical) canonical equations of motion as soon as the Wiener
process is removed. In order to drive the one-world quantum dynamics, each
path requires the information of the drift velocity, which is only to be
provided by the Schr\"{o}dinger function in turn. Hence, the individual
one-world path runs in a stochastic manner and guided by the drift velocity
given by the Schr\"{o}dinger function. The nonlinear relation between the
one-world dynamics (local information) and the Schr\"{o}dinger dynamics
(coherent sum of the possible global events) has been identified as the very
deepest origin of the mystery inherent to quantum mechanics.

In the last half of this article from Sec. \ref{sec:Variational} and Sec.
\ref{sec:ADFpath}, we have analyzed the physical and mathematical processes of
time-propagation of the Schr\"{o}dinger function from the view point of two
classes of path dynamics.

Classical trajectories run on a Lagrange manifold having the symplectic
structure, thereby conserving not only the energy by phase-space volume
elements (Poincar\'{e}-Cartan). Having an anticipation that quantum mechanics
should have the similar structure, we have developed the quantum mechanical
Maupertuis-Hamilton variational principle by mapping the Schr\"{o}dinger
function on a parameter space. Yet, quantum mechanics demands another
variational functional emerging from the principle of flux conservation (or
the equation of continuity). The deterministic paths run on these manifolds of
symplectic structure as a solution of the resultant ordinary differential
equations. Thus the time-propagation of the Schr\"{o}dinger distribution
function turns out to be compatible with the path-dynamics in the parameter
space. The similarity between classical and quantum dynamics on the manifold
of symplectic property comes from their common principle of the conservation
laws. In contrast, the most significant difference between the two is that
quantum mechanics requires the flow dynamics and the relevant variational
functional. Another difference is that even a single trajectory in the
parameter space allows the Schr\"{o}dinger function to simultaneously pass
through two slits in the double-slit experiment.

As the last step, we have studied the mechanism of time-propagation of the
Schr\"{o}dinger function through the intricate relationship between quantum
and classical dynamics. By investigating the necessary role and limitation of
the classical paths, we have captured how the space-time distribution of the
Schr\"{o}dinger function is materialized. Semiclassical mechanics based on the
complex Gaussian function reveals hidden variables in the dynamics.
Furthermore, a Schr\"{o}dinger function running on a non-smooth potential
function of short reach-range shows the wave-like properties based on the
Huygens-like principle, such as diffraction, bifurcation, tunneling, and so
on. For a Gaussian to materialize those dynamics, it needs to split and
proliferate to smaller pieces. This is the final end of the roles of classical
trajectory. We have studied this aspect in terms of the Weierstrass transform.
Such branching paths and Gaussians are summarized in term of a path integral.
Thus the classical paths repeat branching and dissolve into the deep sea of
the genuine quantum dynamics, ending up with full quantum dynamics.

The Schr\"{o}dinger equation allows for physically interesting applications
through approximate solutions to the partial differential equations, even
without fully understanding its essence, the meaning of the wave function, the
physical consistency within the theory, or why imaginary numbers should appear
in its basic structure. The 100 years of quantum mechanics has thus worked
very efficiently. Nevertheless, in reality, many individual scientists still
face fundamental questions on quantum mechanics. The author hopes that the
present article could share a part of answers to those general questions.

\bigskip

\begin{acknowledgments}
The author thanks Prof. Satoshi Takahashi and Dr. Kota Hanasaki for valuable
discussions and comments. This work has been supported by JSPS KAKENHI, Grant
No. JP20H00373 and JP24K01435.
\end{acknowledgments}

% \bibliography{QuantMech}


\end{document}